\documentclass[useAMS]{mn2e}

\usepackage{amssymb,amsmath,psfig,times}
\def\gsim{ \lower .75ex \hbox{$\sim$} \llap{\raise .27ex \hbox{$>$}} }
\def\lsim{ \lower .75ex\hbox{$\sim$} \llap{\raise .27ex \hbox{$<$}} }

\def\sc{Schwarzschild}
\def\beq{\begin{equation}}
\def\eeq{\end{equation}}

\def\sc{Schwarzschild}

\title[FSRQs and BL Lacs]
{The transition between BL Lac objects and Flat Spectrum Radio Quasars}  
\author[G. Ghisellini et al.]
{G. Ghisellini$^1$\thanks{Email: gabriele.ghisellini@brera.inaf.it}, 
F. Tavecchio$^1$, 
L. Foschini$^1$,
G. Ghirlanda$^1$
\\
$^1$INAF -- Osservatorio Astronomico di Brera, Via Bianchi 46, I--23807 Merate, Italy\\
}

\begin{document}  

\maketitle

\begin{abstract}
We study the BL Lac objects detected in the one year all sky survey of the 
{\it Fermi} satellite, with a energy spectral slope $\alpha_\gamma$ in the 
[0.1--100 GeV] band greater than 1.2.
In the $\alpha_\gamma$ vs $\gamma$--ray luminosity plane,
these BL Lacs occupy the region populated by Flat Spectrum Radio Quasars (FSRQs).
Studying the properties of their spectral energy distributions (SED)
and of their emitting lines, we find that several of these BL Lacs have a SED similar to 
FSRQs and that they do have broad
lines of large equivalent width, and should be reclassified as FSRQs
even adopting the current phenomenological definition (i.e. equivalent width EW of the
emitting line greater than 5 \AA).
In other cases, even if the EW width is small, the emitting lines can be as luminous
as in quasars, and again their SED is similar to the SED of FSRQs.
Sources classified as BL Lacs with a SED appearing as intermediate between BL Lacs and FSRQs
also have relatively weak broad emission lines and small EW, and can be considered
as transition sources.
These properties are confirmed also by model fitting, that allows to derive the relevant
intrinsic jet parameters and the jet power. 
This study leads us to propose a physical distinction between 
the two classes of blazars, based on the luminosity of the broad line region measured
in Eddington units. 
The dividing line is of the order of $L_{\rm BLR}/L_{\rm Edd}\sim 5\times 10^{-4}$,
in good agreement with the idea that the presence of strong emitting lines is related
to a transition in the accretion regime, becoming radiatively inefficient below
a disk luminosity of the order of one per cent of the Eddington one.
\end{abstract}
\begin{keywords}
BL Lacertae objects: general --- quasars: general ---
radiation mechanisms: non--thermal --- gamma-rays: theory --- X-rays: general
\end{keywords}

\section{Introduction}

Among the blazars detected by the Large Area Telescope (LAT) onboard the {\it Fermi} 
satellite after 11 months of all sky survey (Abdo et al. 2010a, hereafter A10) 
there are roughly an equal number of sources identified as BL Lac objects 
and Flat Spectrum Radio Quasars (FSRQs).
The corresponding catalog of AGN detected at high Galactic latitude 
($|b| > 10^\circ$) is called First LAT AGN Catalog (1LAC).
In general, the LAT--detected BL Lac objects and Flat Spectrum Radio Quasars (FSRQs) 
separate quite well in the $\gamma$--ray spectral index -- $\gamma$--ray luminosity plane 
($\alpha_\gamma-L_\gamma$, where $\alpha_\gamma$ is the energy spectral index),
in agreement with the early results borne out with the 3--months all sky survey
of {\it Fermi}/LAT, which contained only 1/7 of the blazars in the 1LAC catalog
(Ghisellini, Maraschi \& Tavecchio 2009, hereafter GMT09).
On the other hand, there are a number of sources, classified as BL Lac objects,
located in the region of the plane preferentially ``inhabited" by FSRQs:
these are BL Lacs with a relatively steep spectrum (i.e. $\alpha_\gamma>1.2$).
These ``intruders" has been classified as BL Lac objects on the basis of the 
``historical" distinction among BL Lacs and FSRQs, i.e. 
by means of the equivalent width (EW) of their emission lines (see e.g. Urry \& Padovani 1995).
Objects with a rest frame EW$<$5 \AA\ are called BL Lacs.
This definition has the obvious advantage of being simple and of immediate
use for an observational characterization of the object.
On the other hand, the optical continuum of most blazars is relativistically
enhanced by beaming, and very variable. 
In several cases a small EW does not imply emission lines of 
low luminosity, being simply the result of a particularly
beamed non--thermal continuum. 
On the opposite side, EW greater than
5 \AA\ may be the results of a particularly low state of the beamed
continuum in a source of intrinsically weak lines.
A division based on the EW of emission lines does measure the 
relative importance of the beamed non--thermal continuum and
the underlying thermal emission, but after the discovery that most of the 
non--thermal emission is at $\gamma$--ray energies, we know that 
the optical non--thermal flux very often is a minor contribution
to the total, bolometric, non--thermal output.
Therefore the EW alone is not a good indicator of the relative
importance of the two contributions.

Up to now, we construct samples of BL Lac as well as of FSRQs in order
to study their properties and their possible differences, and adopt
the classical, EW--based, sub--division.
If the aim is to study intrinsically physical properties, this may be 
dangerous, since with the EW classification we may -- for instance --
classify as a BL Lac object a source with very luminous lines,
typical of a FSRQ, only because at the time of the spectroscopic
observations leading to the measurement of the EW the optical
non--thermal flux was particularly intense.
For illustration, let us take the case of PKS 0208--512.
It has an observed MgII emission line of EW$\sim$5 \AA\ (2.5 in the rest frame), 
whose luminosity
is close to $10^{44}$ erg s$^{-1}$, stronger than in some FSRQs.
This object is classified as a BL Lac, but all its physical properties are
resembling FSRQs.

We therefore believe that a new classification scheme is needed,
based on a physical property of the source.
We suggest a division based on the luminosity of the broad emission lines,
normalized to the corresponding Eddington luminosity, 
the natural luminosity--scale.
Normalizing in this way allows to compare objects of different black hole
masses.
This division implies to estimate the black hole mass,
that it is not a direct observable quantity.
On the other hand, in recent years, the establishing of correlations between 
i) the luminosity of the bulge of the host galaxy and the black hole mass 
(Magorrian et al. 1998; Bentz et al. 2009);
ii) the correlation between the dispersion velocity and the black hole mass
(Ferrarese \& Merritt 2000; G\"ultekin et al. 2009), and 
iii) the correlation between the luminosity of the continuum at selected
frequencies and the size of the Broad Line Region (BLR, Vestergaard 2002; 
Decarli et al. 2010 and references therein), 
made the estimate of the black hole mass much more affordable.
Furthermore, in specific cases, very powerful blazars do have their IR--optical--UV
continuum dominated by a thermal component produced by their accretion disk:
modelling it with a standard Shakura--Sunyaev (1973) disk allows
to find both the black hole mass and the accretion rate.

We then investigate if the ``intruder" BL Lacs in the
$\alpha_\gamma$--$L_\gamma$ plane
have intrinsically weak emission lines (in Eddington units)
or if instead their EW is only a consequence of a particularly 
enhanced non--thermal continuum, or else if they
are transition objects, with intermediate values of the broad line
luminosity.

We use a cosmology with $h=\Omega_\Lambda=0.7$ and $\Omega_{\rm M}=0.3$,
and use the notation $Q=10^X Q_x$ in cgs units (except for the black hole masses,
measured in solar mass units).

\begin{table*} 
\centering
\begin{tabular}{lllllll}
\hline
\hline
{\it Fermi} Name        &Coord (J2000.0) &Alias      &$z$    &$\Gamma_\gamma$       &$F_{\rm \gamma}$ &$\log L_{\rm \gamma}$\\
\hline 
1FGL J0058.0+3314       &00 58 32.07  +33 11 17.2 &GB6 0058+3311  &1.371  &2.33$\pm$0.11  &3.13  &47.36 \\
1FGL J0112.0+2247       &01 12 05.82  +22 44 38.8 &TXS 0109+224         &0.265  &2.23$\pm$0.05  &7.81  &45.99 \\
{\it 1FGL J0210.6-5101} &02 11 13.18  +10 51 34.8 &{\it PKS 0208--512}  &1.003  &2.37$\pm$0.04  &14.59 &47.69  \\
1FGL J0522.8-3632       &05 22 57.98 --36 27 30.9 &PKS 0521--36         &0.055  &2.60$\pm$0.06  &11.54 &44.45 \\
{\it 1FGL J0538.8-4404} &05 38 50.35 --44 05 08.7 &{\it PKS 0537--441}  &0.892  &2.27$\pm$0.02  &37.77 &48.00  \\
1FGL J0557.6-3831       &05 58 06.47 --38 38 31.7 &PMN 0558--3839       &0.302  &2.32$\pm$0.17  &1.74  &45.44 \\
1FGL J0757.2+0956       &07 57 06.64  +09 56 34.9 &PKS 0754+100         &0.266  &2.39$\pm$0.08  &4.86  &45.73  \\
1FGL J0811.2+0148       &08 11 26.71  +01 46 52.2 &PKS 0808+019         &1.148  &2.45$\pm$0.12  &2.97  &47.08 \\
1FGL J0831.6+0429       &08 31 48.88  +04 29 39.1 &PKS  0829+046        &0.174  &2.50$\pm$0.07  &7.35  &45.39  \\
1FGL J0854.8+2006       &08 54 48.87  +20 06 30.6 &OJ 287               &0.306  &2.38$\pm$0.07  &7.03  &45.18  \\
1FGL J0910.7+3332       &09 10 37.04  +33 29 24.4 &TON  1015            &0.354  &2.32$\pm$0.14  &2.00  &45.66  \\
1FGL J1000.1+6539       &09 58 47.25  +65 33 54.8	&TXS 0954+658         &0.367  &2.51$\pm$0.16  &2.59  &45.69  \\
1FGL J1012.2+0634       &10 12 13.35  +06 30 57.2 &PMN 1012+0630        &0.727  &2.30$\pm$0.2   &1.51  &46.55  \\
1FGL J1027.1-1747       &10 26 58.52 --17 48 58.5 &BZB 1026--1748$^*$   &0.114  &2.32$\pm$0.29  &1.22  &44.62  \\
{\it 1FGL J1058.1-8006} &10 58 43.40 --80 03 54.2 &{\it PKS 1057--79}   &0.581  &2.45$\pm$0.1   &6.26  &46.66 \\
1FGL J1150.2+2419       &11 50 19.21  +24 17 53.9 &B2 1147+24$^*$       &0.2?   &2.25$\pm$0.12  &2.08  &45.17  \\
1FGL J1204.3-0714       &12 04 16.66 --07 10 09.0 &WGA 1204.2--0710$^*$ &0.185  &2.59$\pm$0.23   &2.07 &44.99 \\
1FGL J1341.3+3951       &13 41 05.10  +39 59 45.4 &B2 1338+40           &0.172  &2.45$\pm$0.21  &1.29  &44.94  \\
1FGL J1522.6-2732       &15 22 37.68 --27 30 10.8 &PKS 1519--273        &1.294  &2.25$\pm$0.08  &4.94  &47.55  \\
1FGL J1558.9+5627       &15 58 48.29  +56 25 14.1 &TXS 1557+565$^*$     &0.3    &2.24$\pm$0.13  &2.91  &45.73  \\
{\it 1FGL J1751.5+0937} &17 51 32.82  +09 39 00.7 &{\it PKS 1749+096}   &0.322  &2.29$\pm$0.05  &12.22 &46.43  \\
1FGL J1800.4+7827       &18 00 45.68  +78 28 04.0 &{\it S5  1803+78}    &0.68   &2.35$\pm$0.07  &6.24  &46.94 \\
1FGL J1807.0+6945       &18 06 50.68  +69 49 28.1 &3C 371               &0.05   &2.60$\pm$0.08  &7.70  &44.29 \\
1FGL J2006.0+7751       &20 05 31.00  +77 52 43.2 &S5 2007+77           &0.342  &2.42$\pm$0.16  &3.00  &45.81  \\
{\it 1FGL J2202.8+4216} &22 02 43.29  +42 16 40.0 &{\it BL LAC}         &0.069  &2.38$\pm$0.04  &16.81 &44.97  \\
1FGL J2217.1+2423       &22 17 00.83  +24 21 46.0 &B2 2214+24           &0.505  &2.63$\pm$0.12  &4.97  &46.36    \\
1FGL J2243.1-2541       &	22 43 26.36 --25 44 27.0 &PKS 2240--260        &0.774  &2.32$\pm$0.09  &3.44  &46.75  \\
1FGL J2341.6+8015       &23 40 54.28  +80 15 16.1 &FRBA J2340+8015      &0.274  &2.21$\pm$0.08  &4.21  &45.83  \\
\hline
{\it 1FGL J0238.6+1637} &02 38 38.93  +16 36 59.3 &{\it PKS 0235+164}   &0.94   &2.14$\pm$0.02  &43.4  &48.24 \\
{\it 1FGL J0428.6-3756} &04 28 40.42 --37 56 19.6 &{\it PKS 0426--380}  &1.111  &2.13$\pm$0.02  &31.5  &48.18 \\
\hline
\hline 
\end{tabular}
\vskip 0.4 true cm
\caption{$F_\gamma$ in the LAT band (0.1--100 GeV) in units of $10^{-8}$ ph cm$^{-2}$ s$^{-1}$.
$L_\gamma$, in the same band, is k--corrected and in units of erg s$^{-1}$.
$^*$: no {\it Swift} observations. 
Sources whose name is in italics are present in Ghisellini et al. (2010a, hereafter G10),
and some of them are present in Tavecchio et al. (2010).
}
\label{sample}
\end{table*}

\begin{figure*}
\vskip -0.6cm \hskip -0.4 cm
\psfig{figure=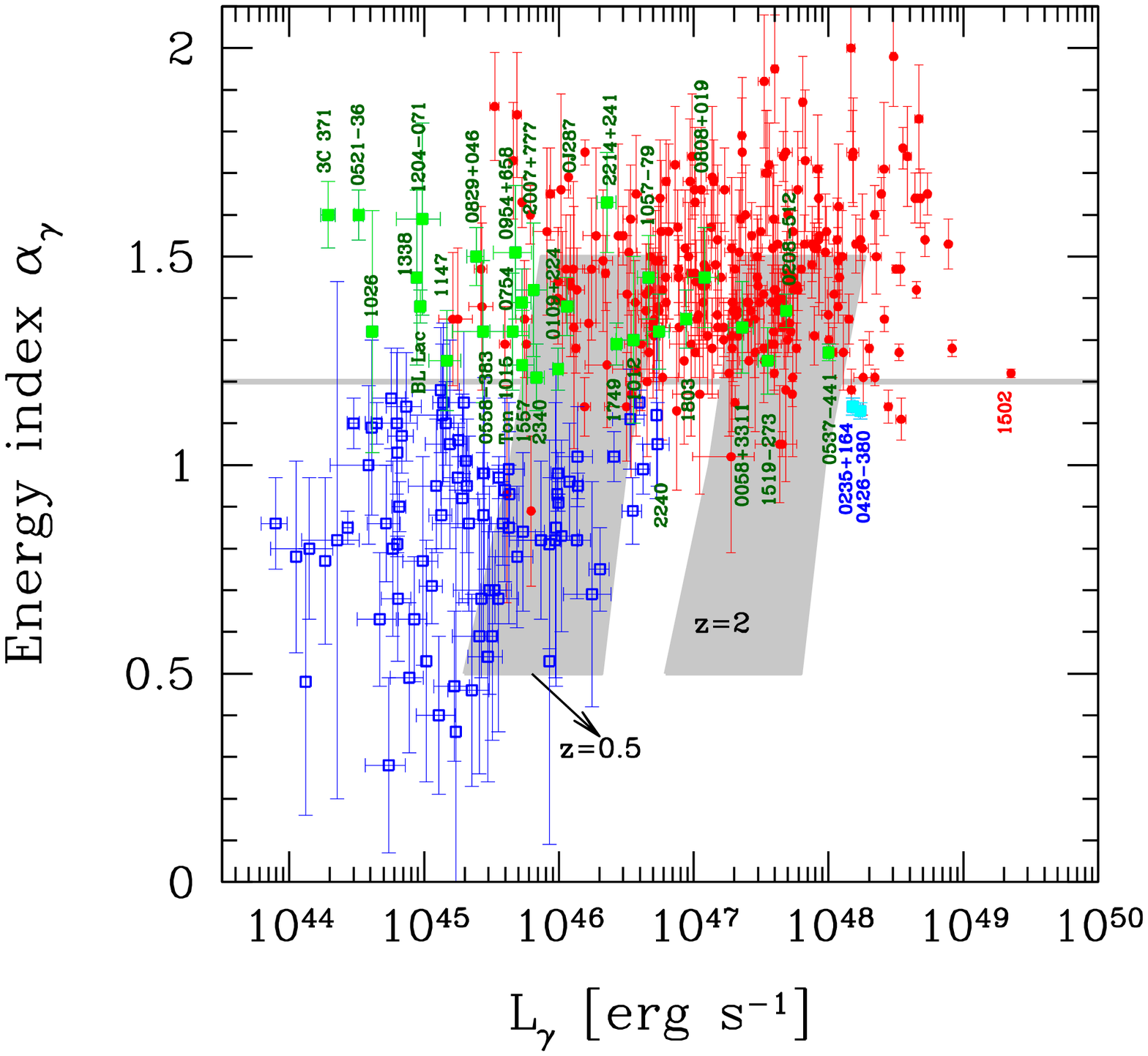,width=18.5cm,height=18.5cm}
\vskip -0.8 cm
\caption{The energy spectral index $\alpha_\gamma$ as a function
of the $\gamma$--ray luminosity $L_\gamma$ in the band [0.1--10 GeV] 
for all blazars with known redshift present in the 1LAC sample.
The filled circles (red in the electronic version) are FSRQs; empty
(blue) squares are BL Lacs with $\alpha_\gamma<1.2$, and fileld
(green) squares are sources classified as BL Lacs in the 1LAC sample
with $\alpha_\gamma>1.2$.
In addition, the two larger (cyan) circles are 0235+164 and 0426--380,
classified as BL Lacs, that have $L_\gamma>10^{48}$ erg s$^{-1}$.
The horizontal grey line marks $\alpha_\gamma=1.2$.
The two grey regions illustrate how the corresponding grey area
shown in Fig. \ref{fluxalpha} would lie assuming a redshift of 0.5 or 2,
as indicated.
}
\label{divide}
\end{figure*}

\section{The Fermi blazars' divide one year after}

Fig. \ref{divide} shows the energy spectral index $\alpha_\gamma$ as a function 
of the (K--corrected, see GMT09) 
$\gamma$--ray luminosity $L_\gamma$ for the FSRQs and BL Lacs on the 1LAC sample
of known redshift.
This figure can be compared with the same one in GMT09 reporting the bright blazars of the LAT 
Bright AGN Sample (LBAS sample (Abdo et al. 2009, hereafter A09) for the 3--months all sky survey.
In that figure there was a specific $\gamma$--ray luminosity dividing
FSRQs and BL Lacs, around a few times 10$^{46}$ erg s$^{-1}$, interpreted
as a consequence of the changing accretion regime of the underlying accretion
disk from radiatively efficient to inefficient, or, in other words, from a standard 
Shakura \& Sunyaev (1973) disk to an ADAF (Advection Dominated Accretion Flow) one.
As remarked in GMT09, the appearance of the dividing luminosity required that all the bright 
blazars in the LBAS sample have black holes of the same mass 
and similar beaming factors.
This is approximately appropriate when considering the brightest sources, but when
the decreased limiting flux allows to explore smaller luminosities both for
FSRQs and BL Lacs, then it is likely that the corresponding black hole mass 
(and/or the beaming factor)
is smaller,
and the dividing luminosity gets spread into a larger range of values (as large as the spread
in black hole masses and beaming factors).
This explains why, in Fig. \ref{divide}, no dividing luminosity is present. 
A comparison with Fig. 1 of GMT09 shows that all the most luminous blazars were
present also in LBAS: the decreased flux limit did not led to discover any new 
more powerful object.
Furthermore, differently from Fig. 1 of GMT09, there is now no trend
between the minimum luminosities and spectral index, although, at low
luminosities, flat $\gamma$--ray spectra BL Lacs are more numerous than
sources with a relatively steep $\alpha_\gamma$.
This is likely due to LAT being more sensitive to flat spectra than to steep
ones (see Fig. 9 of A10).
As already noted in GMT09, a correlation between $\alpha_\gamma$ and $L_\gamma$
is not expected, since at low $\gamma$--ray luminosities we expected the detection 
of FSRQs of lower black hole masses and smaller beaming. 
This fills the top left part of the $\alpha_\gamma$--$L_\gamma$ plane.

The majority of BL Lac sources are characterized by a 
relatively flat $\alpha_\gamma$ ($\alpha_\gamma\lsim 1.2$),
but there are several exceptions.
Some of these BL Lacs, however, have been classified as such
because the equivalent width (EW) of their {\it broad} lines (that
are indeed present) is less than 5 \AA.
We discussed in GMT09 the cases of 
PKS 0537--441, AO 0235+164 and PKS 0426--380, that do have broad lines 
visible in their low emission states
(see Sbarufatti et al. 2005 for PKS 0426--380;
Pian et al. 2002 for PKS 0537--441; Raiteri et al. 2007a for AO 0235+164).
Another example is 0208--512: it was observed to have a MgII broad line with EW$\sim$5\AA\
(2.5\AA\ in the rest frame), but with a very large luminosity ($\sim 10^{44}$ erg s$^{-1}$;
Scarpa \& Falomo 1997).
Therefore 0208--512 (and the other mentioned sources) are FSRQs 
whose non--thermal continuum is enhanced so much
to make the very luminous broad lines to almost disappear, and not BL Lac
objects with genuinely weak lines.
Not appreciating this point may cause some confusion when comparing
FSRQs and BL Lacs.

With respect to the LBAS sample, the number of sources classified as BL Lacs
but of unknown redshift increased: from the source list in A10
there are 159 sources classified as ``BLL" with no redshift (in the ``clean" sample), 
excluding the sources classified as of ``unknown" type.

\begin{figure}
\vskip -0.6cm \hskip -0.4 cm
\psfig{figure=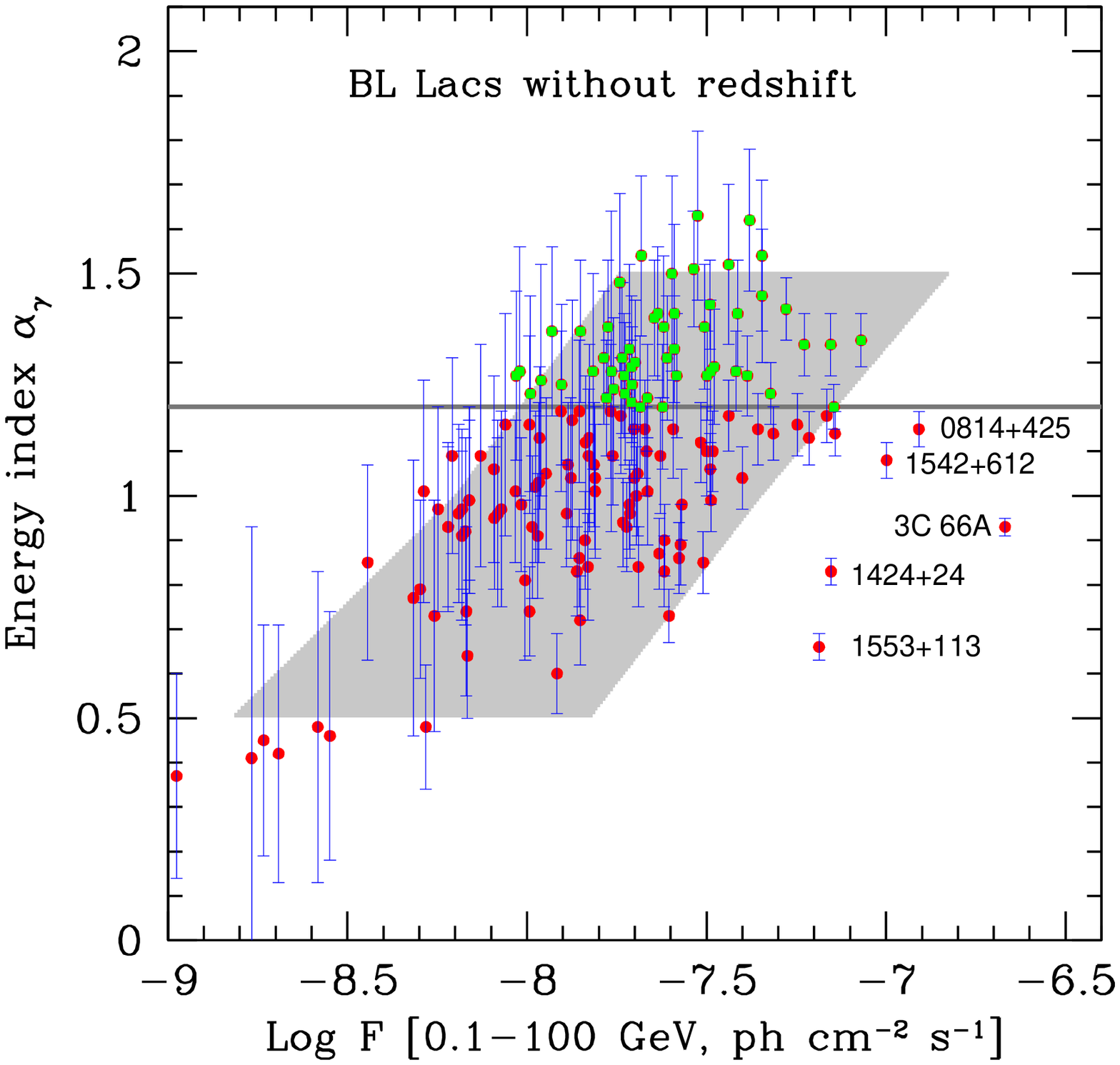,width=9.5cm,height=9.5cm}
\vskip -0.8 cm
\caption{
Energy index $\alpha_\gamma$ vs $\gamma$--ray flux in the
[0.1--100 GeV] energy band for the sources classified in the clean 1LAC 
catalog as ``BLL" (and excluding the ones classified as ``unknown").
The plotted [0.1--100 GeV] photon flux has been calculated from the 
[1--100 GeV] flux listed in A10, using the corresponding spectral index.
The grey region corresponds to the 2 grey areas in Fig. \ref{divide}.
The (green) squares are the 56 sources with $\alpha_\gamma\ge 1.2$
(this limit is shown by the horizontal line).
See text for a brief comment about the labelled sources
that are located outside the grey region.
}
\label{fluxalpha}
\end{figure}

In Fig. \ref{fluxalpha} we show the energy spectral index $\alpha_\gamma$ vs the 0.1--100 GeV 
photon flux of these 159 ``BLL" sources.
Of these 159 sources, 54 have $\alpha_\gamma\ge 1.2$
(80 if we include the sources with ``uncertain" classification).
Fig. \ref{fluxalpha} shows in grey the area where most of the sources are.
Exceptions at large photon fluxes are labelled, and we here briefly comment
about these sources.

PKS 1424+240 and PG 1553+113 have been detected at TeV energies (see
Ong et al. 2009; Teshima et al. 2009 for 1424+240, and 
Aharonian et al. 2006; Albert et al. 2007 for PG 1553+113; see also Prandini et al. 2010),
and very likely also 3C 66A (Acciari et al. 2009), although there can be a contamination
from the nearby 3C 66B radiogalaxy (see the discussion in Tavecchio \& Ghisellini 2009).
The redshift of 3C 66A is uncertain, even if a value of $z=0.444$ is
commonly used.
Due to the TeV detection, these 3 sources cannot lie at very large redshift
although $z$ up to $\sim$0.7 would be possible (see Tavecchio et al. 2010
for PKS 1424+240 modelled assuming $z=0.67$).

For B3 0814+425 the NED database gives $z=0.53$, quoting Sowards--Emmerd et al. (2005)
from SDSS data. 
However, the inspection of the SDSS spectrum does not confirm this redshift 
(nor the other quoted value, $z=1.07$).

Finally, there is no information for the redshift of CRATES J1542+6129 = GB6 J1542+6129.
It has been imaged by the SDSS, but no spectrum is available.

If we assign to all sources in the grey area of Fig. \ref{fluxalpha} a given redshift,
we can see the corresponding region in Fig. \ref{divide}.
We show this for two redshifts: $z=0.5$ and $z=2$, as labelled.
It can be seen that in the case of $z\sim 0.5$ the BLL sources
of unknown redshift would lie in the region already occupied by the other BL Lacs, while
they would be ``outliers" if the redshift is as large as 2
(see also the discussion in A10).

In other words: if the BLL sources in the 1LAC catalog with unknown redshift
will turn out to be at $z\lsim 0.5$--1, then they will fit in the 
phenomenological blazar sequence [i.e. they would be BL Lacs of low and moderate
luminosity, with the majority having a flat $\gamma$--ray slope (i.e. $\alpha_\gamma<$1--1.2); 
Fossati et al. 1998; Donato et al. 2001], 
while they would pose a problem if their redshift is larger.
We will further discuss this point in \S 5).

\subsection{The ``intruder" BL Lacs sample}

We consider all the sources classified as BL Lacs in the ``clean" 1LAC
sample (defined as BL Lacs with $|b|>10^\circ$, detected with a TS significance
larger than 25
(TS stands for Test Statistics, see Mattox et al. 1996 for the definition.
TS=25 approximately corresponds to 5$\sigma$),
whose identification is secure and unique.
We selected the sources of known redshift
with a $\gamma$--ray energy spectral index $\alpha_\gamma$ larger than 1.2, corresponding
to a photon spectral index $\Gamma_\gamma>2.2$.
The resulting 28 BL Lac objects are listed in Tab. \ref{sample}.
At the end of the same table we add the two BL Lacs (0235+164 and 0426--380)
that were ``intruders" because of their extremely large $\gamma$--ray luminosity
(i.e. $L_\gamma>10^{48}$ erg s$^{-1}$), even if their spectral index was
somewhat flatter than $\alpha_\gamma=1.2$.
All these objects are shown and labelled in Fig. \ref{divide}, and
correspond to the filled squares.

We will characterize the SED of all the 30 BL Lacs of Tab. \ref{sample} and 
search for existing data of the luminosity of their broad lines, if present.

\section{Swift observations and analysis}

Several blazars studied in this paper were observed by the {\it Swift} satellite.
These objects are listed in Tab. 2 (X--ray data) and Tab. 3 (optical/UV data).
Even when they were performed during the 11 months of the 1LAC survey,
they correspond to a ``snapshot" of the optical--X--ray state of the source,
while the $\gamma$--ray data are an average over the 11 months.
Given the very rapid blazar variability, the SEDs constructed in this way
should be considered, in all cases, not simultaneous (even when the {\it Swift} UVOT
and XRT data are indeed simultaneous).

The data were screened, cleaned and analysed with the software package
HEASOFT v. 6.8, with the calibration database updated to 30 December 2009.
The XRT data were processed with the standard procedures ({\texttt{XRTPIPELINE v.0.12.4}). 
All sources were observed in photon counting (PC) mode and grade 0--12 
(single to quadruple pixel) were selected. 
The channels with energies below 0.2 keV and above 10 keV were excluded from the fit 
and the spectra were rebinned in energy so to have at least 20--30 counts per bin
in order to apply the $\chi^2$ test. 
When there are no sufficient counts,
then we applied the likelihood statistic in the form reported by Cash (1979). 
Each spectrum was analysed through XSPEC v. 12.5.1n
with an absorbed power law model with a fixed Galactic column density as
measured by Kalberla et al. (2005). 
The computed errors represent the 90\% confidence interval on the spectral parameters. 
Tab. 2 reports the log of the observations and the best fit results of 
the X--ray data with a simple power law model. 
The X--ray spectra displayed in the SED have been properly rebinned to ensure the best 
visualization.

UVOT (Roming et al. 2005) source counts were extracted from 
a circular region 5"--sized centred on the source position, 
while the background was extracted from 
a larger circular nearby source--free region.
Data were integrated with the \texttt{uvotimsum} task and then 
analysed by using the  \texttt{uvotsource} task.
The observed magnitudes have been dereddened according to the formulae 
by Cardelli et al. (1989) and converted into fluxes by using standard 
formulae and zero points from Poole et al. (2008).
Tab. \ref{uvot} lists the observed magnitudes.

\section{Modelling the SED}
\label{modelling}

To model the SEDs of the blazars in this sample we used the same model used in 
Ghisellini et al. (2010a, hereafter G10). 
It is a one--zone, leptonic model, fully discussed in Ghisellini \& Tavecchio (2009).
In that paper we emphasize the relative importance of
the different sources of the seed photons for the inverse Compton
scattering process, and how they change as a function of the 
distance of the emitting region from the black hole.
Here we briefly summarize the main characteristics of the model.

The source is assumed spherical (radius $R$) and located at a distance
$R_{\rm diss}$ from the central black hole. 
The emitting electrons are injected at
a rate $Q(\gamma)$ [cm$^{-3}$ s$^{-1}$] for a finite time equal to the 
light crossing time $R/c$. 
The shape of $Q(\gamma)$ we adopt is assumed to be a smoothly broken power law
with a break at $\gamma_{\rm b}$:
\begin{equation}
Q(\gamma)  \, = \, Q_0\, { (\gamma/\gamma_{\rm b})^{-s_1} \over 1+
(\gamma/\gamma_{\rm b})^{-s_1+s_2} }
\label{qgamma}
\end{equation}
The emitting region is moving with a  velocity $\beta c$
corresponding to a bulk Lorentz factor $\Gamma$.
We observe the source at the viewing angle $\theta_{\rm v}$ and the Doppler
factor is $\delta = 1/[\Gamma(1-\beta\cos\theta_{\rm v})]$.
The magnetic field $B$ is tangled and uniform throughout the emitting region.
For the inverse Compton process, besides the synchrotron seed photons 
(produced internally to the jet), we take into account 
several sources of radiation produced externally to the jet:
i) the broad line photons, assumed to re--emit 10\% of the accretion luminosity
from a shell--like distribution of clouds located at a distance 
$R_{\rm BLR}=10^{17}L_{\rm d, 45}^{1/2}$ cm;
ii) the IR emission from a dusty torus, located at a distance
$R_{\rm IR}=2.5\times 10^{18}L_{\rm d, 45}^{1/2}$ cm and reprocessing
10\%--30\% of the accretion luminosity;
iii) the direct emission from the accretion disk, including its X--ray corona.
We also consider the starlight contribution from the inner region of the host galaxy
and the cosmic background radiation, but these photon sources are unimportant in our case.
All these contributions are evaluated in the blob comoving frame, where we 
calculate the corresponding inverse Compton radiation from all these components, 
and then transform into the observer frame.

We calculate the energy distribution $N(\gamma)$ [cm$^{-3}$]
of the emitting particles at the particular time $R/c$, 
when the injection process ends. 
Our numerical code solves the continuity equation which includes injection, 
radiative cooling and $e^\pm$ pair production and reprocessing. 
Ours is not a time dependent code: we give a ``snapshot" of the 
predicted SED at the time $R/c$, when the particle distribution $N(\gamma)$ 
and consequently the produced flux are at their maximum.

The accretion disk component is calculated assuming a standard
optically thick geometrically thin Shakura \& Sunjaev (1973) disk.
The emission is locally a black body.
The temperature profile of the disk is given e.g. in Frank, King \& Raine (2002).
In our sources, the optical--UV continuum is almost always dominated
by the beamed non--thermal emission.
On the other hand, when detected, the broad emission lines
allow to estimate the luminosity of the accretion disk $L_{\rm d}$.
In these cases we have assumed the value of $L_{\rm d}$ derived
from the emission lines.

By estimating the physical parameters of the source we can calculate
the power that the jet carries  in the form of radiation
($P_{\rm r}$), magnetic field ($P_{\rm B}$), relativistic electrons
($P_{\rm e}$) and cold protons ($P_{\rm p}$) assuming one proton per
electron.
These powers are calculated according to:
\begin{equation}
P_{\rm i} \, =\, \pi R^2 \Gamma^2 c U^\prime_{\rm i}
\label{power}
\end{equation}
where $U^\prime$ is the energy density of the $i_{\rm th}$ component
in the comoving frame.


\begin{table*}
 \centering
 \begin{minipage}{140mm}
  \begin{tabular}{llcccccc}
\hline
Name & Obs Date & Exp & $N_{\rm H}$ & $\Gamma$ & $F_{0.2-10\rm keV}^{\rm obs}$ & $\tilde{\chi^2}$/Cash & d.o.f. \\ 
{}   & DD/MM/YYYY & [ks] & [$10^{20}$~cm$^{-2}$] & {} & [$10^{-13}$~erg~cm$^{-2}$~s$^{-1}$] & {} & {}\\
\hline
0058+3311      & 21-08-2009 & 8.7 & 4.89 & 1.1$\pm$0.4 & 3.3$\pm$0.5 & -/47 & 36\\
0521--36        & 2005-2010\footnote{Data from different observations were integrated: 26-05-2005, 02-02-2008, 07-02-2008, 08-02-2008, 13-02-2008, 05-03-2010, 08-03-2010, 19-06-2010, 23-06-2010, 05-07-2010, 09-07-2010, 13-07-2010.} & 32.4 & 3.32 & 1.61$\pm$0.02 & 203$\pm$2 & 1.2/-- & 280\\
0754+100      & 2007-2010\footnote{Data from different observations were integrated: 18-05-2007, 27-02-2010.} & 20.3 & 2.21 & 1.63$\pm$0.05 & 54$\pm$1 & 0.95/-- & 65\\
0808+019      & 2007-2009\footnote{Data from different observations were integrated: 20-12-2007, 23-12-2007, 14-09-2008, 19-09-2009.} & 19.6 & 3.84 & 2.2$\pm$0.3 & 4.2$\pm$0.3 & 0.2/-- & 3\\
0829+046      & 2006-2010\footnote{Data from different observations were integrated: 23-10-2006, 12-12-2007, 18-09-2009, 20-09-2009, 10-12-2009, 13-12-2009, 11-01-2010, 08-02-2010.} & 27.9 & 2.41 & 1.56$\pm$0.08 & 19.4$\pm$0.6 & 0.65/-- & 30\\
0954+658      & 2006-2010\footnote{Data from different observations were integrated: 04-07-2006, 28-03-2007, 10-01-2008, 11-01-2008, 15-01-2008, 09-01-2009, 01-11-2009, 05-11-2009, 12-12-2009, 23-01-2010, 05-03-2010, 12-03-2010.} & 57.6 & 5.47 & 1.89$\pm$0.07\footnote{Best fit with a broken-power law model (ftest $>99.99$\%): $\Gamma_1=1.1\pm0.1$, $E_{\rm break}=1.3\pm 0.1$~keV, $\Gamma_2=1.89\pm0.07$. In the table is indicated $\Gamma_2$ only.} & 33.6$\pm$0.5 & 1.0/-- & 112\\
1012+0630    & 2010\footnote{Data from different observations were integrated: 24-05-2010 (two observations), 25-05-2010.} & 4.8 & 1.97 & 2.0$\pm$0.3 & 4.6$\pm$0.6 & --/47 & 45 \\
1026--1748   & 24-07-2010 & 0.01 & 6.42 & 2(fixed) & $<30$\footnote{Upper limit derived with PIMMS with fixed photon index equal to 2.} & --/-- & --\\
1204--0710   & 09-08-2010 & 5.1 & 2.02 & 2.8$\pm$0.3 & 10.0$\pm$0.9 & 0.99/- & 4\\
1338+40     & 2008-2009\footnote{Data from different observations were integrated: 15-10-2008, 21-12-2009.} & 11.1 & 0.822 & 1.88$\pm$0.04 & 89$\pm$2 & 1.3/-- & 65\\
1519--273    & 20-01-2010 & 2.2 & 9.11 & 1.4$\pm$0.8 & 4.2$\pm$1.2 & --/19 & 11\\
1807+698    & 2007-2009\footnote{Data from different observations were integrated: 01-03-2007, 15-04-2007, 22-01-2009, 09-11-2009, 11-11-2009.} & 36.3 & 4.11 & 1.84$\pm$0.05 & 31.8$\pm$0.6 & 0.96/-- & 76\\
2007+77     & 14-06-2009 & 6.3 & 8.39 & 1.4$\pm$0.2 & 22$\pm$1 & 1.08/-- & 9\\
2214+24B    & 2010\footnote{Data from different observations were integrated: 25-01-2010, 27-01-2010.} & 8.0 & 5.75 & 1.9$\pm$0.2 & 10.7$\pm$0.8 & 0.67/-- & 7\\
2240-260   & 2008-2009\footnote{Data from different observations were integrated: 25-12-2008, 22-09-2009.} & 5.5 & 1.35 & 1.9$\pm$0.3 &  5.8$\pm$0.8 & --/55 & 66\\
2340+8015  & 2009\footnote{Data from different observations were integrated: 05-09-2009, 08-09-2009, 09-09-2009.} & 10.8 & 14.2 & 2.5$\pm$0.2 & 9.0$\pm$0.6 & 0.76/-- & 9\\
\hline
\end{tabular}
\caption{
Summary of XRT observations.
The observation date column indicates the date of a single snapshot or the years during
which multiple snapshots were performed. 
The corresponding note reports the complete set of observations integrated. 
The column ``Exp" indicates the
effective exposure in ks, while $N_{\rm H}$ is the Galactic absorption
column in units of [$10^{20}$ cm$^{-2}$] from Kalberla et al. (2005).
$\Gamma$ is the photon index of the power law model [$F(E)\propto E^{-\Gamma}$], 
$F_{0.2-10\rm keV}^{\rm obs}$ is the observed (absorbed) flux.
The two last columns indicate the
results of the statistical analysis: the last column contains the degrees
of freedom, while the last but one column displays the reduced $\tilde\chi^2$ or the value of the
likelihood (Cash 1979).}
\end{minipage}
\label{xrt}
\end{table*}

\begin{table*}
\centering
\begin{tabular}{lllllllll}
\hline
\hline
Source     &Date             &$A_V$  &$v$   &$b$             &$u$            &$uvw1$         &$uvm2$           &$uvw2$\\
\hline
0058+3311  &2009--08--21     &0.195  &...   &...    &21.57$\pm$0.24 &...   &...   &...	\\
0521--36   &2008--02--08     &0.130  &15.19$\pm$0.03 &15.90$\pm$0.03  &14.45$\pm$0.03 &15.66$\pm$0.03 &15.70$\pm$0.05 &15.81$\pm$0.04	\\
0754+100   &2010--02--27     &0.075  &17.19$\pm$0.05 &17.72$\pm$0.03  &17.08$\pm$0.03 &17.32$\pm$0.04 &17.22$\pm$0.04 &17.45$\pm$0.03	\\
0808+019   &2007--12--20/23  &0.109  &...	  &...    &...   &...	   &18.73$\pm$0.04 &18.98$\pm$0.04	\\
0829+046   &2018--11--01     &0.108  &15.66$\pm$0.03 &16.15$\pm$0.03  &15.44$\pm$0.03 &15.60$\pm$0.03 &15.57$\pm$0.04 &15.74$\pm$0.04	\\
0954+658   &2009--01--09     &0.380  &17.76$\pm$0.07 &18.41$\pm$0.05  &17.73$\pm$0.05 &18.11$\pm$0.04 &18.23$\pm$0.06 &18.37$\pm$0.04	\\
1012+0630  &2010--05--18     &0.074  &18.73$\pm$0.31 &19.01$\pm$0.18  &18.53$\pm$0.18 &18.69$\pm$0.16 &18.46$\pm$0.13 &18.62$\pm$0.10	\\
1026--1748 &...              &...    &...   &...    &...   &...	   &...   &...	\\
1204--0710 &2010--08--01     &0.070  &16.42$\pm$0.05 &17.05$\pm$0.04  &16.27$\pm$0.04 &16.17$\pm$0.03 &16.04$\pm$0.03 &16.15$\pm$0.03	\\
1338+40    &2009--12--21     &0.025  &$>$19.43 &$>$20.43  &19.83$\pm$0.29 &$>$20.72 &$>$20.90 &$>$21.30	\\
1519--273  &2010--01--20     &0.788  &...	  &...    &19.30$\pm$0.10 &...	   &...   &$>$20.92	\\
1807+698   &2009--01--16     &0.119  &14.98$\pm$0.02 &15.66$\pm$0.01  &15.24$\pm$0.02 &15.46$\pm$0.02 &15.53$\pm$0.02 &15.66$\pm$0.02	\\
2007+77    &...              &...    &...   &...    &...   &...	   &...   &...	\\
2214+24    &2010--01--21     &0.205  &...	  &...    &...   &...	   &...   &17.85	$\pm$0.03\\
2240--260  &2008--12--19     &0.070  &17.21$\pm$0.08 &17.89$\pm$0.07  &16.96$\pm$0.05 &17.14$\pm$0.06 &17.14$\pm$0.06 &17.28$\pm$0.04	\\
2340+8015  &2009--09--03     &0.871  &17.44$\pm$0.07 &18.05$\pm$0.05  &17.47$\pm$0.05 &18.07$\pm$0.05 &18.46$\pm$0.07 &18.39$\pm$0.05	\\
\hline
\hline
\end{tabular}
\caption{Summary of \emph{Swift}/UVOT observed magnitudes. Lower limits are at $3\sigma$ level.}
\label{uvot}
\end{table*} 

\begin{table*} 
\centering
\begin{tabular}{lllllllll}
\hline
\hline
Name           &Emiss. Lines &EW      &Ref     &$L_{\rm BLR}$ &$\log {M_{\rm BH}\over M_\odot}$ &Ref   &${L_{\rm BLR}\over L_{\rm Edd}}$, 
$\left[\log {M_{\rm BH}\over M_\odot}\right]$  &SED \\
               &[2]          &[3]     &[4]     &[5]           &[6]                       &[7]   &[8]                        &[9]\\
\hline 
0058+3311      &...           &...      &...     &...     &... &...    &...                                    &FS  \\
0109+224       &...           &...      &...     &...     &... &...    &...                                    &IBL \\
0208--512      &MgII          &5$\pm$5  &Sc97    &3.7e45  &9.21 &Fa04                             &1.8e-2 [9.2]               &FS   \\
0521--36       &H$\alpha$     &40.7     &Sb06    &4.8e42  &8.52, 8.68, 8.62   &Wo05, Fa03, Fa03b  &9.3e-5 [8.6] &LBL \\
               &              &         &         &       &8.65, 8.33, 8.71   &Ba03, Li06, Fa04  & &        \\
0537--441      &Ly$\alpha$, SiIV, CIV &11.4$\pm$0.7 &Pi05  &6.9e44  &8.74, 8.71 &Wa04, Fa04    &1.0e-2 [8.8]   &FS  \\
0558--3839     &...           &...      &...     &...     &... &...   &...                                     &HBL   \\
0754+100       &[OII], [OIII] &1.1      &Ca03    &...     &... &...    &...                                    &LBL    \\
0808+019       &CII], MgII, [OIII] &5.1 &Sb05    &4.2e43  &... &...    &1.0e-3 [8.5]                           &FS  \\
0829+046       &H$\alpha$     &3.2$\pm$0.8 &SDSS &3.7e42  &8.46, 8.82 &Wo05, Fa03b   &4.5e-5 [8.8]             &LBL  \\
0851+202       &H$\beta$      &1.1      &St89    &6.8e42  &8.79, 8.92 &Wa04, Fa04    &8.3e-5 [8.8]             &LBL \\
0907+3341      &...           &...      &...     &...     &... &...    &                                       &HBL?  \\
0954+658       &H$\alpha$     &2.6      &La96    &2.8e42  &8.53 &Fa04    &6.8e-5 [8.5]                         &LBL   \\
1012+0630      &MgII          &1.2      &Sb05    &7.8e42  &... &...    &1.9e-4 [8.5]                           &LBL   \\
1026--1748     &...           &...      &...     &...     &... &...    &                                       &LBL   \\
1057--79       &MgII, [OIII], [NeIII] &4.24 &Sb09 &5.8e43 &... &...    &7.0e-4 [8.8]                           &LBL   \\
1147+24        &...           &...      &...     &...     &... &...    &                                       &IBL? \\
1204--071      &[OII], [OIII] &...      &La01    &$<$9.5e42  &... &...    &$<$1.2e-4 [8.8]                     &HBL    \\
1338+40        &...           &...      &...     &...     &... &...    &                                       &LBL   \\
1519--273      &MgII          &1.4      &Sb05    &3.4e43  &... &...    &4.2e-4 [8.8]                           &LBL  \\
1557+565       &...           &...      &...     &...     &... &...    &                                       &IBL?   \\
1749+096       &H$\alpha$ H$\beta$, [OII], [OIII] &12.5 &Wh88 &5e43  &8.66 &Fa03b    &7.7e-3 [8.7]             &LBL    \\
1803+78        &MgII, H$\beta$ &2.8     &Re01    &7.1e44  &7.92, 8.57 &Ba06, Wa04   &1.4e-2 [8.6]              &LBL \\
1807+698       &H$\alpha$, [OIII] &6.3  &La96    &1.0e42  &8.49, 8.82, 8.95 &Wo05, Fa03, Fa03b   &1.6e-5 [8.7] &LBL \\
               &                  &     &        &        &8.51, 8.52       &Ba03, Wa04          &             &    \\
2007+77        &[OII], [OIII]     &1.2  &St89    &...     &8.80             &Fa03b               &...          &LBL   \\
2200+420       &H$\alpha$, [OIII] &7.3  &Ve95    &3.3e42  &8.77, 8.35 &Fa03b, Wa04    &5.0e-5 [8.7]            &LBL   \\
2214+24        &...               &...  &...     &...     &... &...    &...                                    &LBL    \\ 
2240--260      &MgII, [OII]       &2.5  &St93    &2.9e43  &... &...    &5.6e-4 [8.6]                           &LBL  \\
2340+8015      &...               &...  &...     &...     &... &...    &...                                    &HBL   \\
\hline
0235+164       &MgII, H$\delta$, H$\gamma$ &15.7$\pm$1.2 &Ra07 &1.0e44 &$>$10.22 &Wa04    &7.7e-4 [9.0]        &FS  \\
0426--380      &MgII, CIII], [OII]         &5.7          &Sb05 &1.1e44 &...         &...        &3.4e-3 [8.6]  &FS  \\ 
\hline
1101+384       &H$\alpha$        &...   &CG97    &4.9e41   &8.29, 8.52, 8.61 8.97 &Ba03, Fa03, Wu02  &1.2e-5 [8.5] &HBL \\
1652+398       &H$\alpha$        &1.1   &St93    &1.6e42   &9.21, 8.78, 8.98      &Ba03, Fa03, Fa03b &1.3e-5 [9.0] &HBL\\
2005--589      &H$\alpha$        &...   &St93    &1.5e41   &8.89, 8.57            &Ca03, Wa08        &2.7e-6 [8.5] &HBL\\

\hline
\hline 
\end{tabular}
\vskip 0.4 true cm
\caption{
Emission lines, BLR total luminosities, black hole masses, and BLR luminosities in units of Eddington ones.
Col [3] reports the maximum observed equivalent width EW in \AA.
In Col. [8] the number in parenthesis is the value of the black hole mass used. 
When the black hole mass is unknown, we have assumed $\log M_{\rm BH}/M_\odot=8.5$.
For 0235+164 and 0426--380 we have used $\log M_{\rm BH}/M_\odot=9$ in agreement with our 
previous estimates (Ghisellini et al. 2009) derived from fitting the SED.
The last column gives the classification according to the appearance of the SED shown 
in Figs. \ref{f1}--\ref{f7} and of the presence/absence of prominent broad lines. 
Question marks mean that the classification is uncertain.
17/30 (57\%) have broad lines; 6/30 (20\%) are ``pure" FS; 
17/30 (57\%) are LBL; 3/30 (10\%) are IBL; 4/30 (13\%) are HBL.
The last three entries are BL Lacs present in G10 for which we found
data for the broad emission lines. All these three are HBL.
References for emission lines:
Ca03: Carangelo et al. 2003;
CG97: Celotti, Padovani \& Ghisellini 1997;
La96: Lawrence et al. 1996;
La01: Landt et al. 2001;
Pi05: Pian, Falomo \& Treves 2005; 
Ra07: Raiteri et al. 2007;
Re01: Rector \& Stocke 2001;
Sc97: Scarpa \& Falomo 1997;
Sb05: Sbarufatti et al. 2005;
Sb06: Sbarufatti et al. 2006;
Sb06: Sbarufatti et al. 2009;
SDSS: http://cas.sdss.org;
St89: Stickel et al. 1989;
St93: Stickel et al. 1993;
Ve95: Vermeulen et al. 1995;
White et al. 1988.
References for the black hole masses:
Ba03: Barth et al. 2003;
Fa03: Falomo et al. 2003a;
Fa03b: Falomo, Carangelo \& Treves 2003b;
Fa04: Fan \& Cao 2004;
Li06:Liu et al. 2006;
Wa04: Wang, Luo \& Ho 2004;
Wa08: Wagner 2008;
Wo05: Woo \& Urry 2005;
Wu02: Wu, Liu \& Zhang 2002.
}
\label{lines}
\end{table*}


\begin{table*} 
\centering
\begin{tabular}{llllllllllllllll}
\hline
\hline
Name   &$z$ &$R_{\rm diss}$ &$M$ &$R_{\rm BLR}$ &$P^\prime_{\rm i}$ &$L_{\rm d}$ &$B$ &$\Gamma$ &$\theta_{\rm v}$
    &$\gamma_{_0}$ &$\gamma_{\rm b}$ &$\gamma_{\rm max}$ &$s_1$  &$s_2$  &$\gamma_{\rm c}$  \\
~[1]      &[2] &[3] &[4] &[5] &[6] &[7] &[8] &[9] &[10] &[11]  &[12] &[13]  &[14]   &[15]  &[16]\\
\hline   
0058+3311       &1.371 &66 (550)  &4e8 &77  &0.015   &0.6 (0.01)      &1    &13   &3   &1   &20    &5e3   &0    &2.5  &23 \\   
0109+224        &0.265 &95  (450) &7e8 &46  &1.3e--3 &0.21 (2e--3)    &1.1  &12.2 &3   &1   &1.5e3 &4e4   &1.1  &2.5  &802  \\   
{\it 0208--512} &1.003 &180 (600) &1e9 &424 &1.7e--2 &18 (0.12)       &3    &13   &3   &1   &200   &8e3   &1    &2.9  &8  \\   
0521--36        &0.055 &45  (500) &3e8 &19  &8e--3   &0.036 (8e--4)   &2    &5    &12  &1   &8e3   &9e3   &1    &2.5  &229  \\   
{\it 0537--441} &0.892 &99 (550)  &6e8 &251 &0.03    &6.3 (0.07)      &3.8  &13   &3   &1   &80    &3e3   &1    &2.2  &13\\
0558--3839      &0.302 &120 (800) &5e8 &27  &8e--4   &0.075 (1e--3)   &2.5  &10   &3   &1   &4e3   &9e5   &--1  &2.8  &217  \\   
0754+100        &0.266 &72  (400) &6e8 &46  &7.5e--3 &0.2 (2.3e--3)   &1.3  &15   &5   &1   &150   &7e3   &1.7  &2.5  &451 \\   
0808+019        &1.148 &54  (600) &3e8 &67  &4.5e--3 &0.45 (0.01)     &7.0  &13   &3   &1   &250   &4e3   &1    &2.8  &19 \\   
0829+046        &0.174 &75  (500) &5e8 &19  &1.2e--3 &0.038 (5e--4)   &0.55 &14   &3   &1   &350   &2e4   &0.75 &2.8  &241 \\   
{\it 0851+202}  &0.306 &90 (600)  &5e8 &26  &4.5e--3 &0.067 (9.e--4)  &1    &10   &3   &70  &5e3   &2e4   &1.7  &3.4  &779 \\
0907+3341       &0.354 &90 (600)  &5e8 &2.7 &9e--4   &7.5e--4 (1e--5) &1.5  &10   &3   &1   &4e3   &5e4   &1    &2.6  &647  \\   
0954+658        &0.367 &50 (550)  &3e8 &17  &5e--3   &0.029 (6.5e--4) &0.7  &14   &3.3 &1   &450   &1.5e4 &1.3  &3.2  &2.2e3  \\   
1012+0630       &0.727 &36 (400)  &3e8 &28  &1e--3   &0.08 (1.8e--3)  &2.7  &12   &3   &1   &500   &7e3   &0.75 &2.7  &241 \\   
1026--1748      &0.114 &75 (500)  &5e8 &8.7 &5.5e--4 &7.5e--3 (1e--4) &0.5  &15   &7   &1   &7e3   &4e4   &1.2  &2.5  &4.2e3 \\  
{\it 1057--79}  &0.569 &180 (1e3) &6e8 &67  &0.01    &0.45    (5e--3) &0.4  &12   &3   &1   &4e3   &4e5   &1.3  &3.6  &1.7e3 \\
1147+24         &0.2?  &68 (450)  &5e8 &25  &1e--3   &0.06    (8e--4) &1.0  &11   &4   &1   &100   &5e4   &1    &2.3  &1.6e3 \\  
1204--071       &0.185 &90 (600)  &5e8 &8.7 &1.2e--3 &7.5e--3 (1e--4) &0.8  &14   &5   &100 &100   &6e4   &0    &2.35 &1.9e3  \\   
1338+40         &0.172 &120 (800) &5e8 &27  &0.014   &0.075   (1e--3) &0.85 &13   &6.5 &30  &50    &5e3   &0    &2.8  &951 \\   
1519--273       &1.294 &68 (450)  &5e8 &58  &1.8e--3 &0.34 (4.5e-3)   &4.0  &18   &2   &1   &200   &3.5e3 &0    &2.4  &38  \\   
1557+565        &0.3   &90 (600)  &5e8 &8.7 &3.3e--3 &7.5e--3 (1e--4) &0.5  &15   &4   &1   &100   &6e4   &0    &2.4  &3.5e3  \\   
{\it 1749+096}  &0.322 &105 (700) &5e8 &77  &2.5e--3 &0.6  (8e--3)    &1.5  &10   &3   &1   &100   &4e3   &0.9  &2.2  &257\\
{\it 1803+784}  &0.680 &60 (500)  &4e8 &268 &4.5e--3 &7.2 (0.12)      &8.7  &12   &3   &1   &80    &2.5e3 &0    &2.2  &16\\
1807+698        &0.051 &120 (800) &5e8 &11  &1.4e--3 &0.011 (1.5e--4) &0.25 &16   &5   &15  &550   &9e3   &1.7  &2.4  &9e3 \\   
2007+77         &0.342 &54  (450) &4e8 &36  &1.5e--3 &0.132 (2.2e--3) &1.6  &10   &3   &1   &250   &3e3   &1    &2.5  &651 \\   
{\it 2200+420}  &0.069 &75 (500)  &5e8 &18  &3e--3   &0.034 (4.5e--4) &0.6  &17   &3   &80  &500   &1e6   &2.2  &3.5  &4.1e3 \\
2214+24         &0.505 &45 (500)  &3e8 &37  &1e--3   &0.14  (3e--3)   &5.0  &15   &3   &1   &300   &7e3   &1    &2.9  &81 \\  
2240--260       &0.774 &108 (900) &4e8 &55  &2e--3   &0.3  (5e--3)    &0.8  &17   &3   &100 &100   &1.2e4 &0.5  &2.2  &780 \\   
2340+8015       &0.274 &105 (700) &5e8 &8.7 &2.3e--3 &7.5e--3 (1.e--4) &0.4 &12   &4   &1   &600   &1.7e5 &0    &2.6  &4.3e3  \\   
\hline
{\it 0235+164}  &0.94  &150 (500) &1e9 &122 &0.018   &1.5 (0.01)      &1.7  &15   &3   &1   &800   &4e3   &0    &2.5  &45 \\
{\it 0426--380} &1.112 &60 (500)  &4e8 &134 &8.5e--3 &1.8 (0.03)      &4.3  &17   &2.3 &1   &250   &5e3   &0    &2.3  &13 \\
\hline
\hline 
\end{tabular}
\vskip 0.4 true cm
\caption{List of parameters used to construct the theoretical SED.
Not all of them are ``input parameters" for the model, because $R_{\rm BLR}$
is uniquely determined from $L_{\rm d}$, and the cooling energy $\gamma_{\rm c}$ 
is a derived parameter.
Col. [1]: name;
Col. [2]: redshift;
Col. [3]: dissipation radius in units of $10^{15}$ cm and (in parenthesis) in units of \sc\ radii;
Col. [4]: black hole mass in solar masses;
Col. [5]: size of the BLR in units of $10^{15}$ cm;
Col. [6]: power injected in the blob calculated in the comoving frame, in units of $10^{45}$ erg s$^{-1}$; 
Col. [7]: accretion disk luminosity in units of $10^{45}$ erg s$^{-1}$ and
        (in parenthesis) in units of $L_{\rm Edd}$;
Col. [8]: magnetic field in Gauss;
Col. [9]: bulk Lorentz factor at $R_{\rm diss}$;
Col. [10]: viewing angle in degrees;
Col. [11] and [13]: minimum, break and maximum random Lorentz factors of the injected electrons;
Col. [14]: and [15]: slopes of the injected electron distribution [$Q(\gamma)$] below and above $\gamma_{\rm b}$;
Col. [16]: values of the minimum random Lorentz factor of those electrons cooling in one light crossing time.
The total X--ray corona luminosity is assumed to be in the range 10--30 per cent of $L_{\rm d}$.
Its spectral shape is assumed to be always $\propto \nu^{-1} \exp(-h\nu/150~{\rm keV})$.
}
\label{para}
\end{table*}

\subsection{Constraints on the accretion luminosity and black hole mass}

For calculating the luminosity of the broad lines, we have followed 
Celotti, Padovani \& Ghisellini (1997),
namely we have assumed that if we set the Ly$\alpha$ line contribution equal to 100, 
the total $L_{\rm BLR}$ is 555.76, and the relative weight of the 
H$\alpha$, H$\beta$, MgII and CIV lines is 77, 22, 34 and 63, respectively (e.g. Francis et al. 1991).
The information found are summarized in Tab. \ref{lines}, reporting also, when
available, the estimate of the black hole mass.
When only one emission line is seen (as in the majority of cases, see Tab. \ref{lines}) 
the estimate of the entire BLR luminosity is uncertain.
Furthermore, the detection of the most prominent line, the Ly$\alpha$ one,
for relatively nearby objects is not possible from the ground, and requires
ultraviolet observations from space. 
Pian, Falomo \& Treves (2005) have studied a small sample of blazars 
spectroscopically observed with the Space Telescope, and compared the relative
strength of the UV lines with the compilation of Francis et al. (1991).
They found that the weights of MgII and CIV are 19 and 53 (setting the line Ly$\alpha$=100),
somewhat less than in Francis et al. (1991).
Therefore the estimates given here for our blazars are uncertain by at least a factor 2.
Despite this uncertainty, the knowledge of the BLR luminosity gives an important
constrain to the model, since it indicates the luminosity of the accretion disk, 
that we set to $L_{\rm d}\sim 10 L_{\rm BLR}$.
This is especially valuable when we do not have any sign of thermal emission in the optical--UV,
often dominated by the non--thermal continuum.
In these cases we have also chosen a value for the black hole mass consistent with what
found in the literature.

\subsection{Results of the modelling}

In Fig. \ref{f0} -- \ref{f7} of the Appendix we show the SED of the considered BL Lacs and the fitting model.
The parameters for the modelling are listed in Tab. \ref{para}, and the derived jet powers in Tab. 6.
One key property of the majority of our sources is to have a relatively low luminosity 
accretion disk. 
If the size of the BLR is connected with $L_{\rm d}$ (we assume $R_{\rm BLR}\propto L_{\rm d}^{1/2}$)
then this implies very compact sizes of the BLR, both in absolute terms and in units of the \sc\ radius.
On the contrary, the dissipation region is always at a few or several hundreds of \sc\ radii, and
in 24/30 cases we have $R_{\rm diss}>R_{\rm BLR}$.
This is in agreement with what found in G10, but here this issue can be treated in more detail
thanks to the knowledge, for some of the sources, of the luminosity of some broad lines
and the black hole mass.
In Fig. \ref{f0} -- \ref{f7} of the Appendix we show, separately, the 
contribution of the synchrotron self--Compton (SSC, long dashed
line) and of the External Compton (EC, dot--dashed line) components.
We find a variety of cases, from sources whose high energy bump is
completely dominated by the EC component (see e.g. 0058+3311; 0208--512; 1803+784),
or by a pure SSC (e.g. 0521--365; 0558--3838; 0851+202; 0907+3341; 1026--174; 1057-79; 1147+24; 1204--71;
1557+565; 2340+801), or by the SSC at softer X--ray energies and by the EC at higher energies.
Rarely (see 0954+658 and 2200+420) there is an important contribution by the second order Compton scattering 
in the SSC spectrum, competing with the EC component in the GeV band.
We alert the reader that in some cases (for instance: 0558--3838 and 1028--174)
the paucity of the data points makes the resulting ``fit" very uncertain.

There are 8 sources in common with the sample studied by G10 (i.e. all blazars with redshift in the LBAS catalog).
These sources are indicated in italics both in Tab. \ref{sample} and in Tab. \ref{para}.
With respect to the model parameters adopted in G10, we here have taken into account the luminosity of the 
BLR and the existing estimates of the black hole mass. 

For 0208--512, the black hole mass adopted here is $M=10^9 M_\odot$ (it was $7\times 10^8 M_\odot$ in G10),
$\Gamma=13$ (it was $\Gamma=10$), $P^\prime$ is nearly half than in G10, $L_{\rm d}$ is similar, $B=3$ G (it was 2.05).

Also for 0537--441 we have changed the adopted black hole mass (now $M=6\times 10^8 M_\odot$, it 
was $2\times 10^9 M_\odot$ in G10), halving the accretion disk luminosity.
Note that these changes in black hole mass, that are rather large, are possible in sources whose IR--UV continuum
is dominated by the non--thermal beamed emission. 
Instead, when the accretion disk emission dominates, the estimate of the black hole mass is much less uncertain (see
the discussion in G10 and Ghisellini et al. 2010b).

For 0426--380 the estimated BLR luminosity implied a much reduced accretion disk luminosity with respect to 
what assumed in G10, and this in turn implied a smaller $R_{\rm diss}$, larger magnetic field, and larger $\Gamma$.
Instead, for 1057--79, the smaller value for $L_{\rm d}$ resulted in 
$R_{\rm diss}$ slightly larger than in G10 and in a smaller magnetic field.

For 0851+202 (=OJ 287), 1749+096 and 2200+420 (=BL Lac) we can now fix the value of $L_{\rm d}$ 
(only an upper limit was used in G10, consistent with the values used now). 
For these sources the model parameters are quite similar to the ones used in G10.

For the sources analyzed in this paper and not present in LBAS the parameters are 
similar to the ones derived in G10 for the blazars of similar $\gamma$--ray luminosity.
The location of the dissipation region $R_{\rm diss}$ is a several hundreds \sc\ radii,
the bulk Lorentz factor is in the range 10--15 (with the exception of 0521--365, with $\Gamma=5$),
magnetic field $B\sim$0.25--7 G, black hole masses $M\sim$(3--10)$\times 10^8M_\odot$.
The possible, expected, difference with respect to the blazars in G10 is that sometimes the
viewing angle $\theta_{\rm v}$ is larger than the typical value of 3 degrees (see 0521--365
with $\theta_{\rm v} =12^\circ$).
This is expected because, with the smaller limiting flux of the 1LAC catalog (with respect to LBAS)
we start to see not only the sources maximally Doppler enhanced (i.e. pointing almost exactly at us),
but also the blazars that are slightly misaligned.

The main point that the modelling should help to clarify is why these
intruder BL Lacs have a relatively steep spectrum, more similar to the
slopes of FSRQs than to the rest of BL Lacs.
To answer this issue, consider first
those intruder BL Lacs that are instead FSRQs,
with luminous accretion disk, dissipation regions within the BLR,
and therefore characterized by $\gamma$--ray slopes similar to ``pure" FSRQs.
In these sources radiative cooling is very severe (see the last column
of Tab. \ref{para} reporting the values of $\gamma_{\rm c}$), and 
the $\gamma$--ray flux is always produced by the steep part
of the electron distribution.

The remaining objects, instead, have low luminosity accretion disks,
small $R_{\rm BLR}$, and dissipation locations beyond the BLR
(i.e. $R_{\rm diss}>R_{\rm BLR}$). 
Nevertheless, contrary to BL Lacs showing a flat spectrum (i.e. $\alpha_\gamma<1$),
they also are characterized by a relatively strong radiative cooling, 
in this case mainly due to the synchrotron and the SSC processes.
As a result the electron population producing the $\gamma$--ray flux is 
steep (i.e. $\alpha_\gamma\sim s_2/2$ above $\gamma_{\rm c}$).
Finally, in a few cases (see for instance 1807+698=3C 371), the cooling is weak, with
$\gamma_{c}$ becoming equal to $\gamma_{\rm max}$, and 
the $\gamma$--ray spectrum is steep because
it is produced by most energetic electrons (near the end of the electron energy distribution).

We would like to stress that, apart from finding the reason of the steep $\gamma$--ray spectrum, 
the aim of the present paper is not on the outcomes
of the model fitting, being instead the finding of a new classifying scheme for
the sub--classes of blazars. 
However, knowing the intrinsic physical quantities characterizing the emitting region of the jet
is certainly helpful for our goal, as is the determination of the black hole mass and
accretion rate of those blazars with the IR--UV continuum dominated by the accretion luminosity.

\begin{table} 
\centering
\begin{tabular}{lllll}
\hline
\hline
Name   &$\log P_{\rm r}$ &$\log P_{\rm B}$ &$\log P_{\rm e}$ &$\log P_{\rm p}$ \\
\hline   
0058+3311    &45.18 &43.41 &44.89 &46.93 \\  
0109+224     &44.05 &43.91 &43.86 &44.99   \\ 
0208--512    &45.40 &45.27 &44.68 &47.06   \\ 
0521--36     &44.23 &42.88 &43.70 &44.93    \\
0537--441    &45.62 &44.95 &45.03 &47.31   \\
0558--3839   &43.88 &44.53 &42.89 &43.43   \\ 
0754+100     &44.52 &43.87 &44.91 &47.18  \\ 
0808+019     &44.76 &44.96 &44.26 &46.39   \\
0829+046     &43.93 &43.71 &43.72 &44.95  \\ 
0851+202     &44.34 &43.48 &44.29 &45.09    \\  
0907+3341    &43.84 &43.83 &43.44 &44.38  \\  
0954+658     &43.91 &42.95 &44.52 &46.13  \\  
1012+0630    &43.93 &43.71 &43.72 &44.95   \\
1026--1748   &43.61 &43.07 &43.60 &44.38  \\
1057--79     &44.75 &43.38 &44.78 &45.96   \\
1147+24      &43.56 &43.32 &43.74 &44.65  \\
1204--071    &43.84 &43.64 &44.03 &44.78  \\
1338+40      &44.33 &43.82 &44.96 &46.27     \\
1519--273    &44.68 &44.95 &44.10 &45.66     \\
1557+565     &44.05 &43.14 &44.50 &45.55 \\
1749+096     &44.04 &43.97 &44.03 &45.61 \\
1803+78      &44.70 &45.16 &44.06 &45.98   \\
1807+698     &43.45 &42.94 &44.43 &45.67  \\
2007+77      &43.57 &43.43 &43.79 &45.21  \\
2200+420     &43.46 &43.34 &44.43 &45.42 \\
2214+24      &43.12 &44.63 &43.91 &45.67      \\
2240--260    &44.41 &43.99 &44.40 &45.16   \\
2340+8015    &43.92 &42.91 &44.13 &44.54    \\
\hline
0235+164     &45.57 &44.72 &44.76 &46.14   \\
0426--380    &45.37 &44.86 &44.38 &46.25   \\
\hline
\hline 
\end{tabular}
\vskip 0.4 true cm
\caption{
Logarithm of the jet power in the form of radiation ($P_{\rm r}$), 
Poynting flux ($P_{\rm B}$),
bulk motion of electrons ($P_{\rm e}$) and protons ($P_{\rm p}$,
assuming one proton per emitting electron). 
Powers are in erg s$^{-1}$.
}
\label{powers}
\end{table}

\begin{table} 
\centering
\begin{tabular}{lllll}
\hline
\hline
Name      &$z$    &Line       &Ref  &$L_{\rm BLR}/L_{\rm Edd}$     \\
\hline 
0820+560  &1.417  &MgII, CIII &SDSS &7.7e-3 [9.3] \\
0917+449  &2.189  &CIV        &SDSS &8.7e-3 [9.8] \\
0954+556  &0.895  &Ly$\alpha$ &Be02 &2.3e-3 [9.1] \\
1013+054  &1.714  &CIV        &SDSS &1.1e-3 [9.5] \\
1030+61   &1.401  &MgII       &SDSS &9.5e-4 [9.5] \\
1055+01   &0.89   &MgII$^a$   &SDSS &4.7e-3 [9.0] \\
1144--379 &1.048  &MgII       &St93 &9.7e-3 [8.5] \\
1156+295  &0.729  &MgII       &SDSS, Pi05 &7.8e-3 [8.7] \\
1226+023  &0.158  &Ly$\alpha$, CIV  &Pi05 &3.3e-2 [8.9] \\
1253--055 &0.536  &Ly$\alpha$, CIV  &Pi05 &2.3e-3 [8.9] \\
1308+32   &0.996  &MgII       &SDSS &8.1e-3 [8.9]   \\
1502+106  &1.839  &CIV        &SDSS &8.2e-3 [9.5]  \\
1510--089 &0.36   &Ly$\alpha$, MgII &Ce97 &4.9e-3 [8.6]   \\
1633+382  &1.813  &MgII       &St93 &1.4e-2 [9.5]   \\
2141+175  &0.211  &CIV        &Os94, Li06 &3.5e-3 [8.6]   \\
2227--088 &1.559  &CIV        &SDSS &2.2e-2 [9.2]   \\
2230+114  &1.037  &Ly$\alpha$, CIV  &Pi05 &6.4e-2 [8.7]   \\
2251+158  &0.859  &Ly$\alpha$, CIV  &Pi05 &2.3e-1 [8.7]   \\
\hline
\hline 
\end{tabular}
\vskip 0.4 true cm
\caption{
The luminosity of the BLR in units of Eddington
for the $\gamma$--ray FSRQs studied in G10, for which
there is an estimate of the black hole mass $M$.
The specific value used for $\log M$ is indicated in square
brackets in the last column.
$^a$: the SDSS associates the MgII line with the
AIII+CIII line, so the SDSS redshift is wrong.
References:
Be02: Bechtold et al. 2002;
Ce97: Celotti, Padovani \& Ghisellini 1997;
Li06: Liu et al. 2006;
Os94: Osmer et al. 1994;
Pi05: Pian, Falomo \& Treves 2005; 
SDSS: Schneider et al. 2010, Shen et al. 2010
(https://www.cfa.harvard.edu/$\sim$yshen/BH$\_$mass/dr7.htm );  
St93: Stickel et al. 1993.
}
\label{fsrq}
\end{table}

\begin{figure}
\vskip -0.6cm \hskip -0.4 cm
\psfig{figure=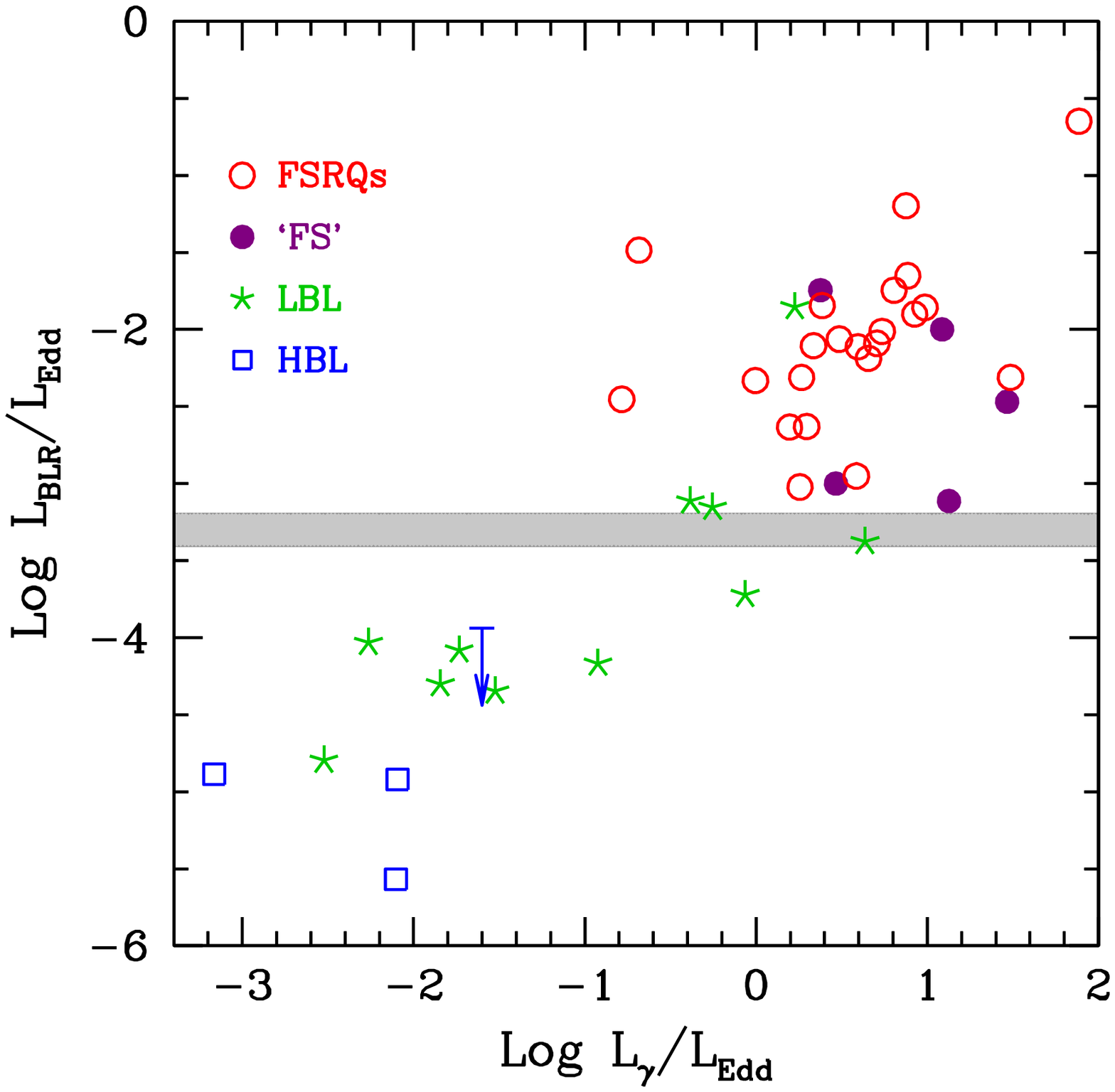,width=9.5cm,height=9.5cm}
\vskip -0.8 cm
\caption{Luminosity of the broad line region (in units
of the Eddington one for sources with at least one broad line 
in their spectrum and with an estimate of the black hole mass)
as a function of the $\gamma$--ray luminosity in units of the Eddington one.
Empty (red) circles: FSRQs studied in G10 with estimates of the black hole mass
and broad emission line luminosity;
full (violet) circles: `FS' sources (i.e. BL Lacs reclassified as FSRQs in 
this paper); (green) stars: LBLs; empty (blue) squares and upper limit: HBLs. 
The three data points for HBLs (Mkn 421, Mkn 510 and 2005--489) do not belong to the
sample studied in this paper (i.e. they have $\alpha_\gamma<1/2$),
but are included for comparison. 
}
\label{f9}
\end{figure}

\section{A new classification for Flat Spectrum Radio Quasars and BL Lac objects}

Tab. \ref{lines} lists the luminosities and EW of the broad lines of our 
intruder BL Lacs.
For several of these objects there are estimates of the mass of their black holes
and therefore we have the luminosity of the BLR in Eddington units.
For the sources with no black hole mass estimate, we use a fiducial, 
average, black hole mass $M=3\times 10^8M_\odot$.

The last column of Tab. \ref{lines} reports a tentative classification of the
object on the base of its SED, 
{\it independent of the presence of the broad emission lines
and of their strength}.
For this SED--based classification we follow following criteria:
\begin{itemize}
\item  
``FS" are those sources, currently classified as BL Lacs, 
whose high energy peak dominates the bolometric
output and the X--ray spectrum belongs to the high energy peak;
\item 
``LBL" (i.e. Low peak BL Lacs, according to the definition of Padovani \& Giommi 1995)
have the synchrotron peak and high energy peak of comparable luminosities,
and the X--ray emission belongs to the high energy peak;
\item 
``IBL'' (i.e. Intermediate BL Lacs) are defined according to the shape
of the X--ray spectrum, steep (i.e. due to the tail of the synchrotron emission) 
at low frequencies and hardening at higher frequencies;
\item 
``HBL" (High peak BL Lacs) have the X--ray emission dominated
by the synchrotron process.
These objects have usually a steep X--ray spectrum, but sometimes,
as in the famous case of the flaring state of Mkn 501 (Pian et al. 1998), the synchrotron
spectrum peaks at so large frequencies to make the X--ray spectrum flat.
\end{itemize}
The above distinctions builds on  the
classification proposed by Padovani \& Giommi (1995; 1996) 
(see also the recent extension in Abdo et al. 2010b) 
on the base of the
position of the synchrotron peak energy, and it is consistent with the 
typical SEDs we have studied in G10 and in Tavecchio et al. (2010).
Of the 30 objects that we considered,
17/30 (57\%) have detected broad lines; 6/30 (20\%) are FS; 
17/30 (57\%) are LBL; 3/30 (10\%) are IBL; 4/30 (13\%) are HBL.

The relation between $L_{\rm BLR}/L_{\rm Edd}$ and the $\gamma$--ray luminosity
(measured in Eddington units as well) is shown in Fig. \ref{f9}, where we plot the
objects with detected broad lines, divided according to the classification from the SED.
To these sources we have also added all the blazars detected in the 3--months
all sky survey of {\it Fermi}/LAT (LBAS, see A09 and G10) for which there are 
estimates of the mass of their black holes and flux measurements of at least one
broad emission line.
These are 3 HBL (i.e. Mkn 421, Mkn 501 and 2005--489; data in Tab. \ref{lines}) 
and 18 FSRQs (data in Tab. \ref{fsrq}).

A clear trend is visible: objects with stronger emission lines
are more luminous in the $\gamma$--ray band (normalizing both luminosities
to the Eddington one).
In this figure we have drawn a dividing line at $L_{\rm BLR}/L_{\rm Edd}=5\times 10^{-4}$,
that separates ``pure FSRQs" and ``FS" from the rest.
The considered sources are still limited in number, but this division is in 
agreement with the idea that the blazars' divide occur for a change in 
the accretion regime.
If we consider a disk luminosity roughly 10 times greater than $L_{\rm BLR}$,
the separation from pure FSRQs to BL Lacs occurs at
$L_{\rm d}\sim 5\times 10^{-3}L_{\rm Edd}$,
approximately where the disk goes from a radiatively efficient to inefficient regime.
This is in remarkable good agreement with what we
suggested earlier (GMT09 for {\it Fermi} blazars
and Ghisellini \& Celotti 2001 for FRI and FRII radio--galaxies).

BL Lac objects that we have reclassified as ``FS" indeed occupy the same region
of FSRQs, confirming that the EW--based classification scheme
sometimes hides the real nature of the source.
``LBLs" are intermediate sources, where the line luminosity
(in Eddington units) decreases (but it does not disappear) as well as
the $\gamma$--ray luminosity.
``HBLs" are at the extreme of the distribution, with very weak lines 
and weak $L_\gamma$.
Consider furthermore that Fig. \ref{f9} shows only the sources with 
{\it measured} broad lines (there is only one exception, 1204--071,
for which we have found an upper limit), and therefore we should not give 
too much weight to the apparent correlation between $L_{\rm BLR}/L_{\rm Edd}$ and
$L_\gamma/L_{\rm Edd}$. 
Consider also that luminosity variations exceeding even two orders of magnitude
are not uncommon for these objects.
Bearing these caveats in mind, we nevertheless believe that
this figure indeed suggests that
the sequence HBL $\to$ IBL (hopefully, no IBL is included, yet) $\to$ LBL $\to$ FSRQ
can be explained as a sequence of strength of the broad lines.

The location of the points seems to suggest a continuity of 
$L_{\rm BLR}/L_{\rm Edd}$ values, rather than a bimodal distribution,
but the number of sources is really too small for a strong claim.
We then leave this issue to future studies: these will give some
insight on the properties of the ionizing flux: is it changing 
dramatically when the disk emits less than a few thousandth of the
Eddington luminosity, or smoothly?

An important caveat is in order: we considered only the sources
of known redshift, and as discussed in \S 2 there are many sources
classified as BL Lacs in the 1LAC sample of unknown redshift.
In order to discuss how these sources can affect the proposed
classification scheme, consider the possibility
that they all are at $z=0.5$ or else at $z=2$.
In the first case, as Fig. \ref{divide} shows, they would
have a low or moderate $\gamma$--ray luminosity.
The fact that their spectrum is featureless implies that their
emission lines are {\it intrinsically} weak, and if their 
black hole mass is similar to the other BL Lacs their 
$L_{\rm BLR}/L_{\rm Edd}$ and $L_\gamma/L_{\rm Edd}$ ratios
locate these objects in the bottom left region of Fig. \ref{f9},
together with the shown sources.

Instead, if $z\sim 2$, these sources would have a $\gamma$--ray
luminosity as large as the powerful FSRQs 
(the subset of these sources having $\alpha_\gamma>1.2$ indeed
occupies the same region occupied by FSRQs in Fig. \ref{divide}).
We have two possibilities:
\begin{itemize}
\item {\it They have powerful lines} and an even more powerful continuum, swamping the emission lines.
In this case, and for average black hole masses,
their $L_{\rm BLR}/L_{\rm Edd}$ is large, as it is $L_\gamma/L_{\rm Edd}$.
They would then occupy the top right part of Fig. \ref{f9}, together with
the other FSRQs.
They would be sources similar to 0208--512, but with an even stronger continuum.

\item {\it They have intrinsically weak lines}, and average black hole masses.
In this case $L_{\rm BLR}/L_{\rm Edd}$ would be small, but $L_\gamma/L_{\rm Edd}$
is large.
They would then occupy the bottom right part of Fig. \ref{f9}, now devoid of sources.
In this case there would be no correlation between $L_{\rm BLR}/L_{\rm Edd}$ and 
$L_\gamma/L_{\rm Edd}$.
These sources would have a powerful non--thermal emission, possibly extremely
beamed (suggesting a very powerful jet), even if their accretion disk is weak.

\end{itemize}

We conclude that
the new classification scheme we are proposing is not affected by 
the possible redshift values of the BL Lacs with unknown $z$.
On the contrary, our physically--based scheme could strengthen
possible problems concerning the blazar sequence.
In fact, in the latter mentioned case (i.e. $z\sim 2$ for the sources of unknown $z$)
we could demonstrate the existence of very powerful sources that are classified
as BL Lacs not only according to the ``old" definition (based on the EW), but also
according to the one we are proposing.

The suggested new classification scheme can have important 
consequences and ramifications, that we discuss below.

\section{Discussion}

The most important result of this study is the suggestion of a new classification 
criterion distinguishing BL Lac objects from FSRQs, based on the luminosity
of the broad emission lines measured in Eddington units.
The critical value we propose is around $L_{\rm BLR}/L_{\rm Edd}\sim 5\times 10^{-4}$.
We base this suggestion on the measurements of the emission line luminosities,
on the estimates of the black hole mass and on the conversion of the 
luminosity of a few lines (often, only one line) into the luminosity
of the ensemble of the broad emission lines.
We find that the $L_{\rm BLR}/L_{\rm Edd}$ ratio is proportional
(with some scatter) to the $\gamma$--ray luminosity measured
in Eddington units.
If we furthermore divide the blazars into sub--categories according
to the properties of their SED, then again we find a trend:
low values of the $L_{\rm BLR}/L_{\rm Edd}$ ratio correspond 
to low power, high energy emitting (i.e. ``blue") BL Lacs.
Vice--versa, high values of $L_{\rm BLR}/L_{\rm Edd}$ 
correspond to powerful ``red" FSRQs.
This is in total agreement with the blazar sequence, and also with the
blazars' divide we have proposed earlier.

We would like to stress that the distinction between FSRQs and BL Lacs we are
proposing here does not imply a dichotomy.
This could be present if the entire accretion disk suddendly
changes regime below
and above a given threshold in $\dot M$, but more likely there should
be a gradual change (namely, some parts of the accretion disks 
might be radiatively efficient and others not).
Then the transition between BL Lacs and FSRQs might be gradual as well.

The first example of how this new classification might be useful
concerns our starting sample of ``intruder" BL Lacs.
Fig. \ref{divide2} shows how the different types of intruder BL Lacs
are located in the $\alpha_\gamma$--$L_\gamma$ plane.
Blazars re--classified as ``FS" all have large $\gamma$--ray luminosities
(filled magenta circles) and fall in the same region of luminous FSRQs.
LBLs span a large range of $L_\gamma$, but not the very high end,
and can be thought as intermediate objects.
HBLs are at significantly lower $\gamma$--ray luminosities.
Of course one is left curious to see what happens for the remaining BL Lacs
of this plot. 
To satisfy this curiosity we plan to systematically study them 
(Sbarrato et al., in preparation).

The other obvious advantage of our proposed classification is that
it is physically based, and will help to construct cleaner samples
of objects aiming to study possible different properties between
BL Lacs and FSRQs, or, rather, between weak and strong line objects.
For instance, this might help to clarify why some BL Lac objects,
observed and imaged in the radio band, appear to be FR II radio--galaxies
(see e.g. Kharb, Lister \& Cooper 2010)
contrary to the accepted scenario of the parent population of blazars,
assigning FR I to BL Lacs and FR II to FSRQs.

A particularly interesting issue that the new classification scheme
will help to clarify is the evolution of blazars.
Hints of no or slightly negative evolution of BL Lacs (or
subcategories of them) and of positive evolution of FSRQs could be associated
with the evolution of the accretion rate in cosmic time.
It is possible that a better understanding may come from 
considering the entire blazar population as a whole,
characterized by larger rates of accretion in the past
(and therefore by a prevalence of FSRQs over BL Lacs)
and a decreased rate of accretion now (with BL Lacs
becoming more numerous).
We believe that this idea, put forward, among others, by Maraschi \& Rovetti (1994)
and by Cavaliere \& D'Elia (2002) is worth pursuing.

\begin{figure}
\vskip -0.6cm \hskip -0.2 cm
\psfig{figure=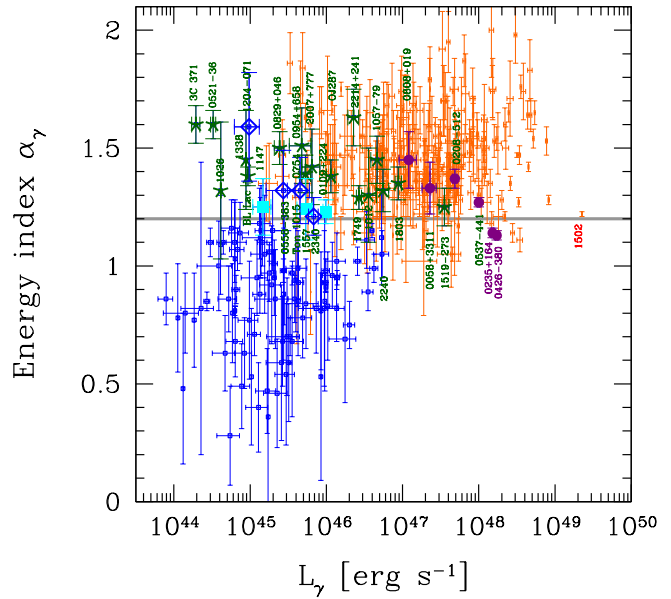,width=9.cm,height=9.cm}
\vskip -0.2 cm
\caption{Same as Fig. \ref{divide}, but with the intruder BL Lacs
reclassified according to Tab. \ref{lines}.
Filled (violet) circles: sources reclassified as FSRQs;
(green) stars: LBLs; (cyan) filled squares: IBLs;	
(blue) diamonds: HBLs.
}
\label{divide2}
\end{figure}

\begin{figure}
\vskip -0.6cm \hskip -0.4 cm
\psfig{figure=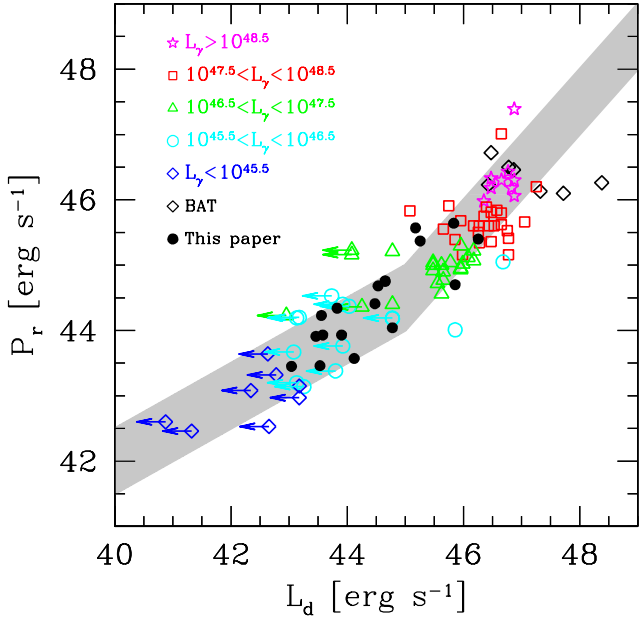,width=9.5cm,height=9.5cm}
\vskip -0.5 cm
\caption{Power of the jet spent in the for of radiation $P_{\rm r}$
as a function of the accretion luminosity $L_{\rm d}$.
Black symbols are the estimates in this paper, BAT points (grey diamonds)
come from the high-redshift blazars present in
the 3 year all sky survey of BAT (and studied in Ghisellini et al. 2010b), the other points and
upper limits come from G10, and are divided according to their $\gamma$--ray
luminosities, as labelled.
The grey stripe indicates what expected if $L_{\rm d}\propto \dot M$ for all
luminosities, while $L_{\rm d}\propto \dot M^2$ and $L_{\rm d} \propto \dot M$ at low 
and high values of $\dot M$, respectively.
}
\label{prld}
\end{figure}

By linking the BLR luminosity to the underlying accretion luminosity we 
suggest that the different ``look" of blazars reflects
primarily their accretion rate in units of the Eddington one.
The $\dot M/\dot M_{\rm Edd}$ ratio controls the radiative efficiency
of the accretion disk, and thus the absolute luminosity of the broad lines,
and (through the BLR size -- disk luminosity relation), their distance from
the black hole.
This in turn regulates the importance of the External Compton process
for the formation of the high energy bump of the SED.
Powerful blazars do have powerful disks, luminous broad lines 
emitted at large distances.
If jet dissipation occurs within the BLR, the External Compton process
is dominant and the cooling severe, resulting in $\gamma$--ray
dominated sources.
Weaker blazars have weak disk and weaker lines emitted closer
to the black hole.
Dissipation in the jet occurs outside the BLR, the main processes
becomes synchrotron and SSC, resulting in more equally shared 
luminosity between the synchrotron and the SSC components.

The other strong evidence we have found in our earlier works
(e.g. Celotti \& Ghisellini 2007; G10; Ghisellini et al. 2010b),
is that the total jet power $P_{\rm jet}$ is proportional to $\dot M$, and indeed
very close to $\dot Mc^2$ {\it independently of the accretion regime}.
The claim that $P_{\rm jet}\sim \dot Mc^2$ depends somewhat to the assumption 
to have one proton per emitting electron.
Although we do have limits to the possible amount of pairs 
(that cannot be energetically important, see Sikora \& Madejski 2000, 
Celotti \& Ghisellini 2008; Ghisellini \& Tavecchio 2010) 
some uncertainty remains on the exact amount of pairs in jets.
An almost model--independent {\it lower limit}
on the jet power is the power spent by the jet to produce the non--thermal 
radiation, $P_{\rm r}$.
It is a lower limit because if $P_{\rm r}$ were comparable to the total jet power, 
$P_{\rm jet}$, then the jet would strongly decelerate, and no superluminal motion would be seen.
Assuming that $P_{\rm r}$ is proportional to $P_{\rm jet}$ implies a
constant efficiency in converting part of the bulk relativistic motion
into relativistic electrons and then radiation.
$P_{\rm r}$ can be derived by Eq. 2 substituting, for $U_{\rm i}$, the energy density 
of the radiation produced by the jet, evaluated in the comoving frame.
The only free parameter entering in the estimate of $P_{\rm r}$ is the bulk Lorentz
factor ($P_{\rm r}\propto L/\Gamma^2$, where $L$ is the bolometric luminosity
of the jet, calculated assuming isotropy).

In Fig. \ref{prld} we show $P_{\rm r}$ as a function of the accretion luminosity $L_{\rm d}$. 
It shows the blazars analyzed in this paper together with the
ones studied in G10 and in Ghisellini et al. (2010b).
For large values of $L_{\rm d}$ and $P_{\rm r}$, the two
quantities are proportional.
Instead, below $L_{\rm d}\sim 10^{45}$ erg s$^{-1}$, the data points and the 
shown upper limits are consistent with $P_{\rm r} \propto L_{\rm d}^{1/2}$.

We interpret this behavior (see also G10 and Ghisellini \& Tavecchio 2008) as  follows.

Assume that $P_{\rm r}$
is  proportional the total jet power $P_{\rm jet}$, in turn {\it always}
proportional to the accretion rate $\dot M c^2$.
Instead, assume that $L_{\rm d}= \eta \dot M c^2$, with $\eta$ being constant
only above a critical luminosity (in Eddington units), while
below this critical value $\eta\propto \dot M$ (Narayan, Garcia \& McClintock 1997).
This means that $\dot M \propto L_{\rm d}$ above some critical value $L_{\rm c}$, 
and $\dot M \propto L_{\rm d}^{1/2}$ below.
Assuming $P_{\rm r} \propto \dot M$ thus implies
\begin{eqnarray}
P_{\rm r} \, &\propto& L_{\rm d}, \qquad\,\,\, L_{\rm d}>L_{\rm c} \nonumber \\
P_{\rm r} \, &\propto& L^{1/2}_{\rm d}, \qquad L_{\rm d}<L_{\rm c} 
\end{eqnarray}

If the jet power is always proportional to $\dot M$, its intrinsic
properties should not change according to $\dot M$ being smaller or
greater than any critical value.
In other words the jet properties should not depend on $\dot M/\dot M_{\rm Edd}$.
Therefore the jet power traces $\dot M$ better than the accretion luminosity,
that strongly depends upon $\dot M/\dot M_{\rm Edd}$.
On the other hand, the ``look" of the jet (i.e. the produced non--thermal SED) 
is strongly influenced by $\dot M/\dot M_{\rm Edd}$ because it is the latter ratio
that rules the strength of the external radiation used as seed for the
formation of the high energy bump of the SED.

To reiterate:
the jet power does depend on $\dot M$ linearly, but there is no dependence of the 
formation, collimation and acceleration mecahnisms
of the relativistic jet in blazars on $\dot M/\dot M_{\rm Edd}$.
Relativistic jets are present for all values of $\dot M/\dot M_{\rm Edd}$.

\section{Summary and conclusions}

In this work we have analyzed a sample of blazars detected by {\it Fermi}/LAT
that have been classified as BL Lac objects and that have an energy 
$\gamma$--ray spectral index $\alpha_\gamma>1.2$.
They therefore occupy a region, in the $\alpha_\gamma$--$L_\gamma$ plane, 
preferentially occupied by FSRQs.
Our intent was to investigate the properties of these objects, to see
if they could be considered as intermediate objects between 
``pure" BL Lacs and ``pure" FSRQs.

Doing so, we collected from the literature the broad emission line data for
a sizeable number of these sources, as well as estimates of their black hole mass.
At the same time, we model their SED with a one--zone leptonic model,
to find out their intrinsic properties and especially to investigate why their
$\gamma$--ray spectrum is rather steep.
Our main results are the following:
\begin{itemize}
\item
From the model fitting, we explain the relatively steep $\alpha_\gamma$
of these blazars as due to a relatively severe cooling of the electron
population. 
The cooling is particularly severe in sources that have strong disks and 
emission lines, but also in the remaining
sources it is fast enough to make the emitting electron distribution steep.

\item 
Some of the considered blazars, classified as BL Lacs, have broad
emission lines as strong as in FSRQs, both in absolute terms and in 
Eddington units.

\item 
There is a trend associating the BLR luminosity
in Eddington units with the $\gamma$--ray luminosity.
Due to the paucity of points, we cannot claim that there
is a strict correlation, yet the indication is that
$L_{\rm BLR}/L_{\rm Edd}$, $L_\gamma/L_{\rm Edd}$ and
the type of the SED (i.e. LBL or HBL) are strongly linked.

\item
From this evidence, we suggest a new classification scheme for dividing
BL Lacs from FSRQs, based on the BLR luminosity in Eddington units:
we propose to set the dividing value at $L_{\rm BLR}/L_{\rm Edd}\sim 5\times 10^{-4}$.

\item
Since the BLR is thought to intercept and reprocess about 10\% of the disk luminosity,
the dividing value corresponds to a disk emitting at the $\sim$0.5\% of the Eddington limit.
This is, approximately, also the value dividing the radiatively efficient from the radiatively inefficient
accretion regimes.

\item 
This work, together with the previous studies we have done on {\it Fermi} (and EGRET)
blazars, confirms that jets are powerful, and that they are born and launched
for all values of the accretion rate (in Eddington units).

\end{itemize}

\section*{Acknowledgments}
We thank the referee, P. Giommi, for useful criticism.
This work was partly financially supported by an ASI (I/088/06/0) grant.
This research made use of the NASA/IPAC Extragalactic Database (NED) 
which is operated by the Jet Propulsion Laboratory, Caltech, under contract 
with NASA, and of the {\it Swift} public data
made available by the HEASARC archive system.

\vskip 1 cm
\appendix{\bf APPENDIX: SPECTRAL ENERGY DISTRIBUTION}

\begin{figure}
\vskip -0.6cm \hskip -0.4 cm
\psfig{figure=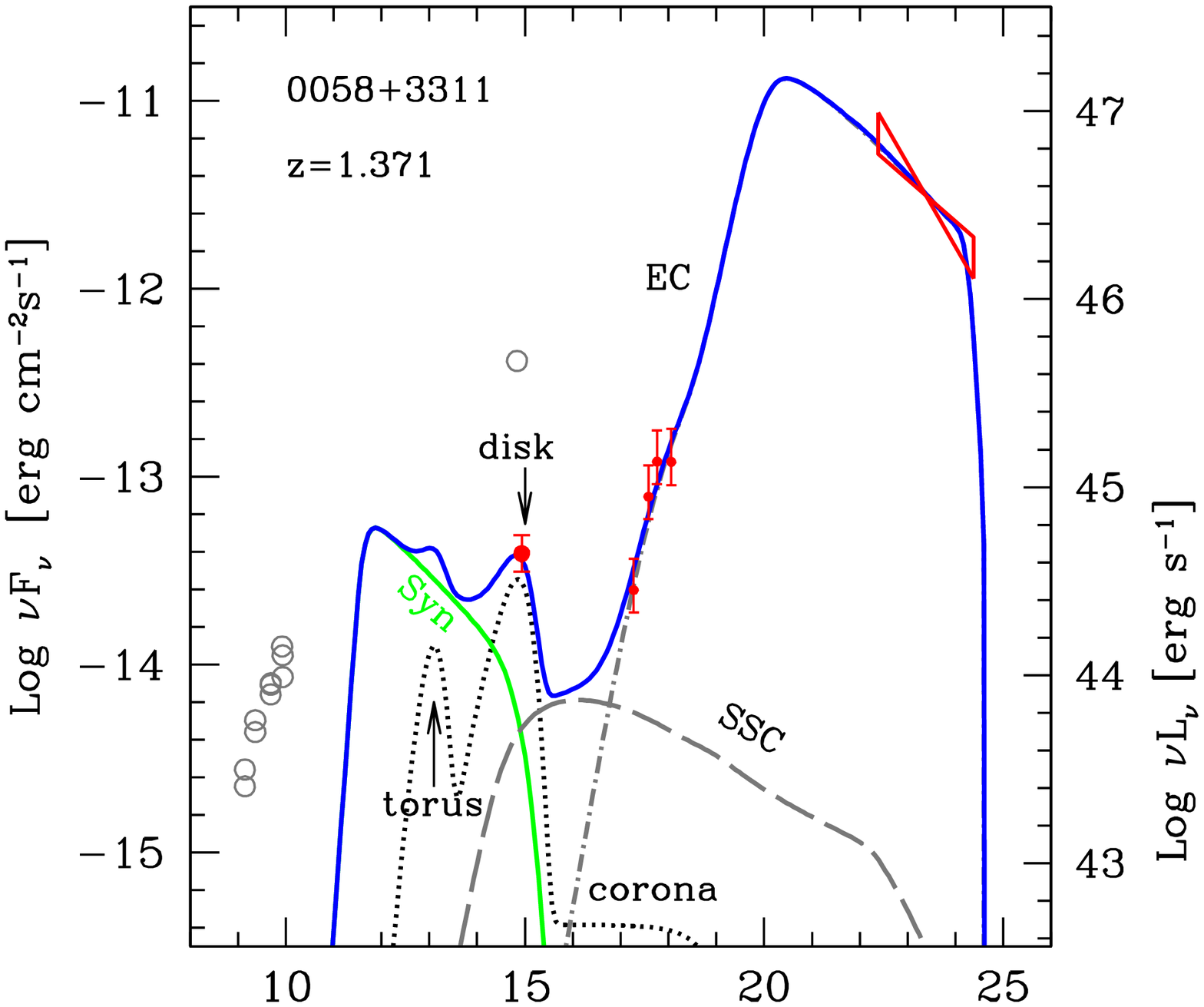,width=9cm,height=6.7cm}
\vskip -1.3 cm \hskip -0.4 cm
\psfig{figure=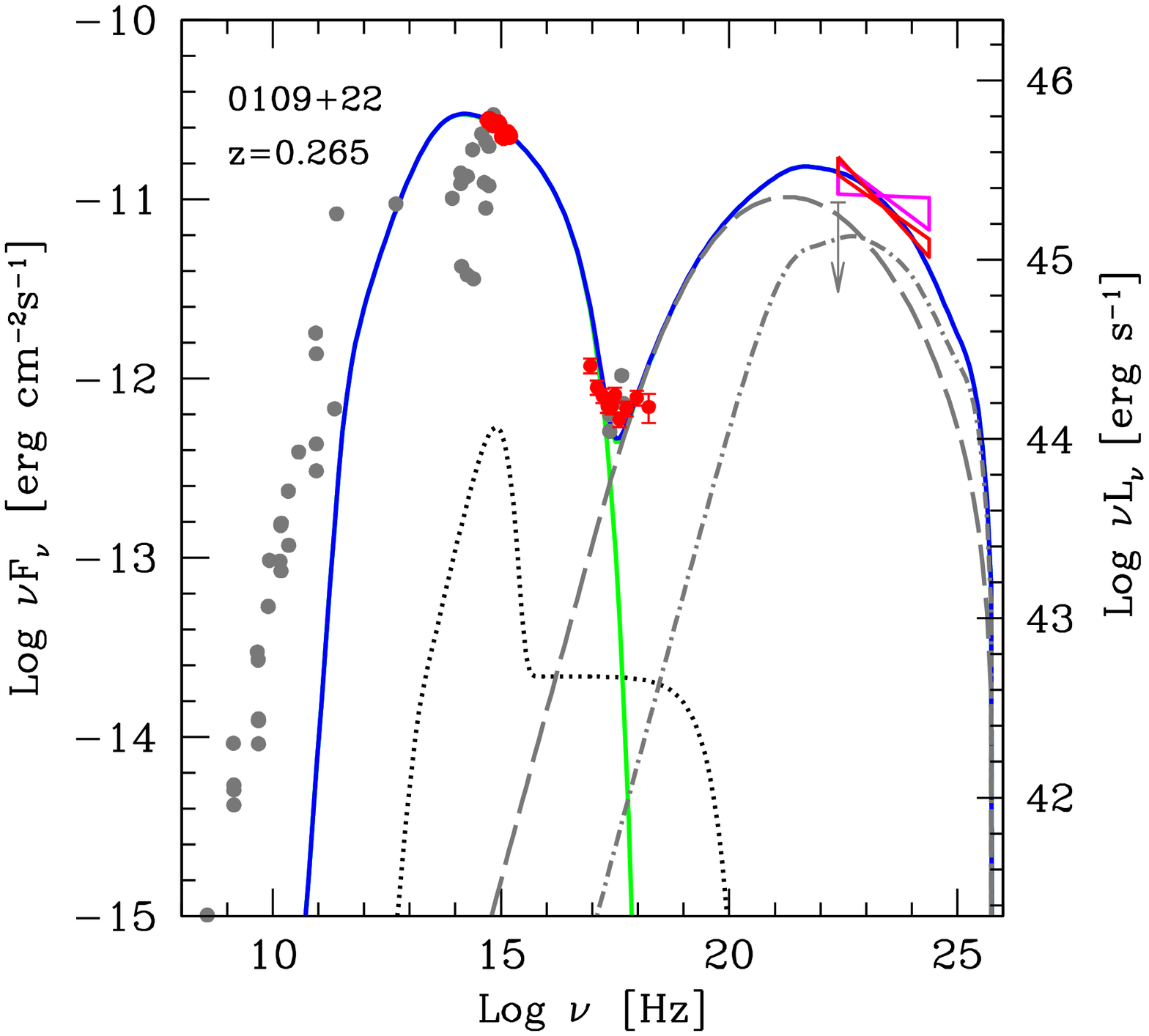,width=9cm,height=6.7cm}
\vskip -0.5 cm
\caption{
SED of 0058+3311 and 0109+22. Darker points (red in the electronic version)
refer to the {\it Fermi}/LAT and {\it Swift} (UVOT and XRT) observations.
The lines are the result of the modelling. 
We label the synchrotron component (green solid line),
the disk, torus and X--ray corona emission (dotted black),
the SSC flux (grey long dashed) and external Compton flux (grey dot--dashed).
The thick (blue) line is the sum of all components.
}
\label{f0}
\end{figure}

\begin{figure}
\vskip -0.6cm \hskip -0.4 cm
\psfig{figure=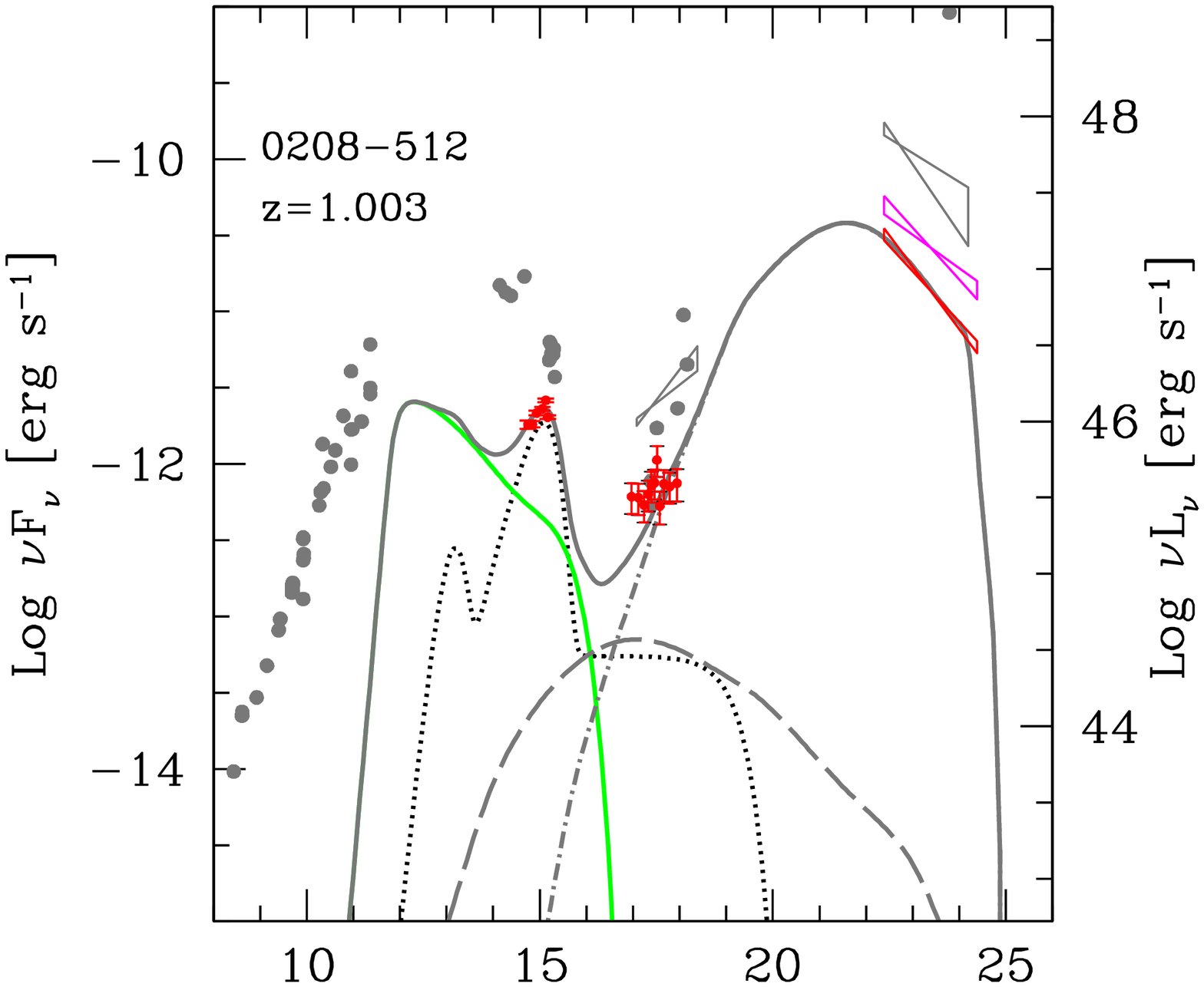,width=9cm,height=6.7cm}
\vskip -1.3 cm \hskip -0.4 cm
\psfig{figure=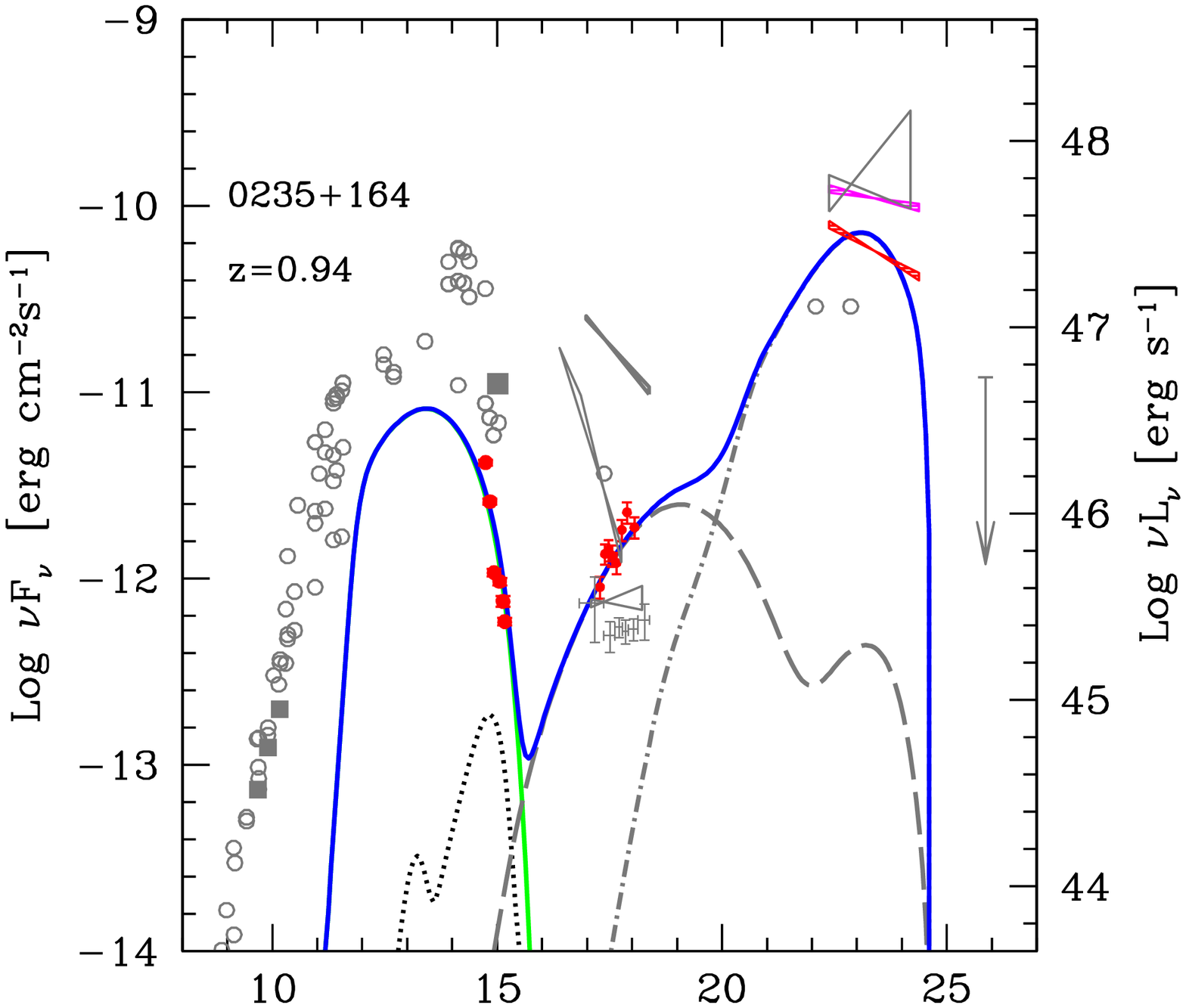,width=9cm,height=6.7cm}
\vskip -1.3 cm \hskip -0.4 cm
\psfig{figure=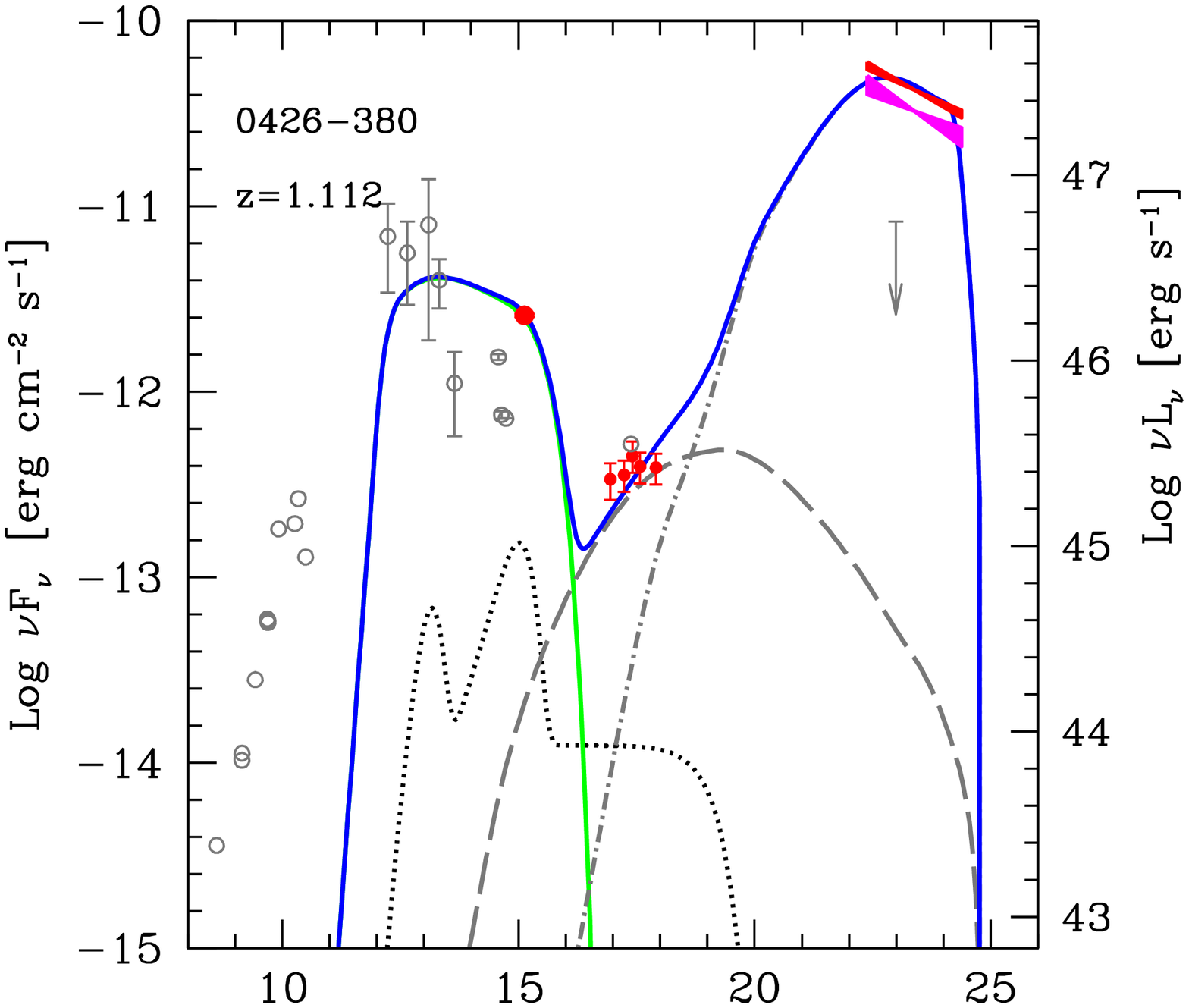,width=9cm,height=6.7cm}
\vskip -1.3 cm \hskip -0.4 cm
\psfig{figure=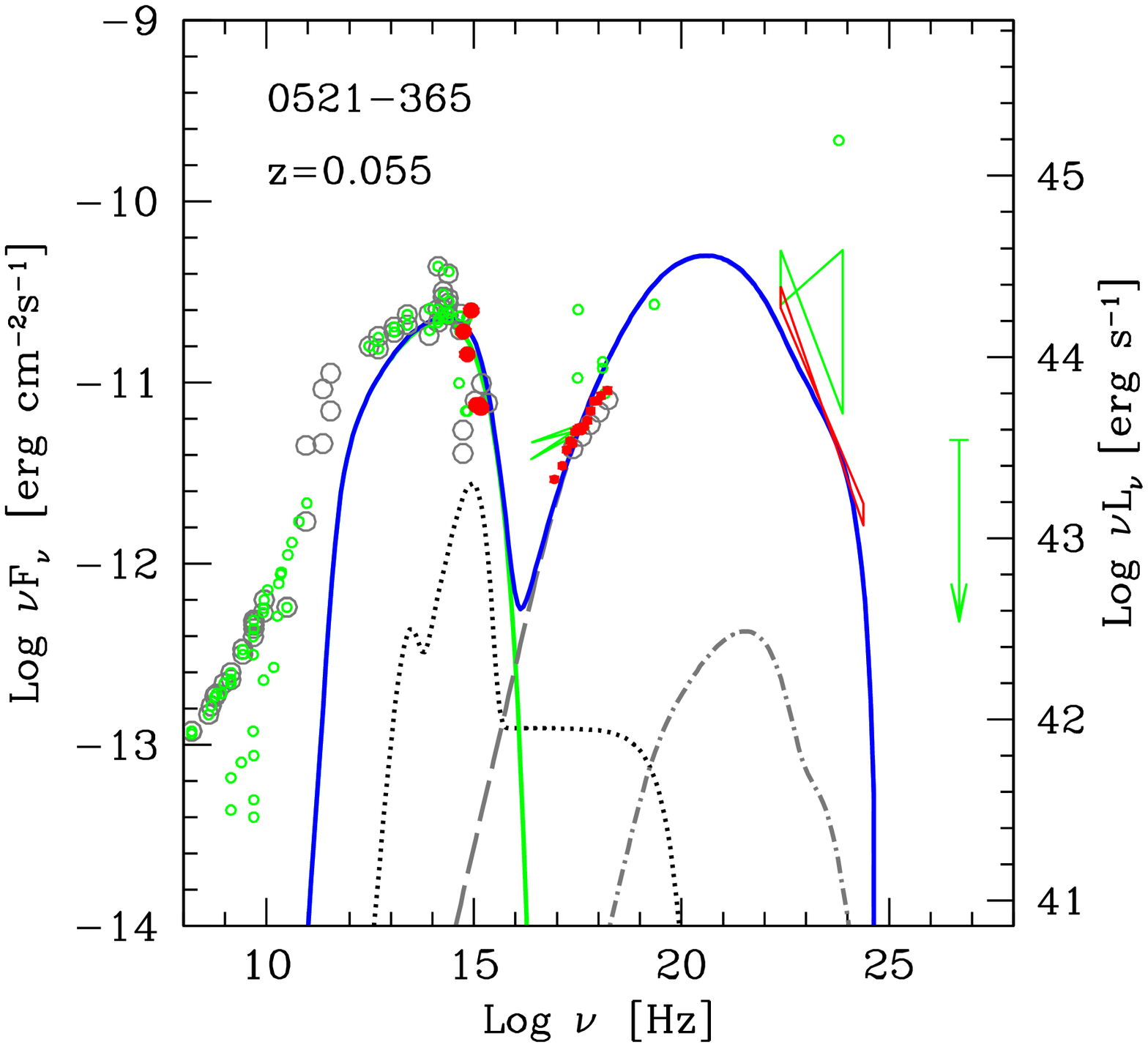,width=9cm,height=6.7cm}
\vskip -0.8 cm
\caption{SED of PKS 0208--512, PKS 0235+164, PKS 0426--380 and PKS 0521--365. 
Symbols and lines as in Fig. \ref{f1}.
}
\label{f1}
\end{figure}

\begin{figure}
\vskip -0.6cm \hskip -0.4 cm
\psfig{figure=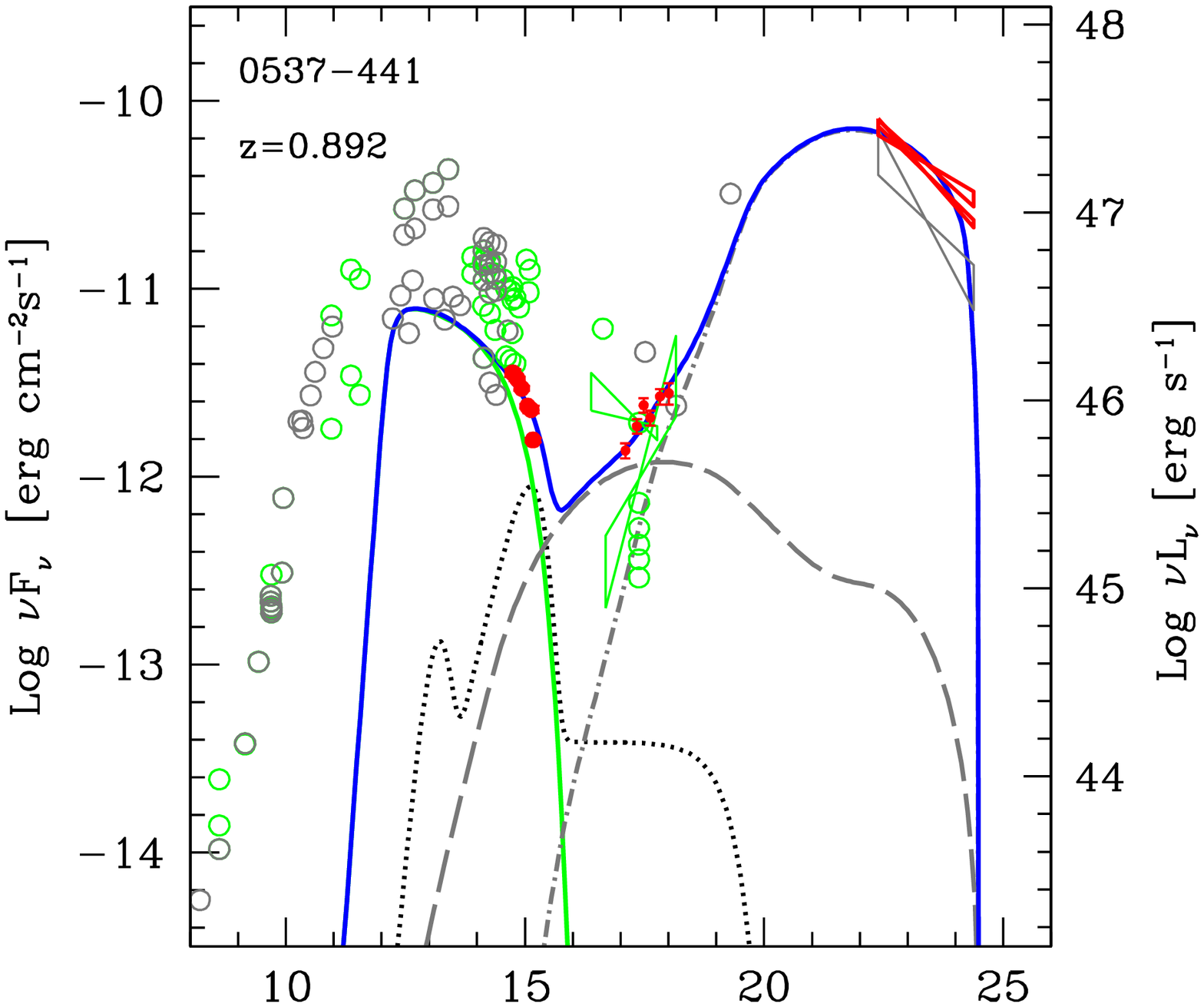,width=9cm,height=6.7cm}
\vskip -1.3 cm \hskip -0.4 cm
\psfig{figure=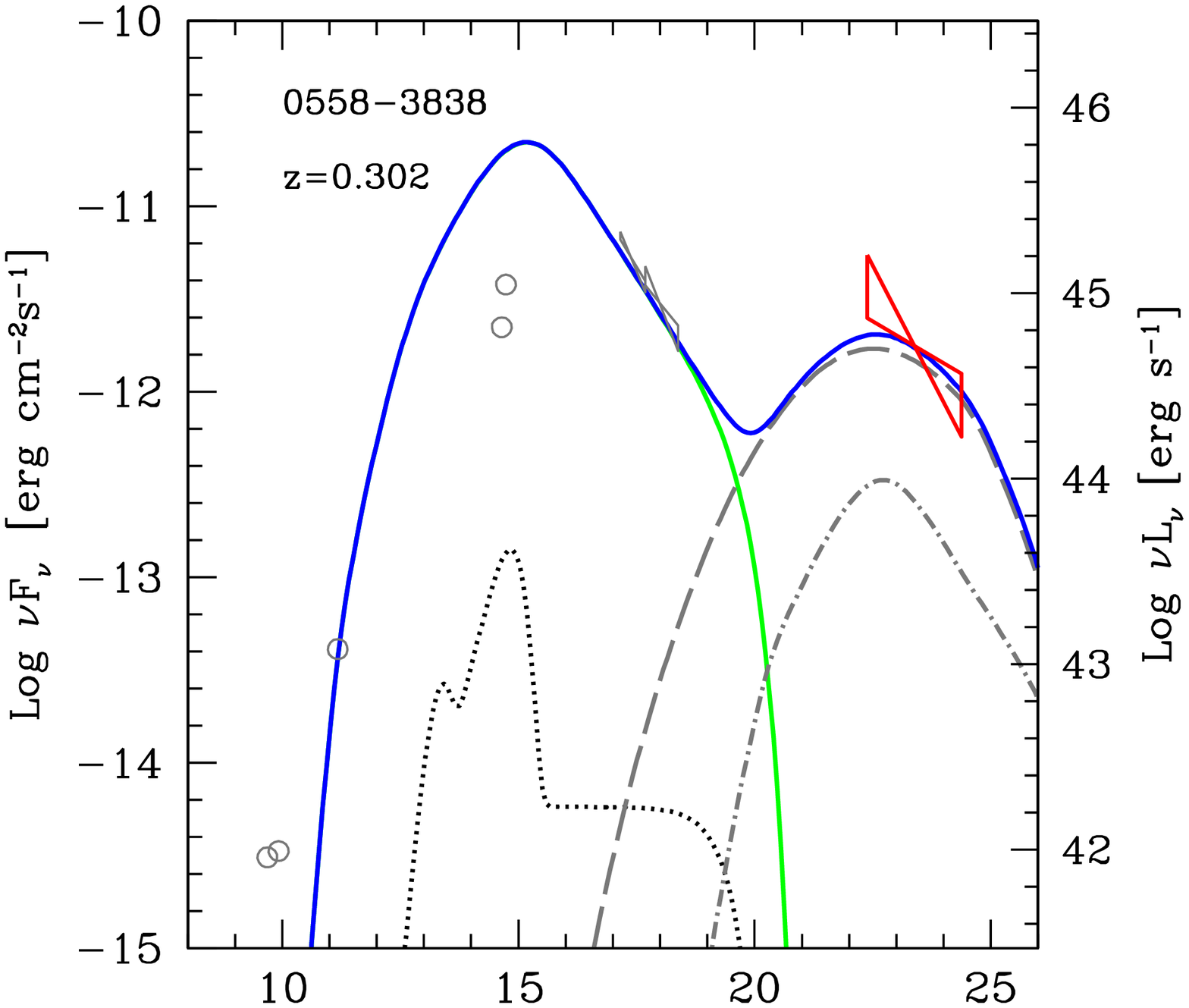,width=9cm,height=6.7cm}
\vskip -1.3 cm \hskip -0.4 cm
\psfig{figure=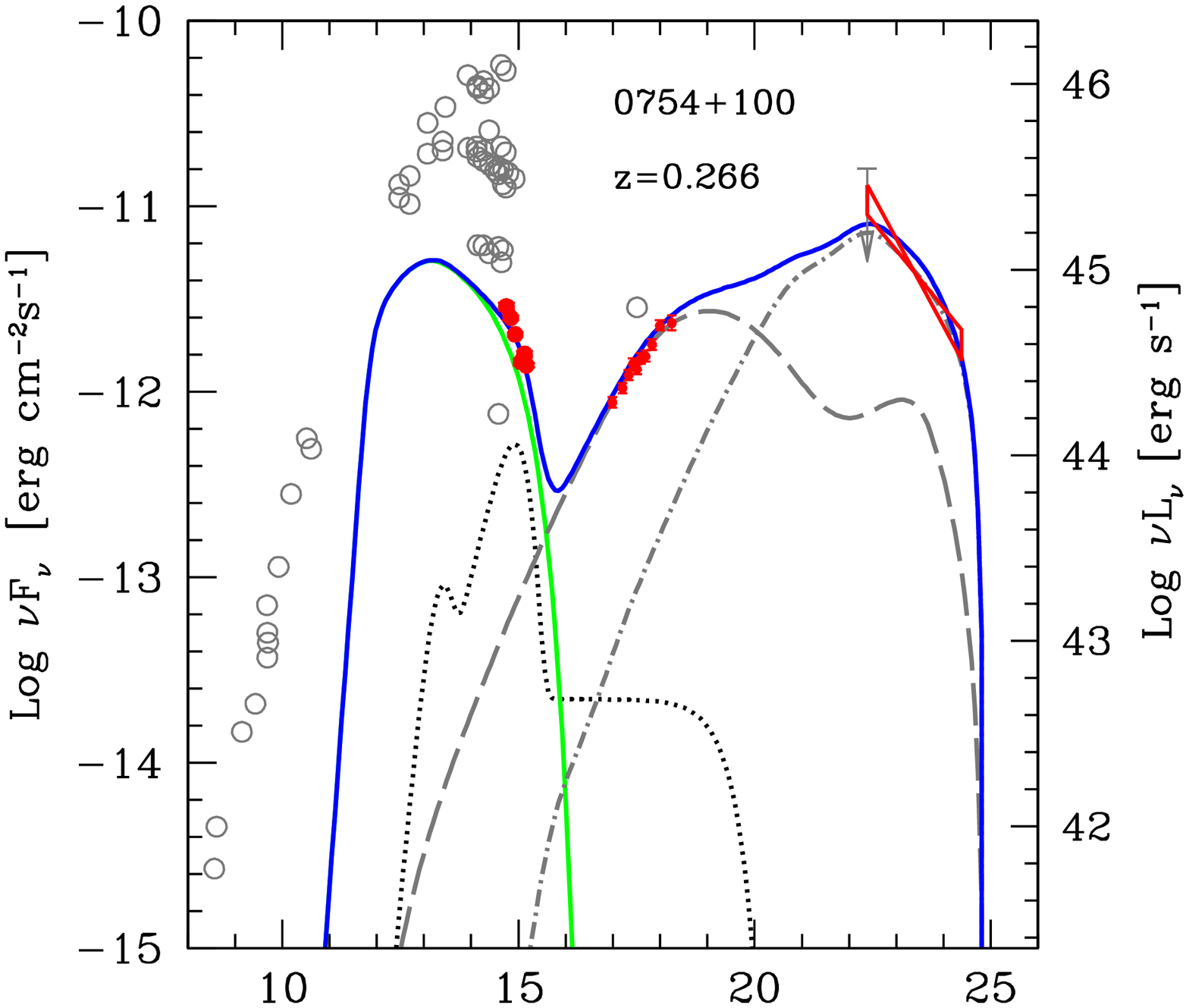,width=9cm,height=6.7cm}
\vskip -1.3 cm \hskip -0.4 cm
\psfig{figure=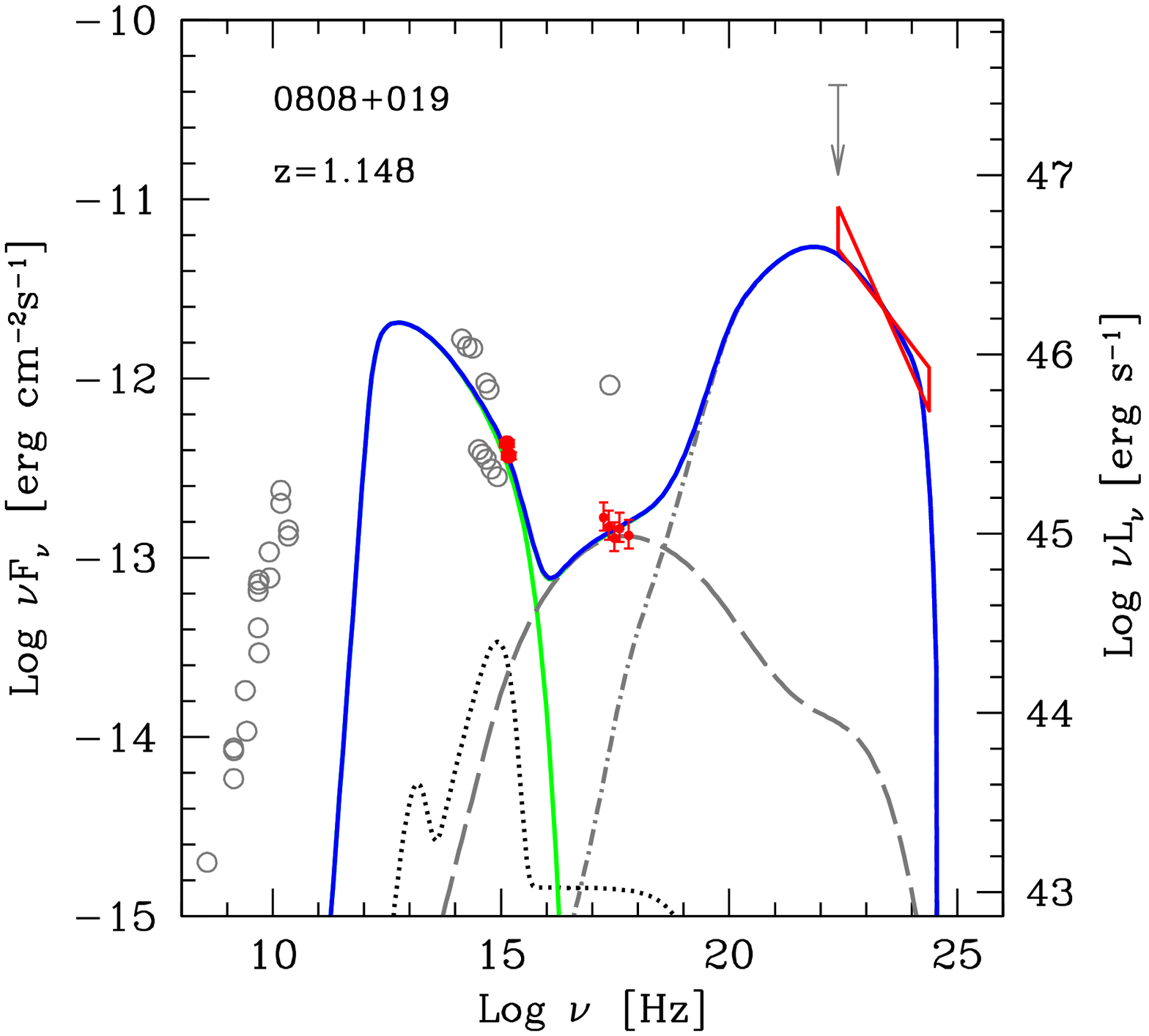,width=9cm,height=6.7cm}
\vskip -0.8 cm
\caption{
SED of PKS 0537--441, PMN 0558--3839, PKS 0754+100 and PKS 0808+019.
Symbols and lines as in Fig. \ref{f1}.
}
\label{f2}
\end{figure}

\begin{figure}
\vskip -0.6cm \hskip -0.4 cm
\psfig{figure=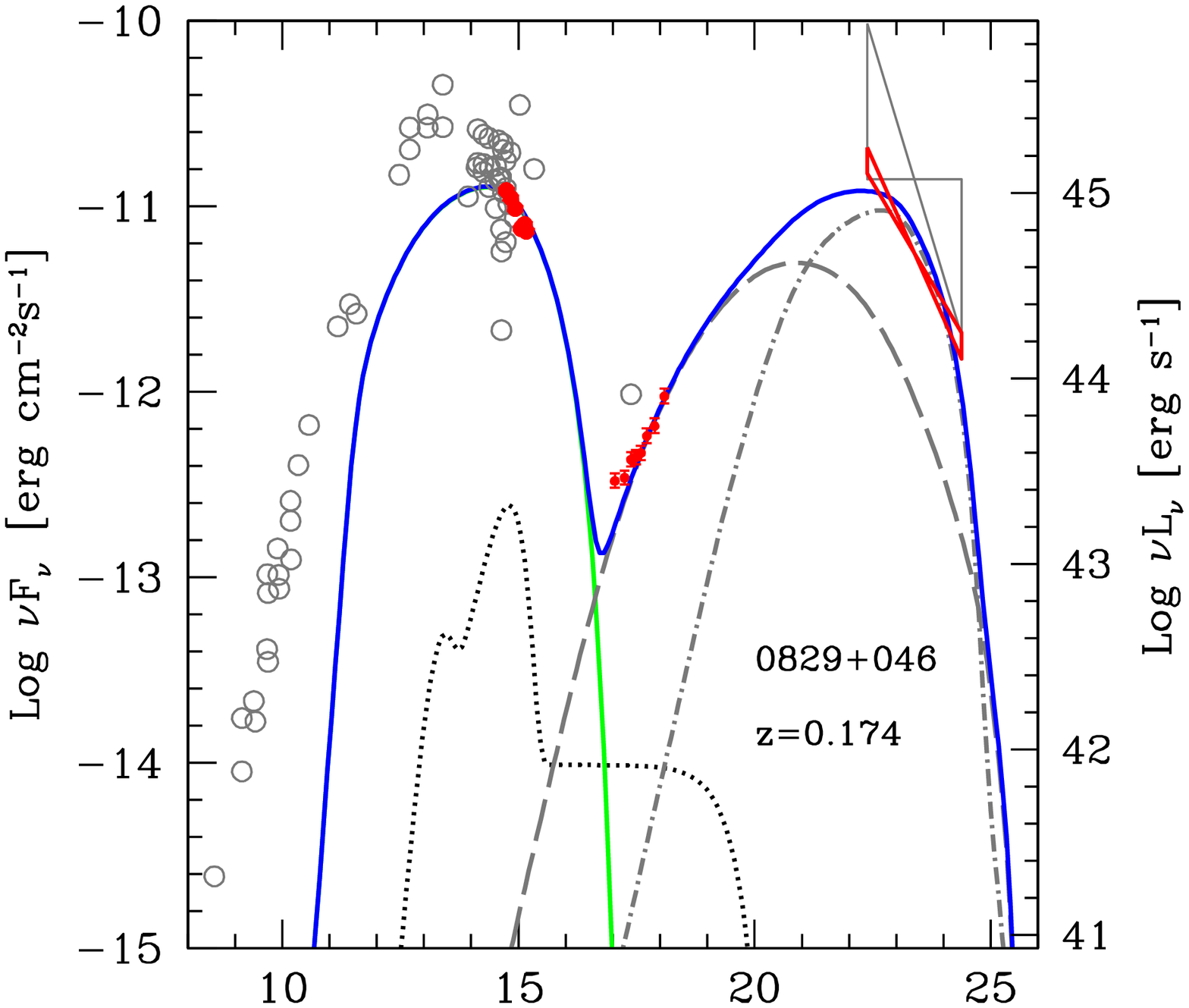,width=9cm,height=6.7cm}
\vskip -1.3 cm \hskip -0.4 cm
\psfig{figure=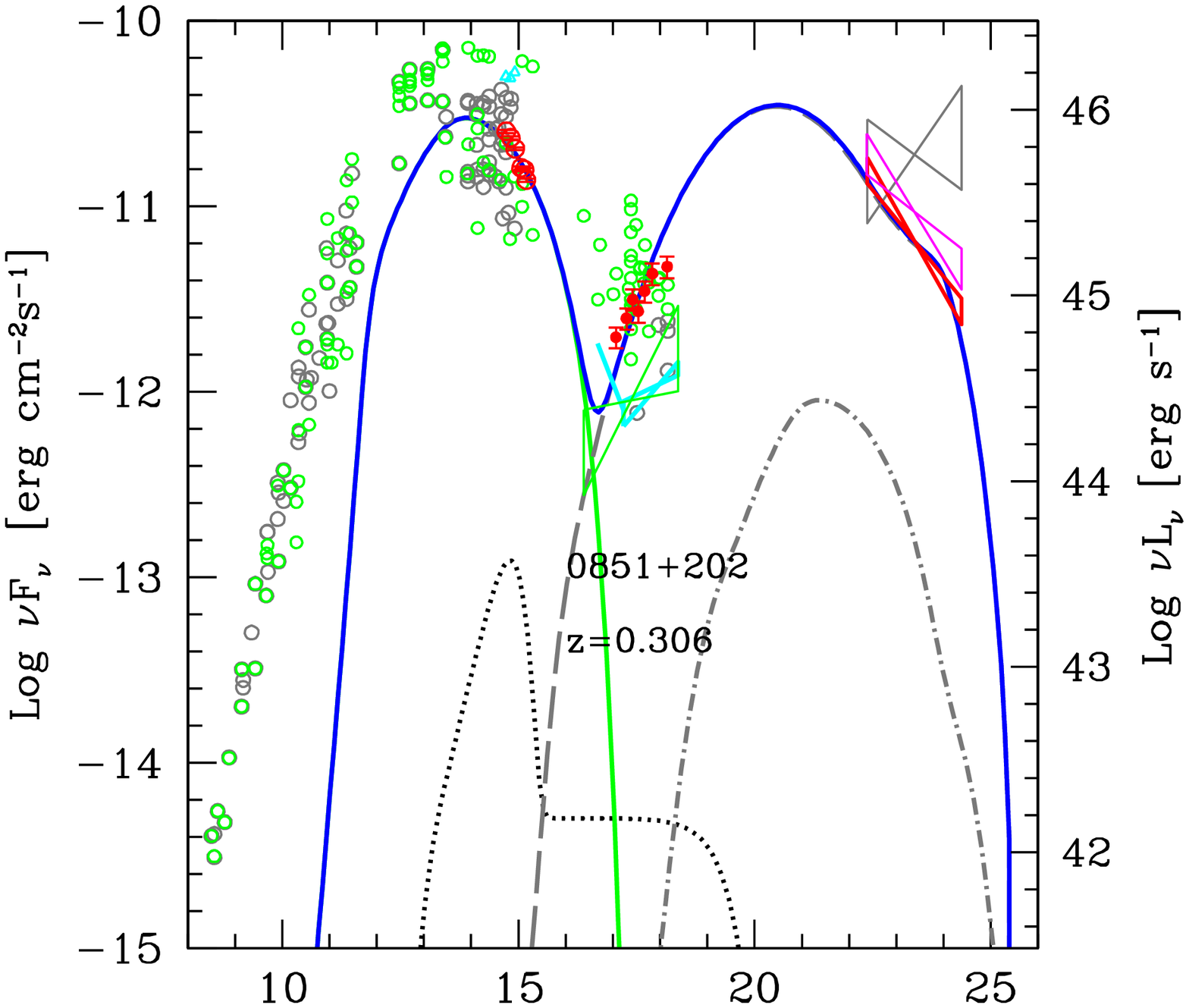,width=9cm,height=6.7cm}
\vskip -1.3 cm \hskip -0.4 cm
\psfig{figure=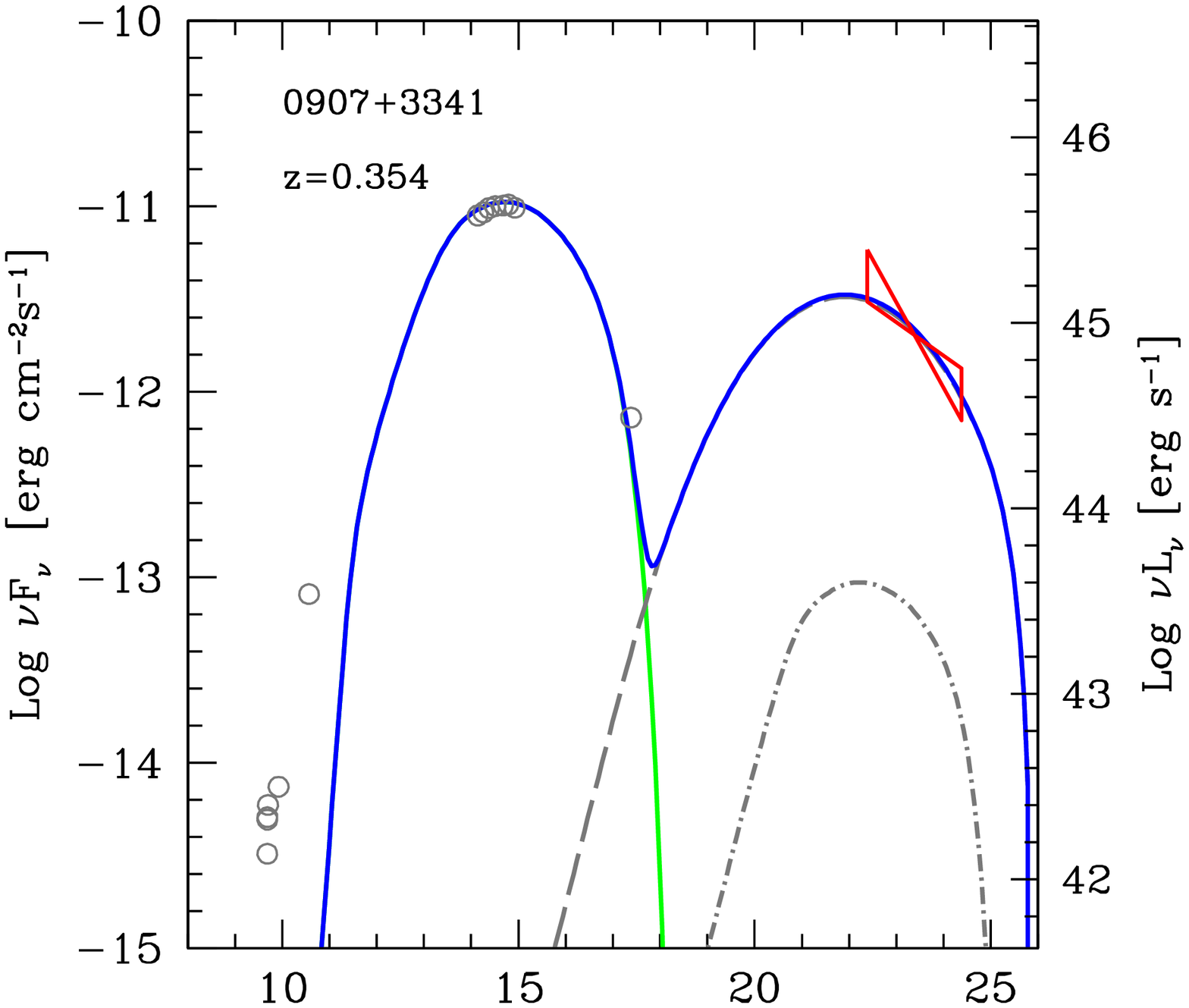,width=9cm,height=6.7cm}
\vskip -1.3 cm \hskip -0.4 cm
\psfig{figure=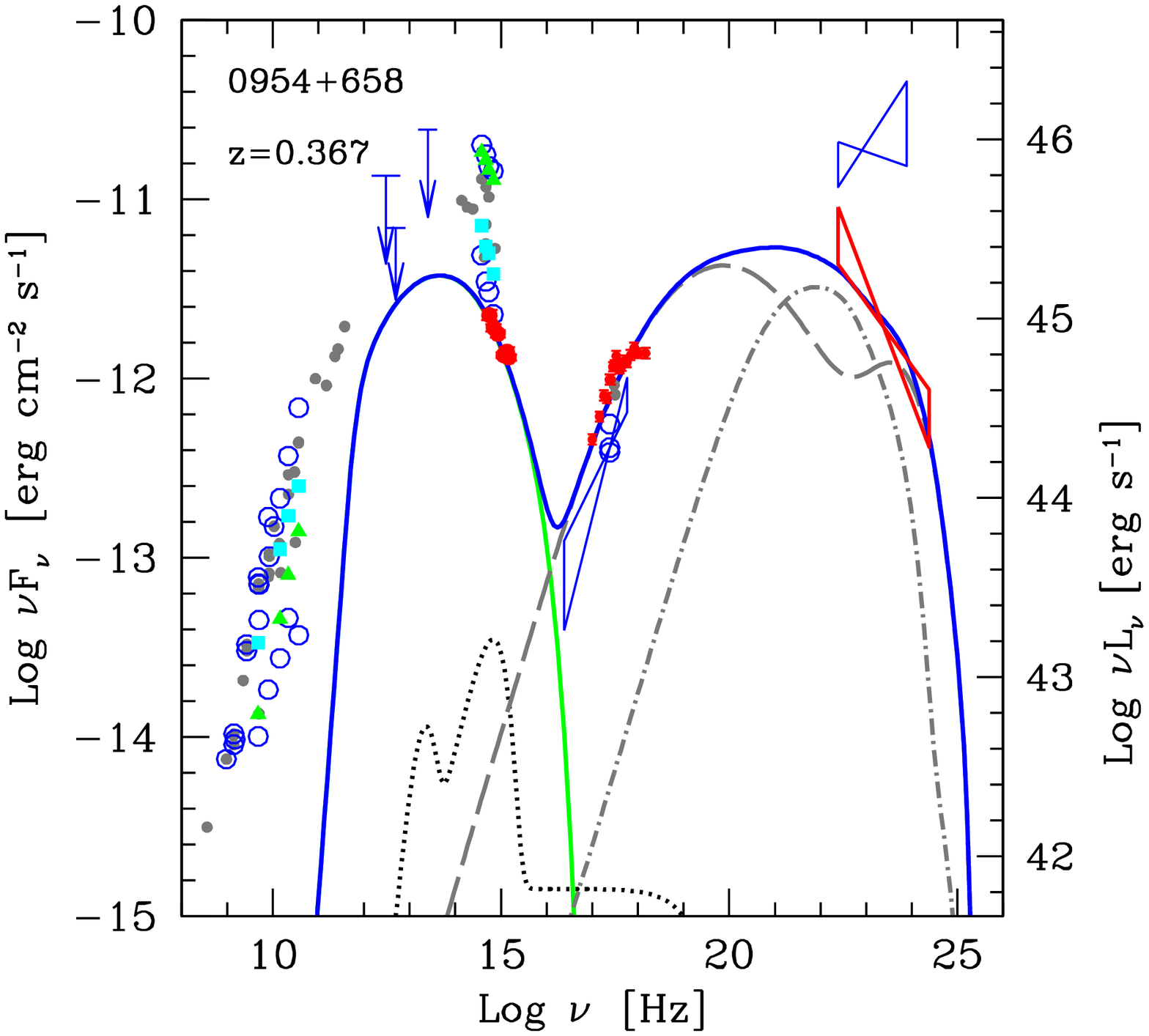,width=9cm,height=6.7cm}
\vskip -0.8 cm
\caption{
SED of PKS 0829+046, 0851+202 (=OJ 287), 0907+3341 (=TON 1015) and
0954+658.
Symbols and lines as in Fig. \ref{f1}.
}
\label{f3}
\end{figure}

\begin{figure}
\vskip -0.6cm \hskip -0.4 cm
\psfig{figure=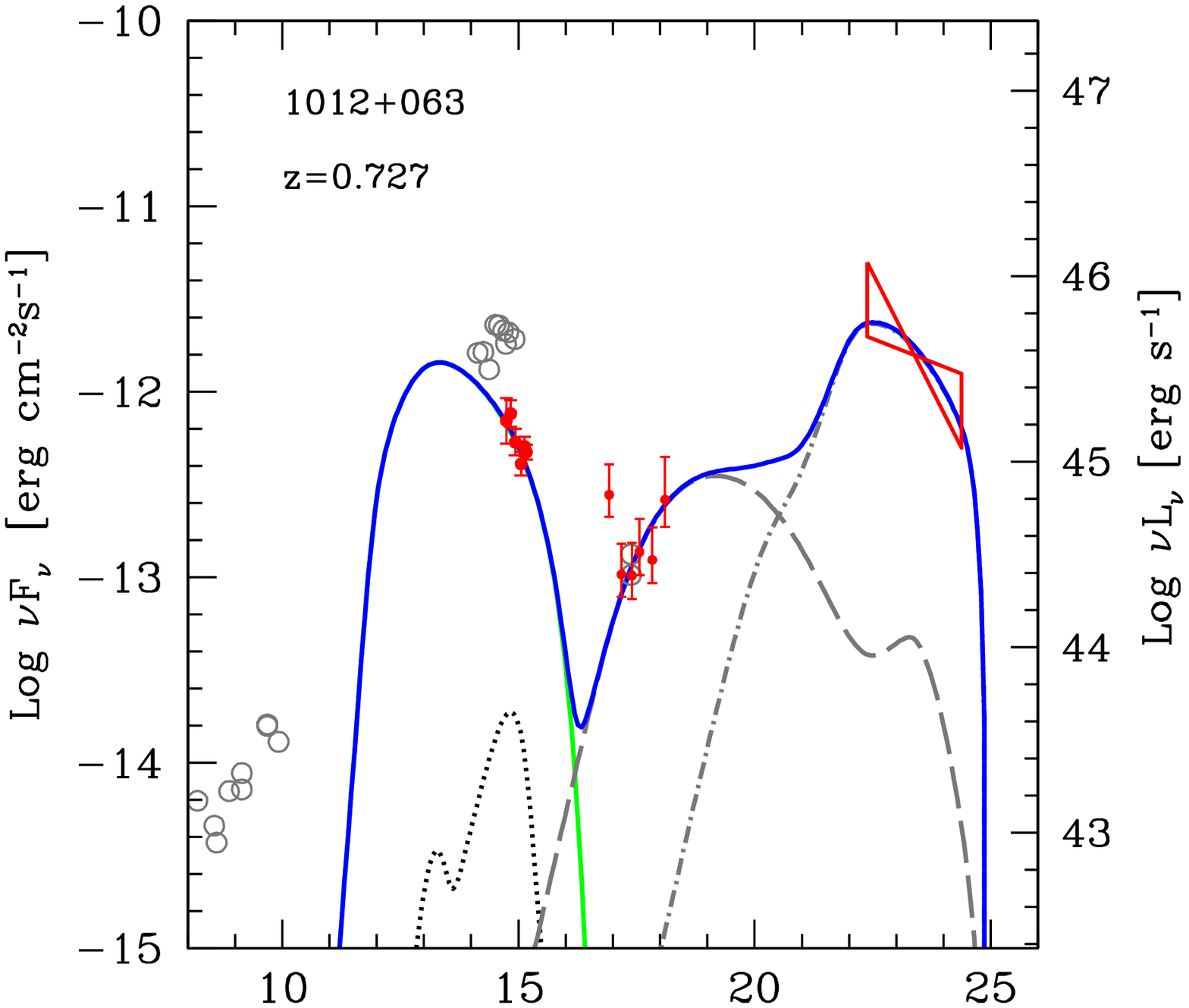,width=9cm,height=6.7cm}
\vskip -1.3 cm \hskip -0.4 cm
\psfig{figure=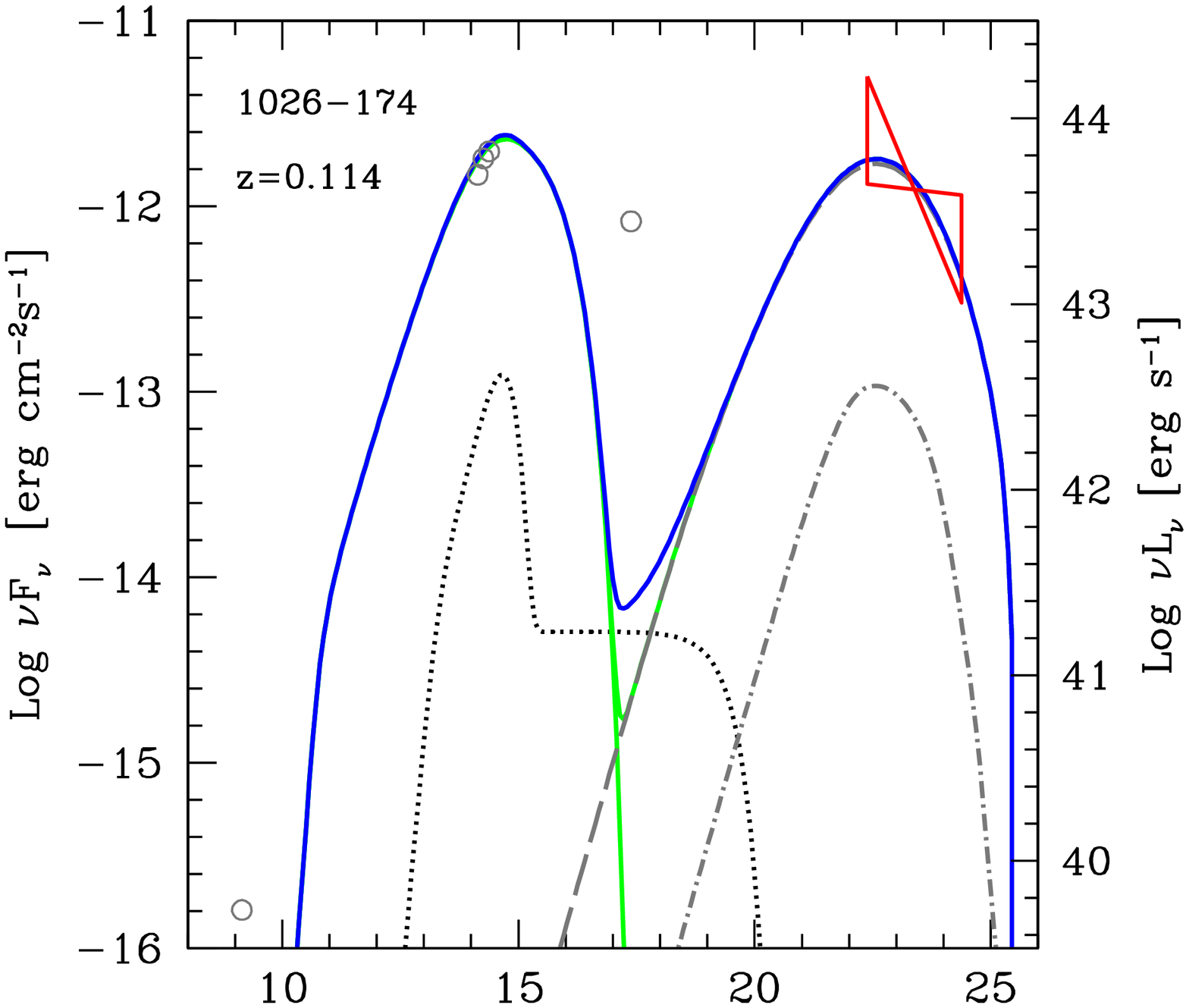,width=9cm,height=6.7cm}
\vskip -1.3 cm \hskip -0.4 cm
\psfig{figure=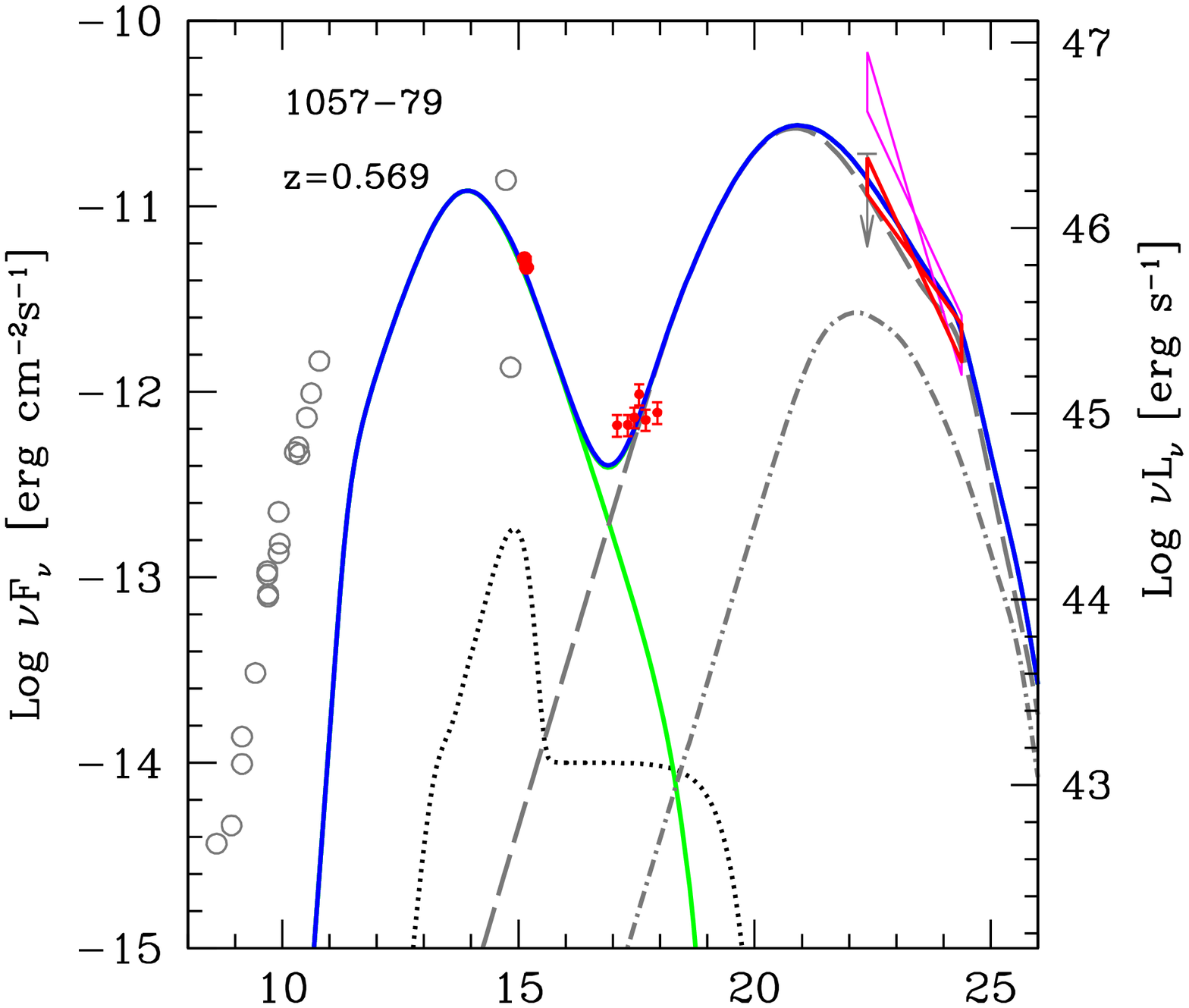,width=9cm,height=6.7cm}
\vskip -1.3 cm \hskip -0.4 cm
\psfig{figure=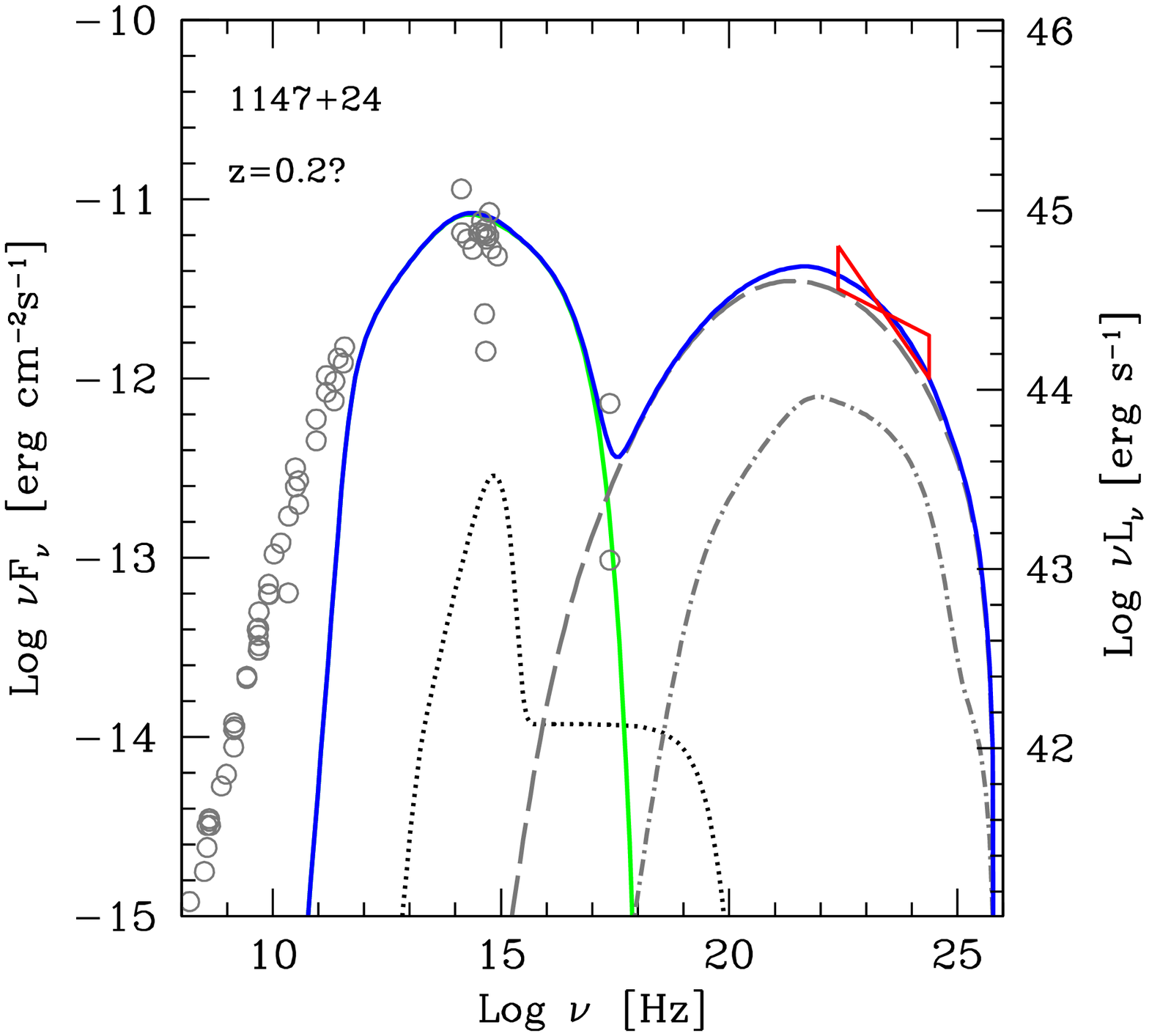,width=9cm,height=6.7cm}
\vskip -0.8 cm
\caption{
SED of 1012+0630, 1026--1748, PKS 1057--79 and B2 1147+24.
Symbols and lines as in Fig. \ref{f1}.
}
\label{f4}
\end{figure}

\begin{figure}
\vskip -0.6cm \hskip -0.4 cm
\psfig{figure=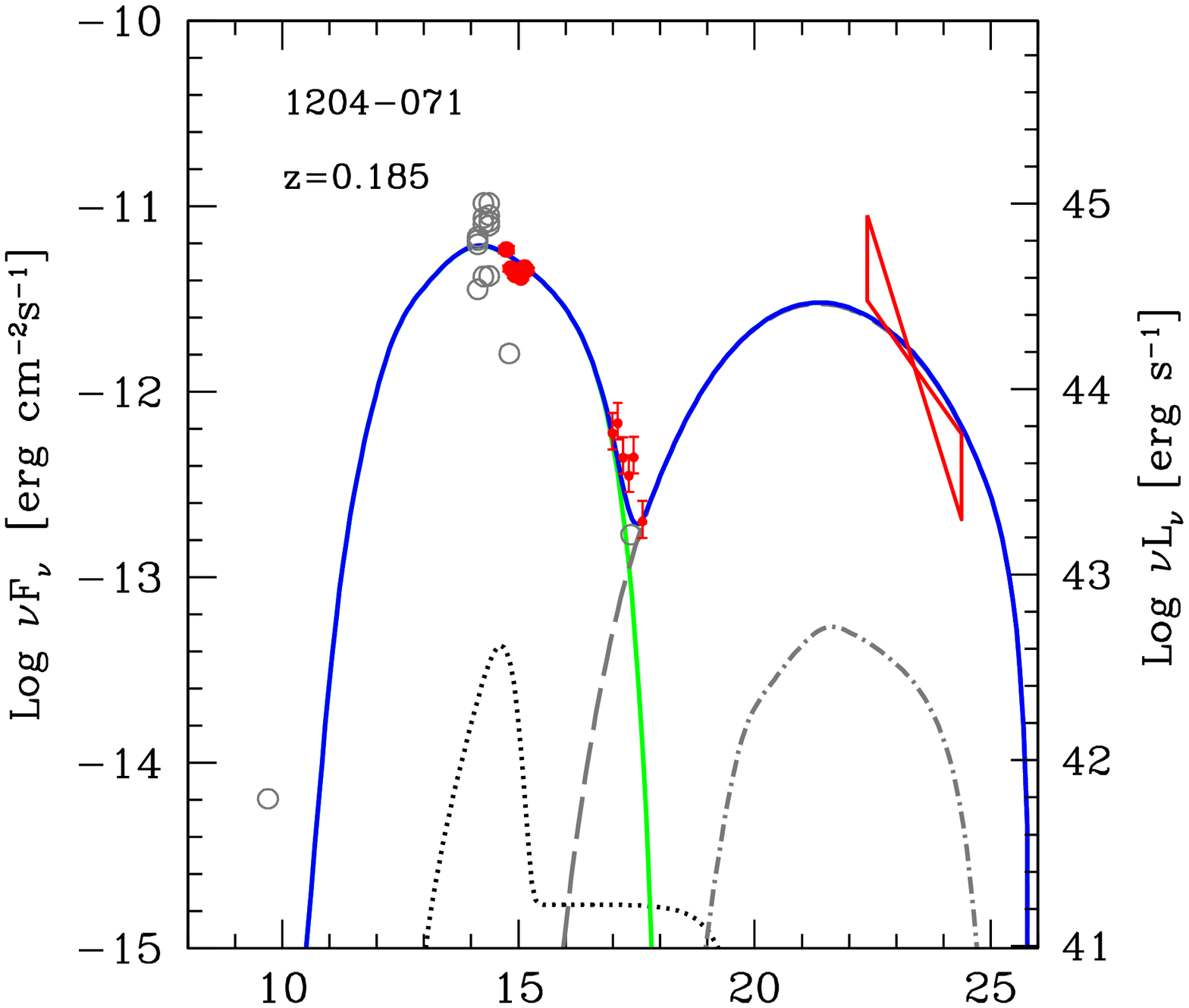,width=9cm,height=6.7cm}
\vskip -1.3 cm \hskip -0.4 cm
\psfig{figure=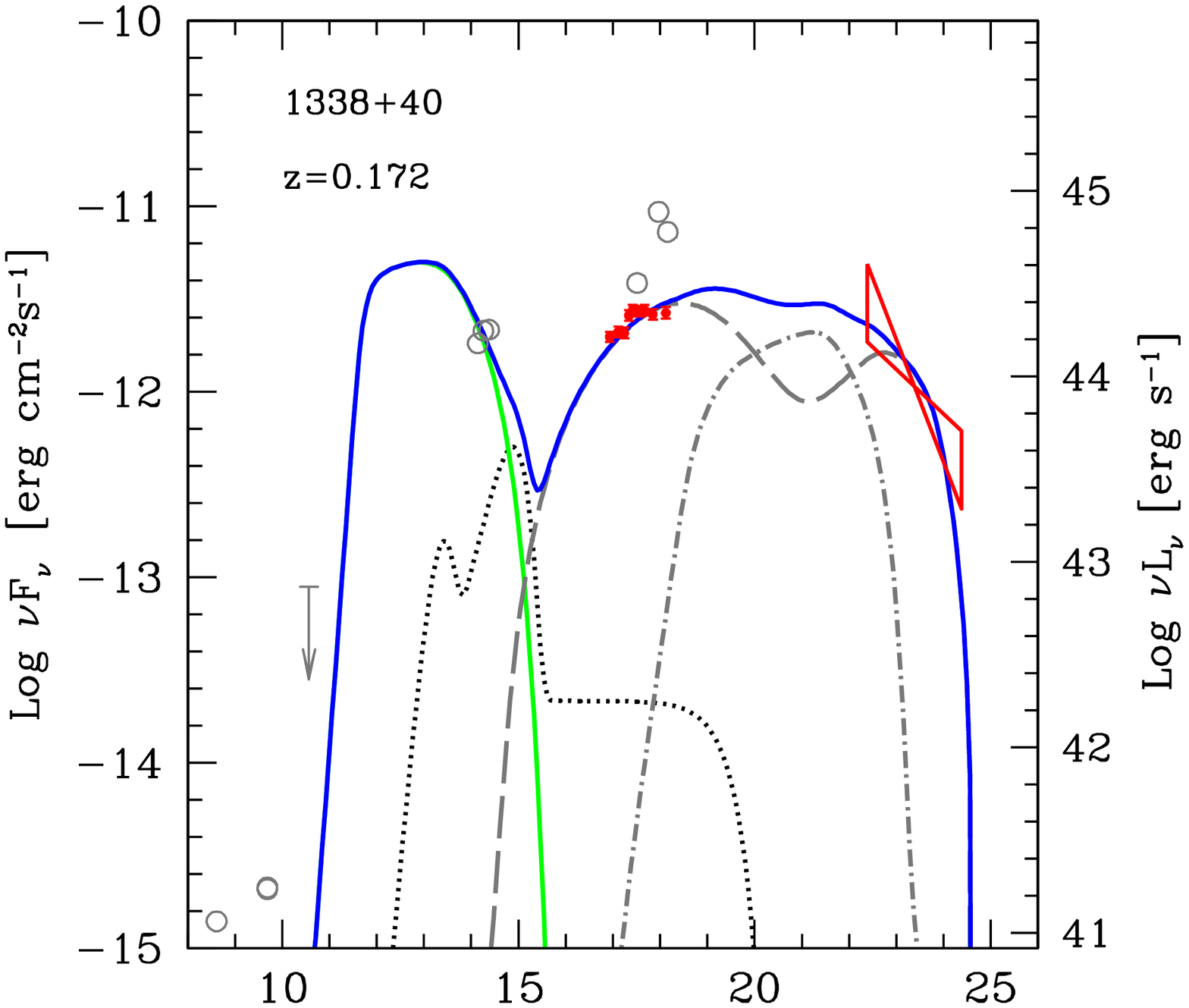,width=9cm,height=6.7cm}
\vskip -1.3 cm \hskip -0.4 cm
\psfig{figure=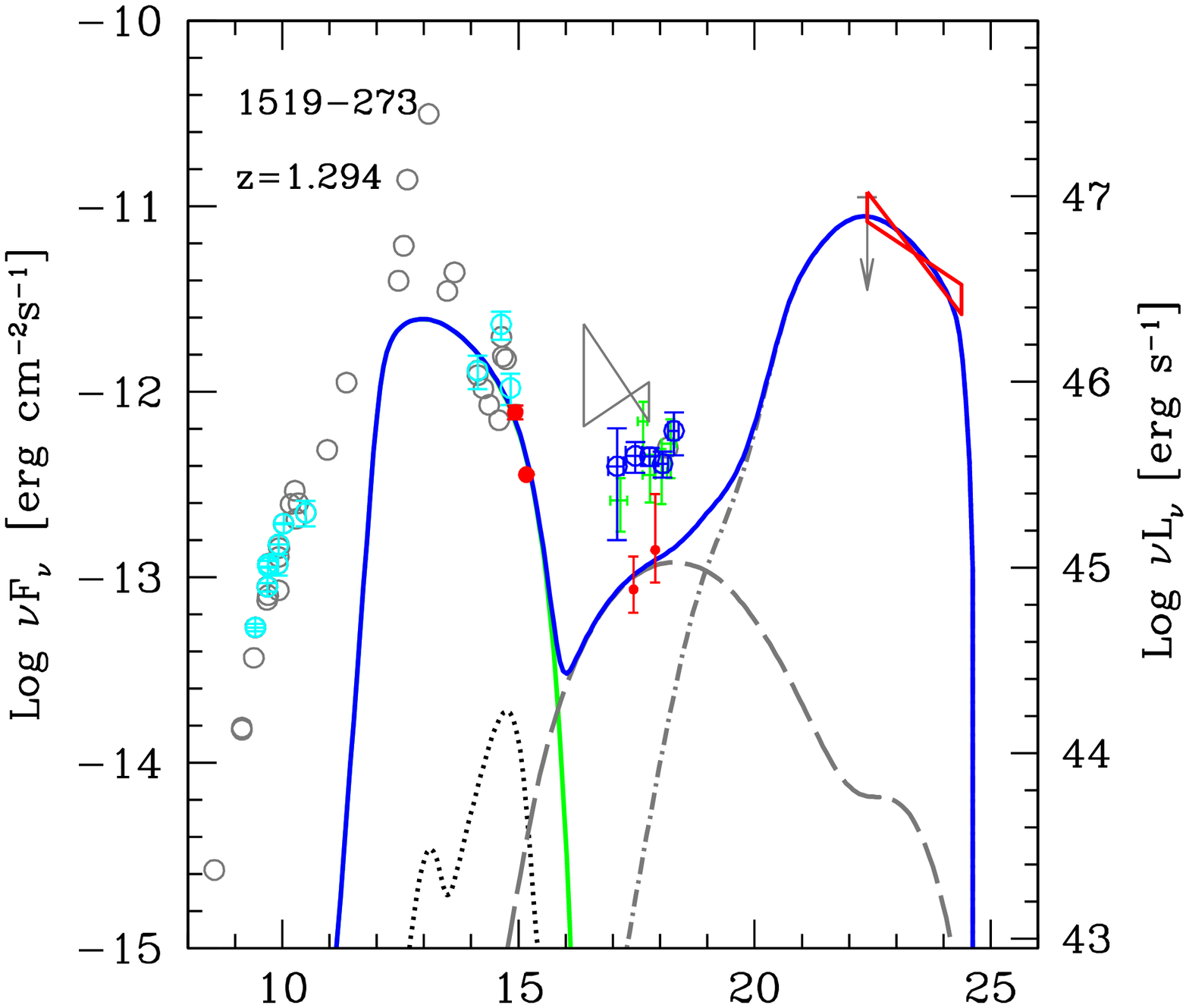,width=9cm,height=6.7cm}
\vskip -1.3 cm \hskip -0.4 cm
\psfig{figure=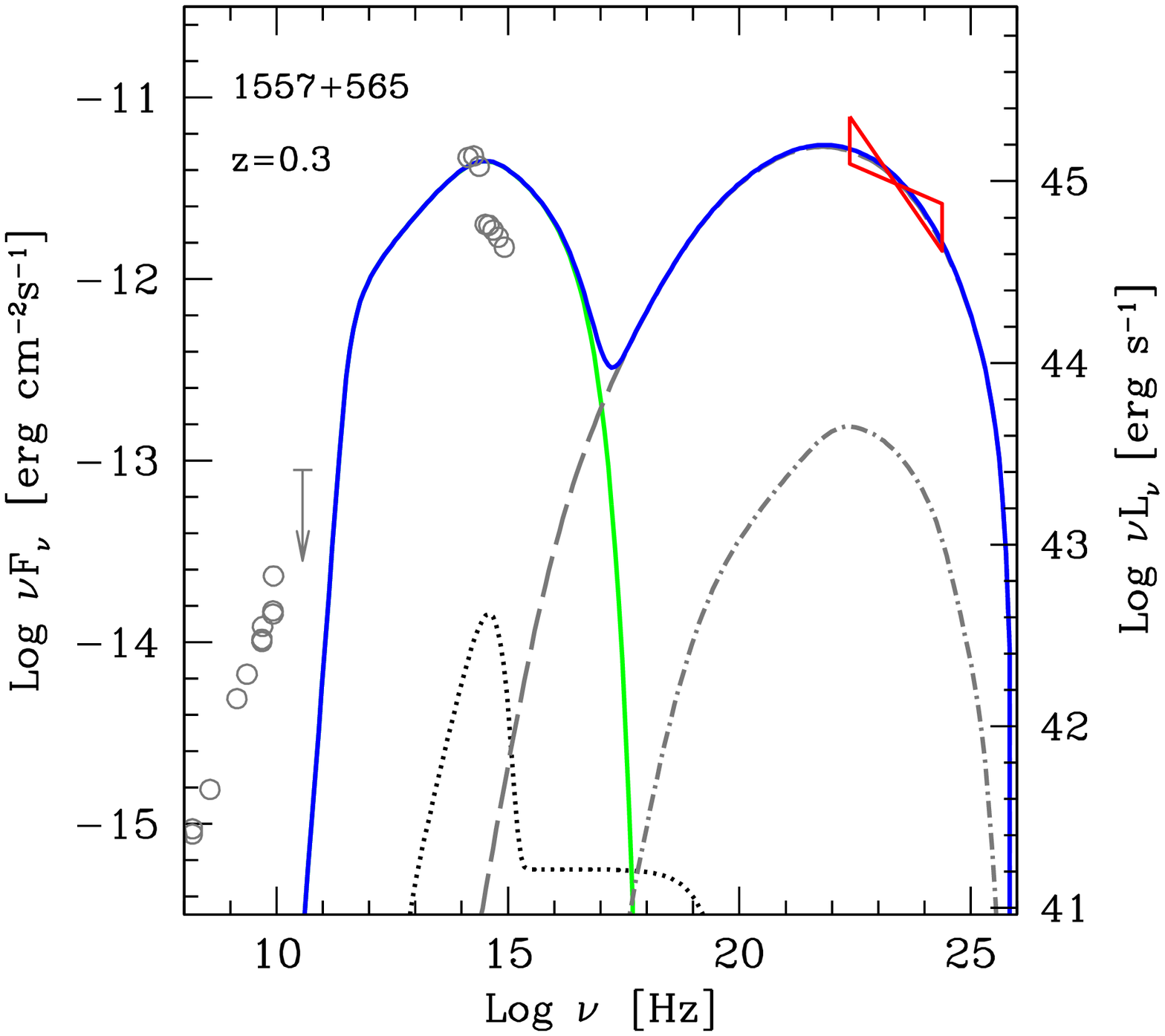,width=9cm,height=6.7cm}
\vskip -0.8 cm 
\caption{
SED of 1204--071, B2 1338+40, PKS 1519--273 and 1557+565.
Symbols and lines as in Fig. \ref{f1}.
}
\label{f5}
\end{figure}

\begin{figure}
\vskip -0.6cm \hskip -0.4 cm
\psfig{figure=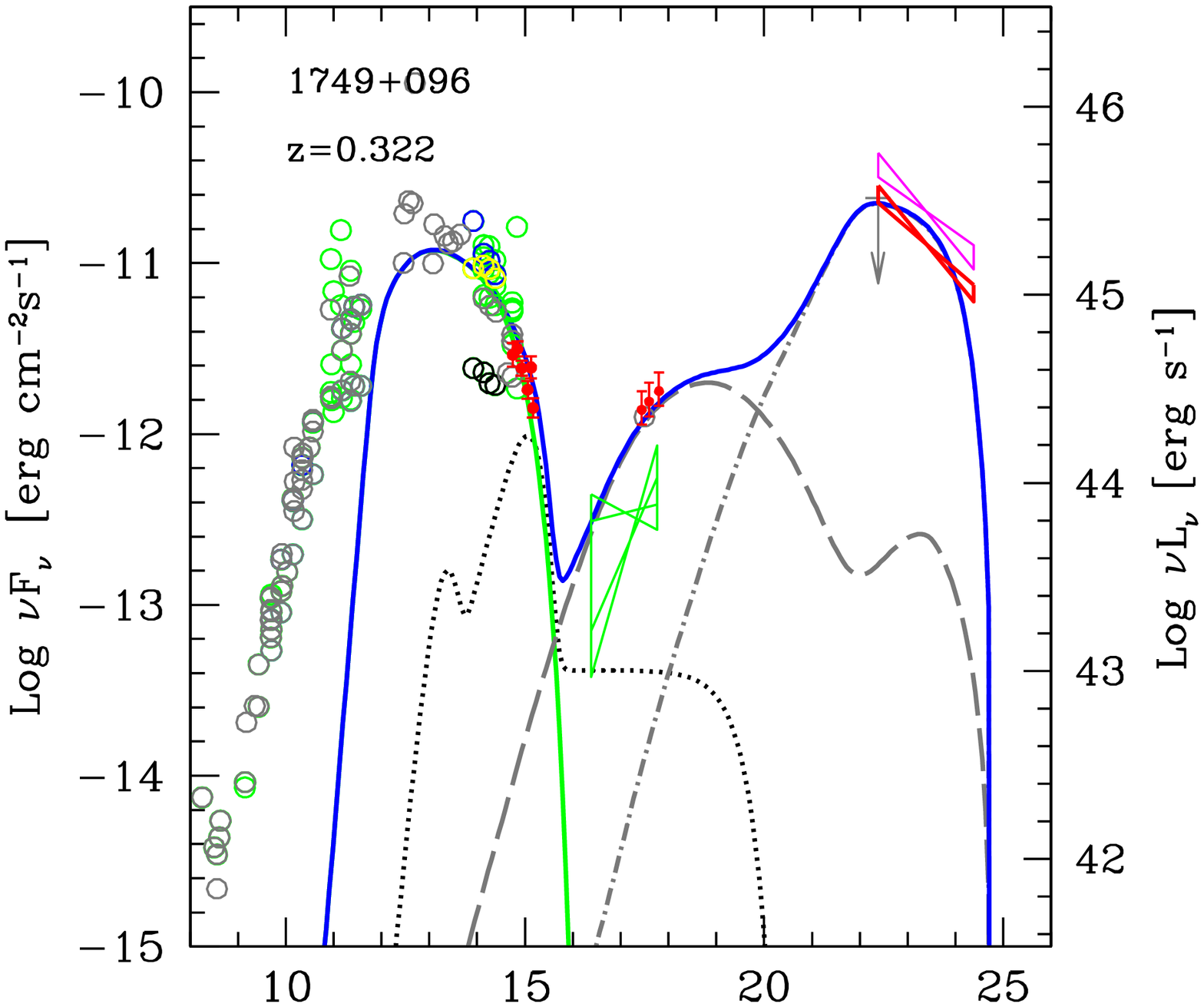,width=9cm,height=6.7cm}
\vskip -1.3 cm \hskip -0.4 cm
\psfig{figure=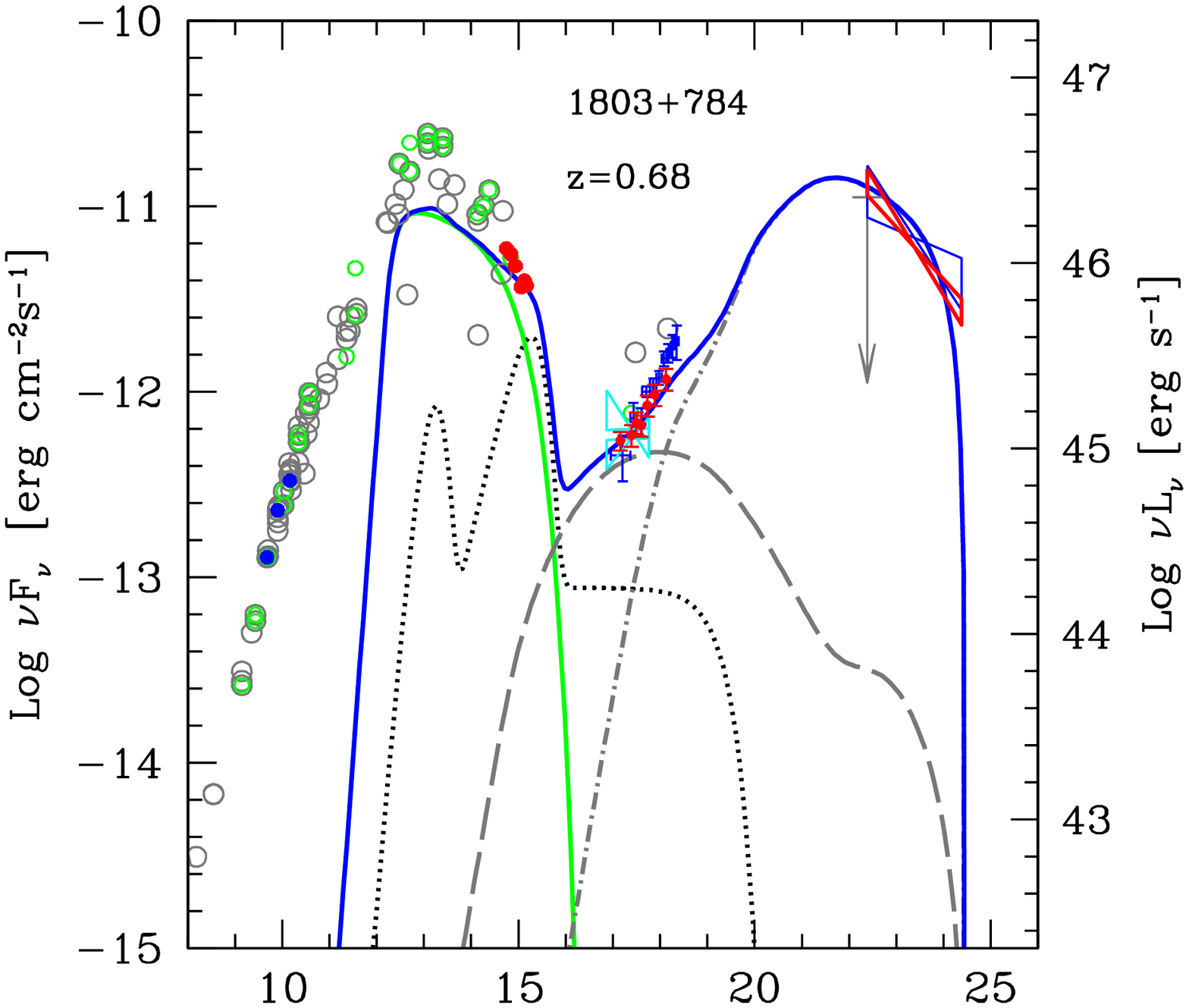,width=9cm,height=6.7cm}
\vskip -1.3 cm \hskip -0.4 cm
\psfig{figure=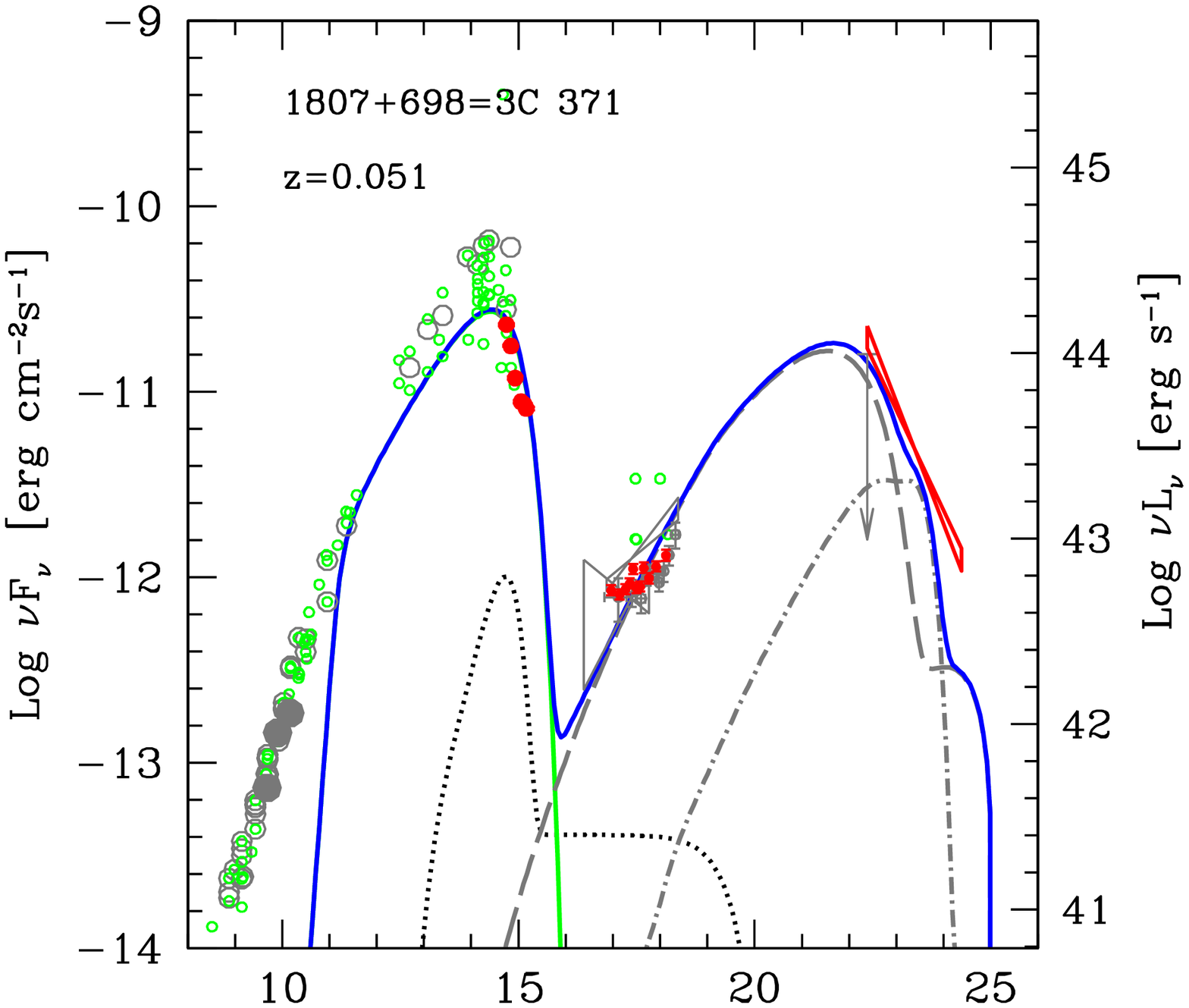,width=9cm,height=6.7cm}
\vskip -1.3 cm \hskip -0.4 cm
\psfig{figure=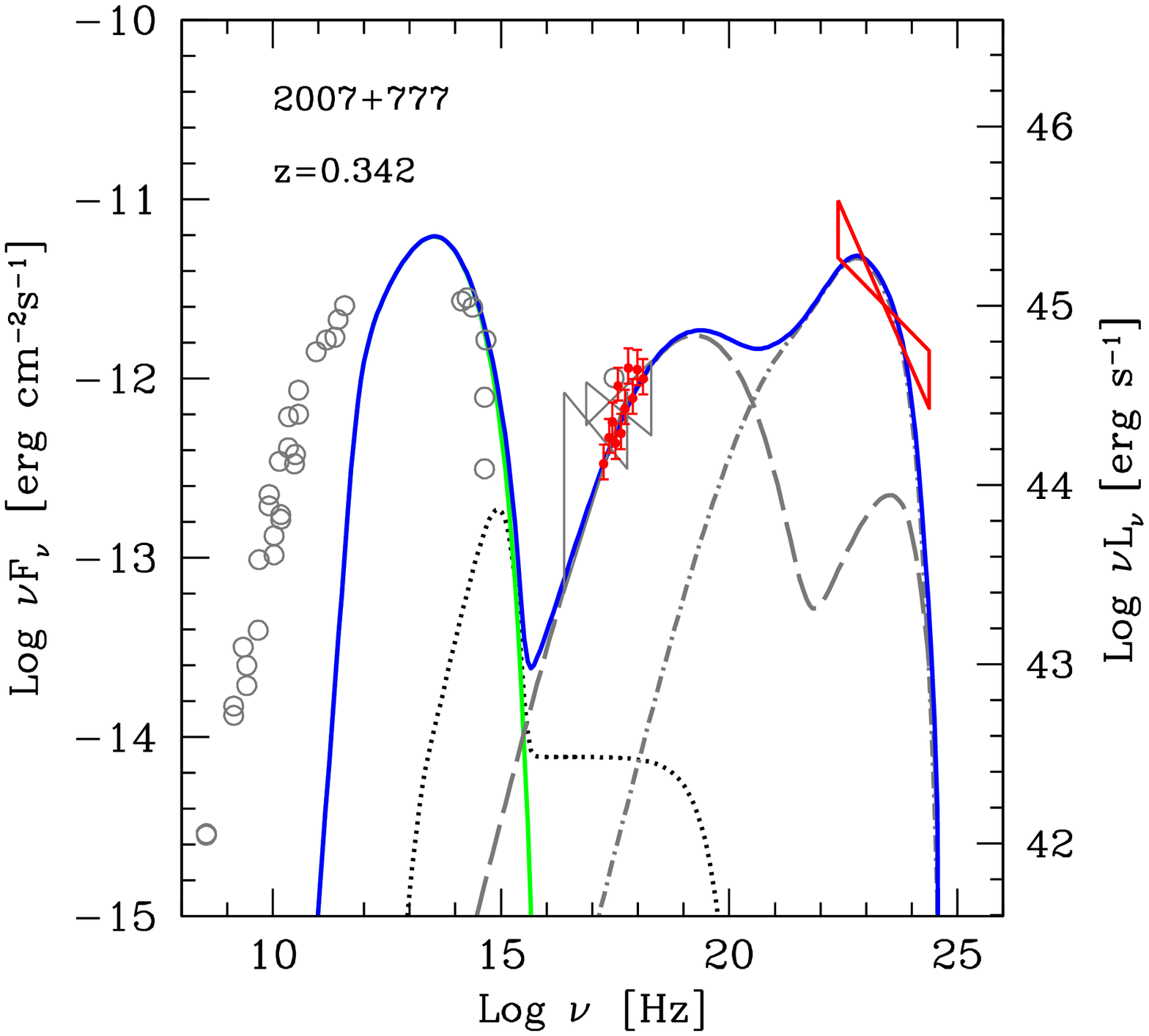,width=9cm,height=6.7cm}
\vskip -0.8 cm
\caption{
SED of PKS 1749+096, S5 1803+78, 1897+698 (=3C 371) and S5 2007+77.
Symbols and lines as in Fig. \ref{f1}.
}
\label{f6}
\end{figure}

\begin{figure}
\vskip -0.6cm \hskip -0.4 cm
\psfig{figure=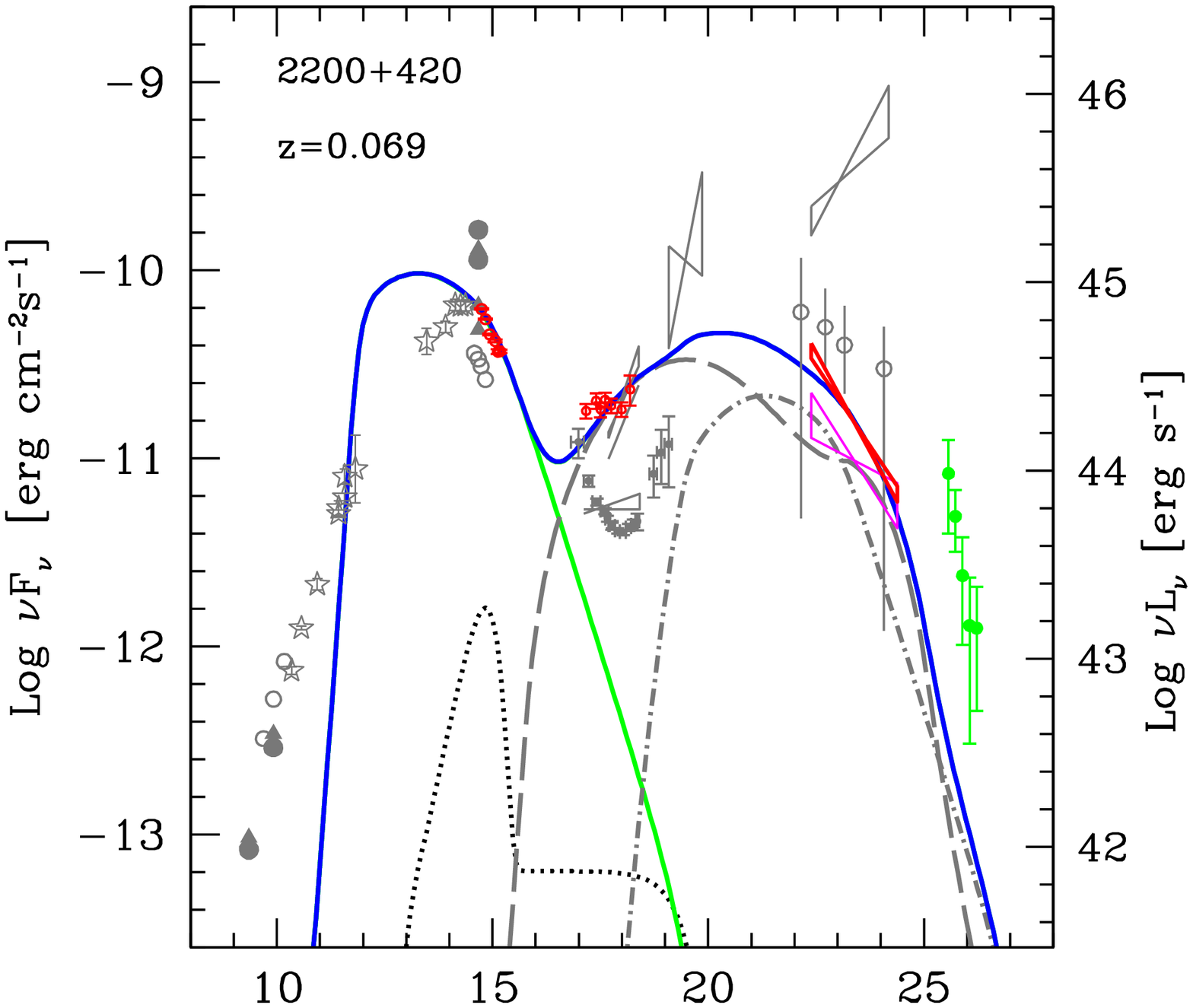,width=9cm,height=6.7cm}
\vskip -1.3 cm \hskip -0.4 cm
\psfig{figure=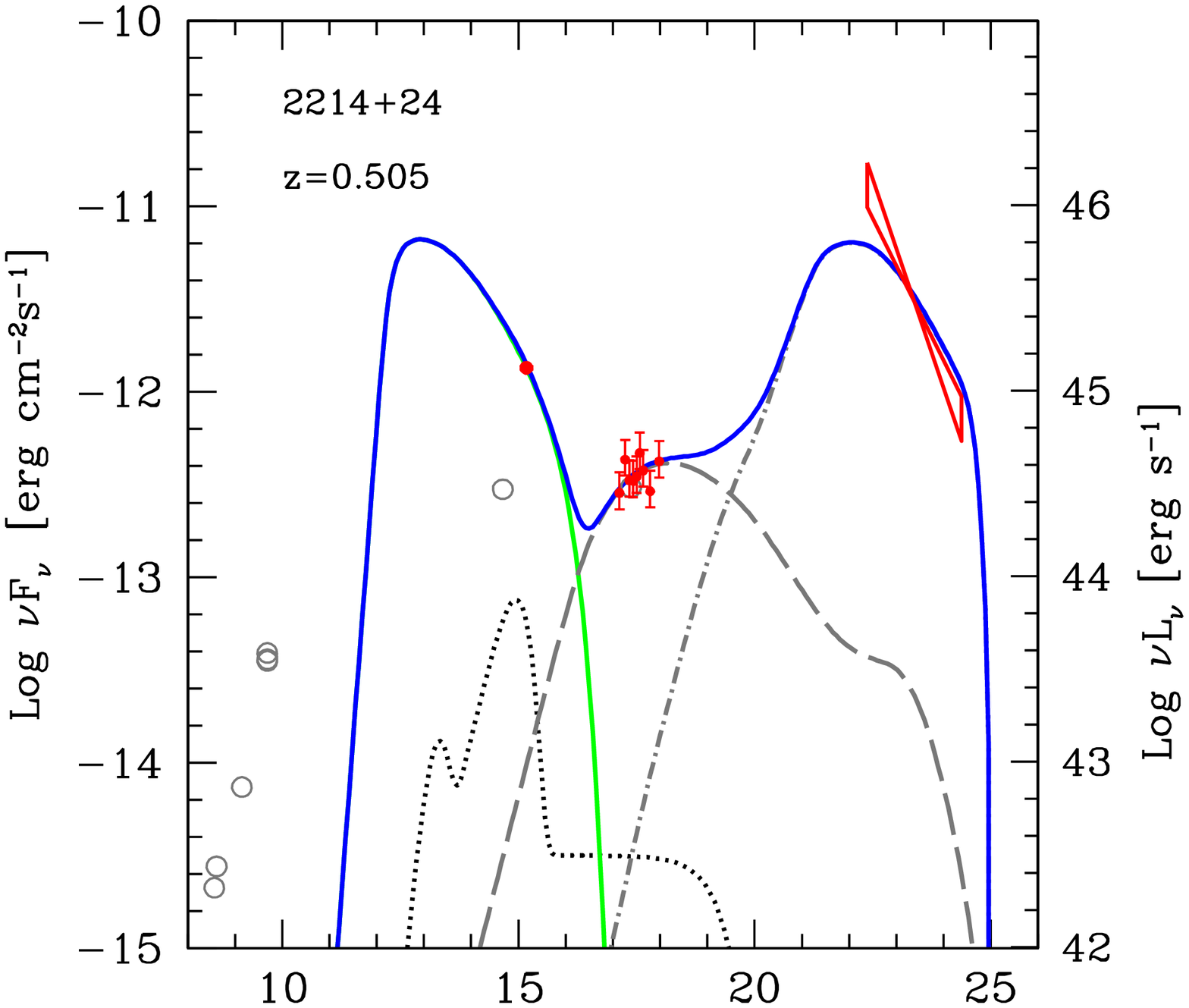,width=9cm,height=6.7cm}
\vskip -1.3 cm \hskip -0.4 cm
\psfig{figure=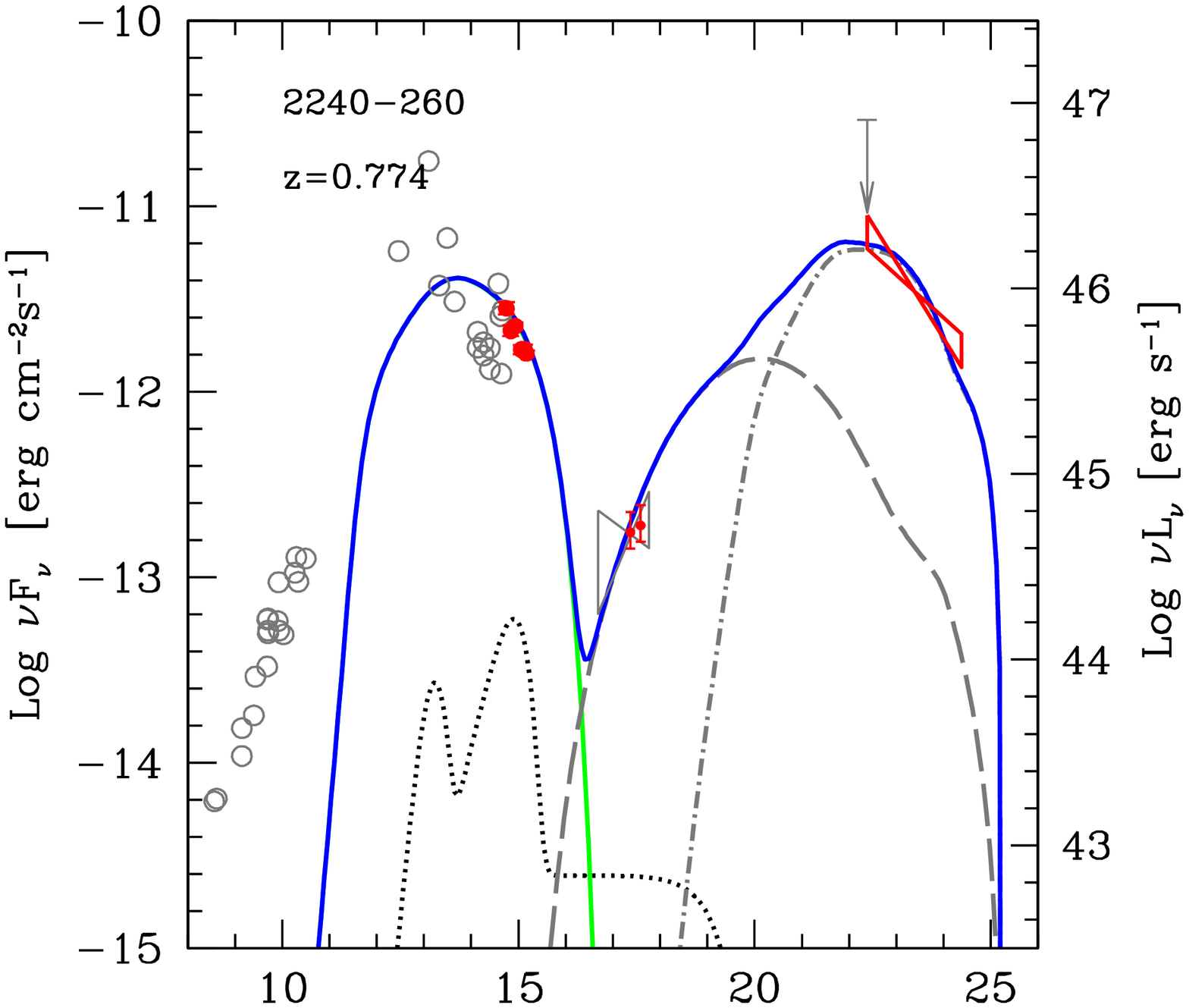,width=9cm,height=6.7cm}
\vskip -1.3 cm \hskip -0.4 cm
\psfig{figure=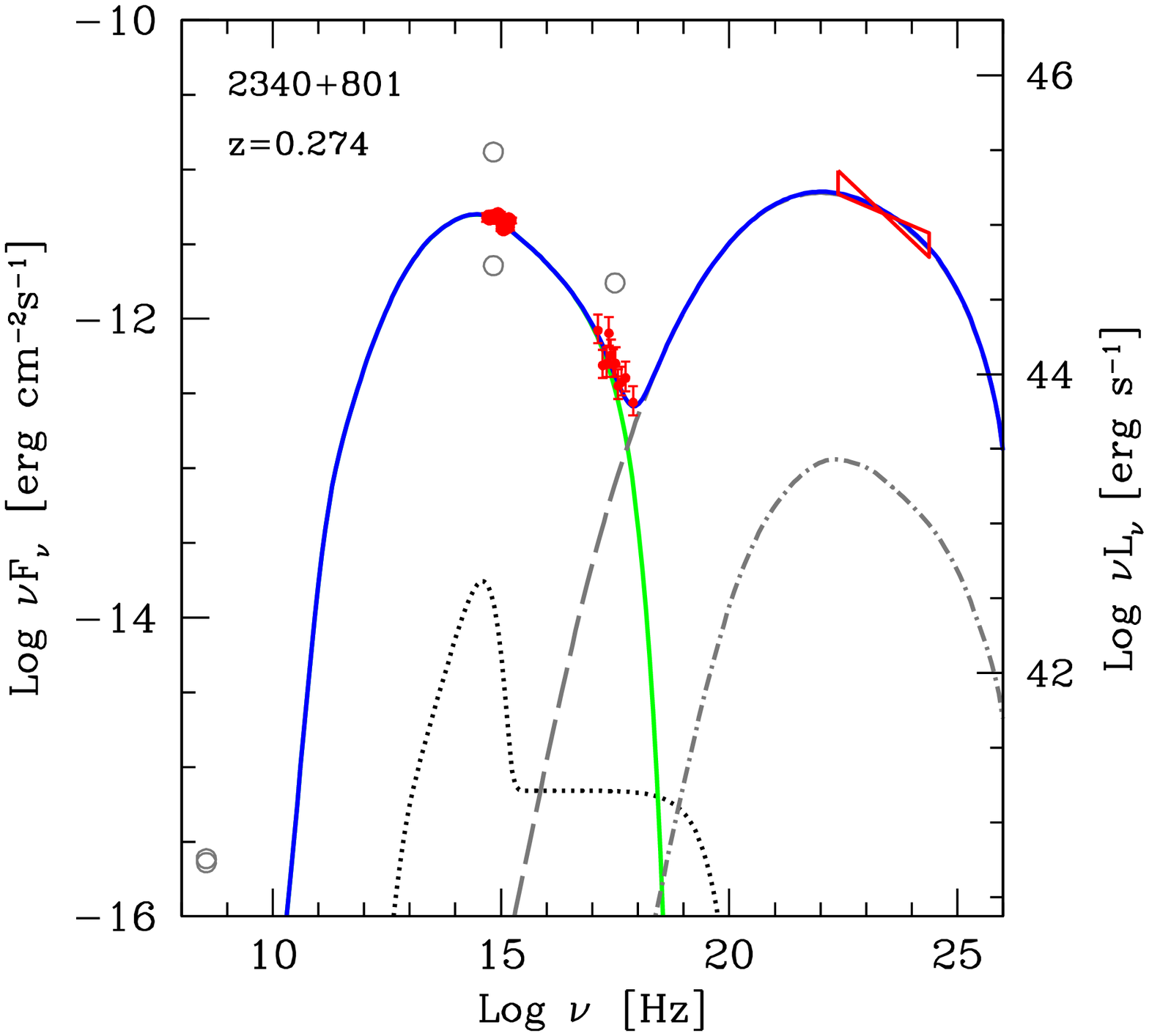,width=9cm,height=6.7cm}
\vskip -0.8 cm
\caption{
SED of 2200+420 (=BL Lac), B2 2214+24, PKS 2240--260 and 2340+8015.
Symbols and lines as in Fig. \ref{f1}.
}
\label{f7}
\end{figure}

\end{document}